\documentclass[pdflatex,sn-mathphys-num]{sn-jnl}

\usepackage{graphicx}%
\usepackage{multirow}%
\usepackage{amsmath,amssymb,amsfonts}%
\usepackage{amsthm}%
\usepackage{mathrsfs}%
\usepackage[title]{appendix}%
\usepackage{xcolor}%
\usepackage{textcomp}%
\usepackage{manyfoot}%
\usepackage{booktabs}%
\usepackage{algorithm}%
\usepackage{algorithmicx}%
\usepackage{algpseudocode}%
\usepackage{listings}%
\usepackage{empheq}

\usepackage{fix-cm}
\usepackage{stackengine}
\usepackage{array} 
\usepackage{epsfig}
\usepackage{graphics}
\usepackage{colortbl}
\usepackage{hhline}
\usepackage{ulem}
\usepackage{tikz}

\makeatletter
\newcommand{\dvec@raise}{-0.1ex} 
\newcommand{\dvec@sep}{-0.5ex}   

\newcommand{\dvec}[1]{\mathpalette\dvec@i{#1}}
\def\dvec@i#1#2{%
  \ooalign{%
    $\m@th#1#2$\cr
    \noalign{\kern\dvec@raise}%
    $\m@th#1\vec{\phantom{#2}}$\cr
    \noalign{\kern\dvec@sep}%
    $\m@th#1\vec{\phantom{#2}}$\cr
  }%
}
\makeatother

\makeatletter
\newcommand{\ddvec@raise}{-0.1ex} 
\newcommand{\ddvec@sep}{-0.4ex}   

\newcommand{\ddvec}[1]{\mathpalette\ddvec@i{#1}}
\def\ddvec@i#1#2{%
  \ooalign{%
    $\m@th#1#2$\cr
    \noalign{\kern\ddvec@raise}%
    $\m@th#1\vec{\phantom{#2}}$\cr
    \noalign{\kern\ddvec@sep}%
    $\m@th#1\vec{\phantom{#2}}$\cr
  }%
}
\makeatother

\newcommand{\cev}[1]{\reflectbox{\ensuremath{\vec{\mkern2mu\reflectbox{\ensuremath{#1}}}}}}

\theoremstyle{thmstyleone}%
\theoremstyle{thmstyletwo}%

\theoremstyle{thmstylethree}%

\begin{document}
	
\title[Covariant reggeization framework for diffraction. Part I: Hadronic tensors in Minkovsky space-time of any dimension]{Covariant reggeization framework for diffraction. Part I: Hadronic tensors in Minkovsky space-time of any dimension}


\author[1]{\fnm{Roman A.} \sur{Ryutin}}\email{roman.riutin@gmail.com}

\affil[1]{\orgdiv{Theoretical Department}, \orgname{
		Institute for High Energy Physics,\\ NRC ``Kurchatov Institute''},
		\orgaddress{\street{Pobeda sq.,1}, \city{Protvino}, \postcode{142281},\\ \state{Moscow reg.}, \country{Russia}}}


\abstract{In this paper we consider the general structure of 
	irreducible tensor representations of the Poincar\'e group of arbitrary 
	space-time dimension $D$
	with multiple sets of Lorentz indices and different ways
	to construct them from basic elements (Lorentz vectors and the metric 
	tensor). Then we apply the same methods to obtain the expansion
	of general hadronic tensors	in terms of these 
	irreducible tensors. We propose to use an effective approach
	in hadronic diffraction, which was usually called covariant reggeization, and
	obtain basic functions and tensors to calculate all the diffractive cross-sections.}





\maketitle	
	


\section*{Introduction}

In this article we propose the generalization of an approach presented in some previous works~\cite{mySpinParity}-\cite{myVisualization}. Since classical papers and books~\cite{Gribov}-\cite{CollinsBook} had been published, some authors considered the reggeization in terms of irreducible tensors~\cite{CovariantReggeIn}-\cite{CovariantReggeOut} (especially tensor part of the particle propagator of any spin) instead of the usual Regge expansion in terms of Legendre polynomials. 

This work was motivated by several facts:
\begin{itemize}
	\item
	First, study of the spin-parity analysis in the exclusive 
	central diffraction~\cite{mySpinParity},\cite{KMRspinparity}, where 
	the covariant approach yields strictly defined cross-section 
	behavior for fixed spins. However, since 
	calculations for arbitrary spins, as they 
	should be in principle, are quite cumbersome, many 
	authors (see~\cite{Nachtmann} for example) mix 
	integer spins and trajectories in their approach. For 
	high energies this is
	less critical and may give similar results, but for low and 
	intermediate energies significant discrepancies are possible.
	
	\item
	Secondly, the direct application of this approach
	leads to zeros in the cross sections in the limit 
	of small  momentum transfer~\cite{mySDandPomCS}, which
	was demonstrated quite a long time ago for 
	conserved currents~\cite{CloseIn}-\cite{CloseOut}. This is
	particularly visible in single dissociation
	and central production. However, experimental 
	data do not show such a behaviour. Of course, reggeon 
	is a rather complex object, and the requirement
	for current conservation is likely unnecessarily
	strong. However, this behavior near zero momentum 
	transfer squared can be eliminated in various ways, so that the model 
	could be used for data analysis. It was shown 
	that unitarization can smooth out this 
	behaviour~\cite{mySDandPomCS}. Another solution 
	is to assume that the 
	current is conserved in more 
	dimensions; then for $D=4$ it will 
	be effectively non-conserved.
	
	\item
	Third, we would like to have a clear
	understanding of a reggeon from a 
	quantum field theory perspective, and 
	what the obtained cross sections for 
	its scattering by hadrons mean. We can 
	conditionally consider it analogous to an 
	atom with its energy levels and scattering, or
	a string, or simply a mathematical 
	object corresponding to a mixture of 
	various processes. Or we can consider 
	it, among other things, as a reggeized 
	quantum field of arbitrary spin (see 
	section~\ref{section1}). In 
	the last case it leads to rigorous, rather beautiful 
	mathematical consequences, which can 
	be tested experimentally to determine 
	the extent to which they are fulfilled or violated.
	
	\item
	The next question to be solved is whether 
	we can extract reggeon--hadron and 
	reggeon--reggeon cross sections 
	from data, which was made first by 
	Kaidalov~\cite{Kaidalov}, and what is 
	the physical meaning of these cross 
	sections, which can be of the order 
	of hadronic cross sections, as was
	shown in the previous paper~\cite{mySDandPomCS}.
\end{itemize}
   
 In this paper reggeon is assumed to arise from currents of any spin 
 in Minkovsky space-time $\mathbb{M}_D=\mathbb{R}^{1,D-1}$, which are 
 totally symmetric in Lorentz indices (to avoid complications due to tensors with mixed symmetry arising in dimensions greater than four). This simple model should be further extended to more realistic settings (for example, AdS space and so on).
    
The task of this work is to show how we can deal with different irreducible tensor representations of the Poincare group with multiple groups of Lorentz indices, i.e. with different 
multi-reggeon vertexes, and also
with expansions of any Lorentz symmetric tensor in terms of these vertexes. We 
then show how to use the proposed method
to construct the amplitudes of all diffractive processes, using the resulting 
tensor structures as basic elements like in a Lego constructor.

To generalize covariant regggeization, we could use also so called "continuous spin 
representations"~\cite{CSPin}-\cite{CSPout}, since it may have some advantages 
in the case of curved spaces (see, for example~\cite{AdS1}-\cite{conformal2}).

Of course, we can also use the helicity~\cite{helicityIn}-\cite{helicityOut} or the spin-tensor~\cite{HSP5}-\cite{HSP9} formalisms for particles with higher spins, which may simplify some calculations, but in this case the tensor formalism is more clear and conceptually understandable approach.

The method described here can be used to obtain the general form of differential cross sections not only for diffractive processes. We limit ourselves to hadron diffraction, since it is currently becoming an increasingly popular tool for studying the structure of hadrons at modern high-energy accelerators. This is the main part of the so called "forward physics" (i.e. we detect hadrons close to the beam direction). Here is the general list of all diffractive processes
and their abbreviations, which are often 
presented in the literature, and 
for which we calculate 
the cross-sections from general principles 
of quantum field theory (QFT):
\begin{figure}[t!]
	\centering
	\includegraphics[width=0.6\textwidth]{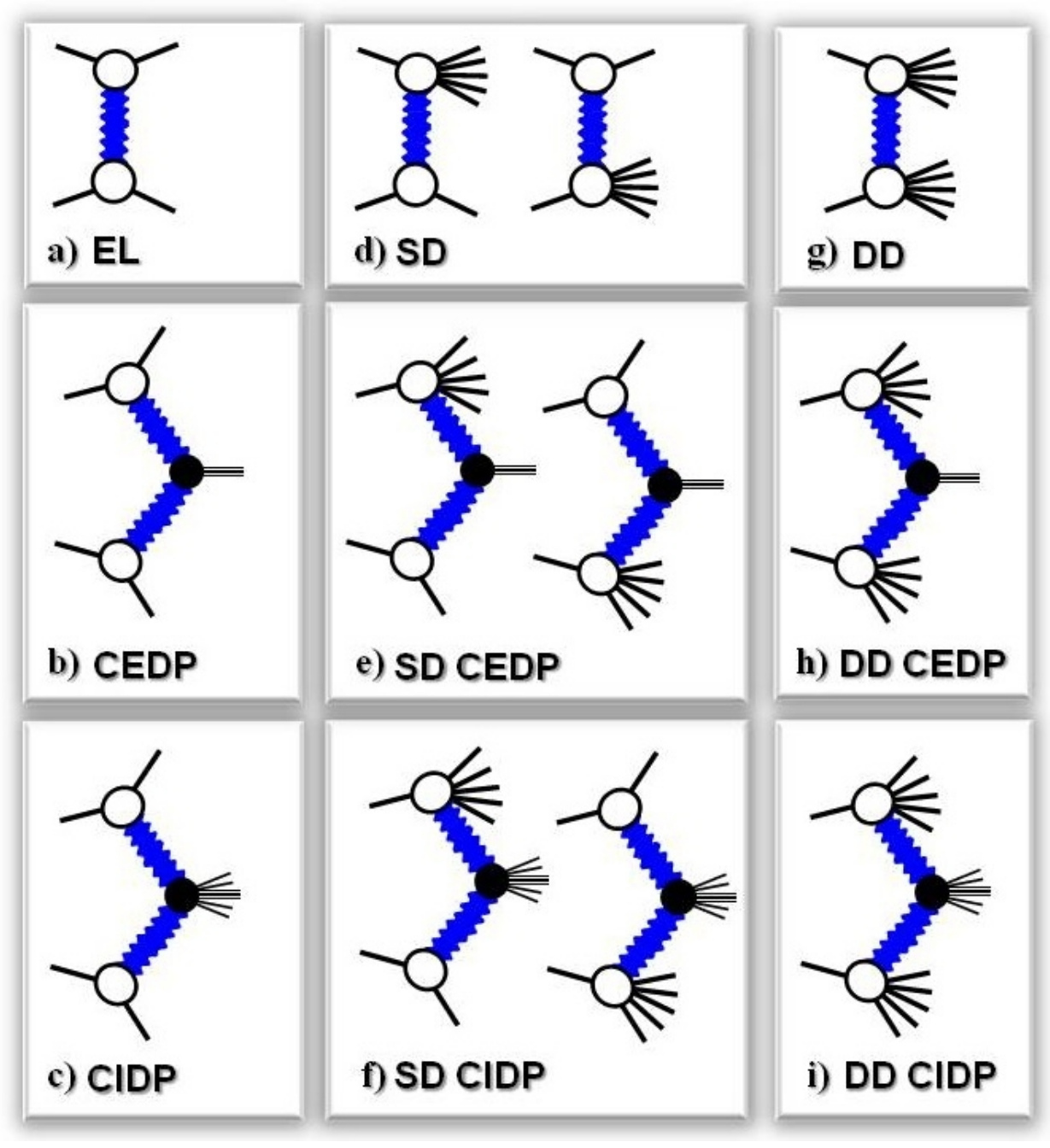}
	\caption{\label{fig:alldiffprocs} Diffractive processes}
\end{figure}
\begin{itemize}
\renewcommand\labelitemi{$\bullet$}
\item both initial hadrons remain intact (forward hadrons):
\begin{itemize} 
\item \textbf{EL}: elastic scattering, $h_1+h_2\rightarrow h_1+h_2$ (Fig.~\ref{fig:alldiffprocs}a);
\item \textbf{CEDP}: central exclusive diffractive production (Fig.~\ref{fig:alldiffprocs}b),\\ $h_1+h_2\rightarrow h_1+\rm{LRG}+\rm{X}_c+\rm{LRG}+h_2$;
\item \textbf{CIDP}: central inclusive diffractive production (Fig.~\ref{fig:alldiffprocs}c),\\
 $h_1+h_2\rightarrow h_1+\rm{LRG}+\rm{everything}+\rm{LRG}+h_2$;
\end{itemize}
\item single initial hadron remains intact (forward hadron), other one dissociates:
\begin{itemize}
\item \textbf{SD}: single diffractive dissociation (Fig.~\ref{fig:alldiffprocs}d),\\ $h_1+h_2\rightarrow h_1+\rm{LRG}+\rm{X}_D$; 
\item \textbf{SD CEDP}: central exclusive diffractive production with single dissociation (Fig.~\ref{fig:alldiffprocs}e),\\
$h_1+h_2\rightarrow h_1+\rm{LRG}+\rm{X}_c+\rm{LRG}+\rm{X}_D$;
\item \textbf{SD CIDP}: central inclusive production with single dissociation (Fig.~\ref{fig:alldiffprocs}f),\\
$h_1+h_2\rightarrow h_1+\rm{LRG}+\rm{everything}+\rm{LRG}+\rm{X}_D$;
\end{itemize}
\item both initial hadrons dissociate:
\begin{itemize}
\item \textbf{DD}:  double diffractive dissociation (Fig.~\ref{fig:alldiffprocs}g),\\
$h_1+h_2\rightarrow \rm{X}_D+\rm{LRG}+\rm{Y}_D$;
\item \textbf{DD CEDP}: central exclusive diffractive production with double dissociation (Fig.~\ref{fig:alldiffprocs}h)\\
$h_1+h_2\rightarrow \rm{X}_D+\rm{LRG}+\rm{X}_c+\rm{LRG}+\rm{Y}_D$;
\item \textbf{DD CIDP}: central inclusive production with double dissociation (Fig.~\ref{fig:alldiffprocs}i),\\
$h_1+h_2\rightarrow \rm{X}_D+\rm{LRG}+\rm{everything}+\rm{LRG}+\rm{Y}_D$.
\end{itemize}
\end{itemize}
Here LRG means "large rapidity gap", $\rm{X}_c$ can be a single particle or small number of particles or several 
jets, for example. $\rm{X}_D$ and $\rm{Y}_D$ 
denote dissociation products. EL, SD, DD are single reggeon exchange processes (with one LRG), and others are double reggeon exchange 
ones (two LRGs), as you can see often in artcles of other authors. We 
will use the above abbreviations in this paper. All the processes are depicted 
in the figure~\ref{fig:alldiffprocs}.

This paper is organized as follows.
\begin{itemize}
\renewcommand\labelitemi{$\bullet$}
\item In section~\ref{section:notations} we explain in detail basic notations.

\item In section~\ref{section1} we consider basic statements of the covariant reggeization framework.

\item In section~\ref{section:ampset} we consider all the irreducible tensors for diffractive processes.

\item In section~\ref{section4} we write expansions of reducible tensors in terms of irreducible ones. And also express amplitudes and
cross-sections for diffractive processes in terms of such quantities.

\item In section~\ref{section:recurrences} we present all the recurrent equations for 
coefficients of irreducible tensors and also their solutions. 

\item In discussions and conclusions we propose further program of the investigation and analyse
the results of this paper.

\item In Appendix~\ref{appendixA} we introduce notations and properties for basic TST structures.

\item  In Appendix~\ref{appendixB} we introduce TS (transverse and symmetric) structures, explain different ways 
to obtain recurrent equations for irreducible tensors, and also propose algorithm for their solution.



\item  In Appendix~\ref{appendixC} we obtain coefficients in the expansions of any symmetric tensor in terms of irreducible ones. 

\item  In Appendix~\ref{appendixD} we give combinatorial relations and rules for generating function method.

\item  In Appendix~\ref{appendixE} we give some examples of irreducible tensors for fixed spins.
\end{itemize}

This work may also have some pedagogical applications, since several useful 
techniques and different approaches are used for calculations to obtain 
final expressions for tensors. Since this paper deals only with mathematical calculations, we 
will not dwell in detail on the experimental aspects of diffractive physics, leaving this for future publications.

\section{Summary of notations}
\label{section:notations}

Let us introduce some basic notations that we use in this paper to avoid misunderstandings in next sections:
\begin{enumerate}
\item Everywhere in the text $D$ denotes the dimension of Minkovsky space-time $\mathbb{M}_D$ and $$g_{\mu\nu} = \mathrm{diag} \{1,-1,-1,...,-1\}$$ is the diagonal 
metric tensor of this space-time.
	
\item For any momentum $v$ in the Minkovsky space-time $\mathbb{M}_D=\mathbb{R}^{(1,D-1)}$ we use general notations:
\begin{equation}
v=\{ v_0,v_1,...,v_{D-1}\},\quad v^2=g_{\mu\nu} v_{\mu} v_{\nu},\quad v\in \mathbb{M}_D
\label{notationsmomenta}
\end{equation}
If $v$ is space-like, for example $v=q_i$ (momentum transfer), then $q_i^2=-Q_i^2$. In other case, when $v$ is time-like, for example $v=p_i$ (real 
particle), then $p_i^2=m_i^2$. 

\item 
When we consider any system of real particles in the final state, we use $M$ letter for its momentum squared, for
example, the mass of the central system $M_c^2=(q_1+q_2)^2$  or the mass of dissociative state $M_{SD}^2=(p+q)^2$. 
 
\item $(\alpha)_{J_i}$ denotes symmetric sets of $J_i$ indexes like 
\begin{eqnarray}
&& (\alpha)_{J_i}\equiv(\alpha_1\alpha_2...\alpha_{J_i}),\nonumber\\
&& \alpha \in\{ \mu,\mu',\nu,\nu',\rho\} \leftrightarrow i \in\{1,1',2,2',3 \}.
\end{eqnarray}
Here we fix the direct correspondence between greek Lorentz indices $\mu,\mu',\nu,\nu',\rho$ 
and the group numbers.  Everywhere $(1),(2),(1'),(2'),(3)$ denote some arbitrary Lorentz indices from the corresponding symmetric 
groups. i. e.
\begin{eqnarray}
q_{(2)}\equiv q_{\nu_i},\quad q_{1(1)}&\equiv& q_{1\mu_i},\quad q_{1(1')}\equiv q_{1\mu'_i},\quad p_{2(2')}\equiv p_{2\nu'_i},
\nonumber\\
T_{(1)}&\equiv& T_{\mu_i},\quad\hspace*{1.5mm} T_{(11)} \equiv T_{\mu_i\mu_j}, 
\nonumber\\
T_{(1)(2')} &\equiv& T_{\mu_i\nu'_j},\quad\!\!\!  T_{(2)(3)} \equiv T_{\nu_i\rho_j},
\nonumber
\end{eqnarray}
and so on, i.e. we take the tensor with some indexes from arbitrary group of indexes. For general tensors with
multiple indexes, for example $H^{\vec{J}}$, groups of indices are indicated by components of $\vec{J}$ (see below). For basic tensors
we have their own definitions~(\ref{def:Grr})-(\ref{def:GHarmrs}) according to the each number of a group. If we have $T_{(\mu)_{J_1},(\nu)_{J_2}}$, symmetrization is applied
only inside each group of indexes.

\item For convenience below we will assume everywhere that the following rules
for numbers that mark index groups in amplitudes with two and four tensor legs are fulfilled:
\begin{equation}
 i\in\{1,1',2,2'\},\qquad 
 (i')'=i,\; (i^{\star})^{\star}=i,\qquad
 1^{\star}=2,\;2^{\star}=1
\label{eq:ijsymbols}
\end{equation}
i. e. $\{i,i'\}$ and $\{i^{\star},{i^{\star}}'\}$
are sets of indexes for different momenta transfers $q_{i}$ and $q_{i^{\star}}$. 
 
 \item
For momenta transfers and indexes we have:
\begin{equation}
 q_{i'(i)}=-q_{i(i)},\quad
 i\in\{1,1',2,2'\}\label{eq:qmomenta}
\end{equation}

\item
When we use the index $\alpha,\beta$, it means that they can be replaced
by one of the indexes $\mu,\mu',\nu,\nu',\rho$ depending on the context, 
and if we use both $\alpha$ and $\beta$ in the same tensor,
it means that they belong to different index groups. 

\item 
When we use momenta $q$ or $p$ without indexes, it means that they can be replaced with the 
corresponding momenta. Depending on the situation, $q$ is replaced by 
momentum transfer and $p$ is replaced by the momentum of the particle in the same vertex. 

\item For summation and free indexes we usually use letters $i$, $j$, $m$, $r$, $s$, $a$, $b$ and
the same letters with additional symbols.

\item Frequently used functions:
\begin{equation}
\theta(x)=\begin{cases} 1, & x\ge 0\\ 0, & x < 0\end{cases}
\label{def:thetax}
\end{equation}
\begin{equation}
\eta(x)=\frac{1+(-1)^x}{2}
\label{def:etax}
\end{equation}
and the Kronecker symbol (scalar and vector) 
\begin{equation}
\delta_{mn}=\begin{cases} 1 & m = n \\ 0 & m \neq n\end{cases}
\qquad
\delta_{\vec{m}\vec{n}}=\prod_{i=1}^s \delta_{m_in_i} 
\label{def:vkronecker}
\end{equation}

\item For vector indexes we have:
\begin{eqnarray}
\|\vec{m}\|&=&\|\vec{n}\|=s,\qquad \vec{m}=\{m_1,...,m_s\},\qquad \vec{n}=\{n_1,...,n_s\},\nonumber\\
|\vec{m}| &\equiv& \sum_{r=1}^{s} m_r, \qquad 
\vec{m}\,! \equiv \prod_{r=1}^{s} m_r!,\qquad\quad\;\;
\vec{m}^{\vec{n}} \equiv \prod_{r=1}^{s} m_r^{n_r}
\nonumber\\
\mathbb{C}_{\vec{m}}^{\vec{n}} &\equiv& 
\frac{\vec{m}\,!}{\vec{n}\,!\,(\vec{m}-\vec{n})!} ,\qquad
\sum_{\vec{m}=\vec{0}}^{\vec{n}} =\sum_{m_1=0}^{n_1}\dots\sum_{m_s=0}^{n_s}
,
\quad
\prod_{\vec{m}=\vec{0}}^{\vec{n}} =\prod_{m_1=0}^{n_1}\dots\prod_{m_s=0}^{n_s},
\nonumber\\
(c)_{\vec{n}} &\equiv& \prod_{r=1}^{s} (c)_{n_r}
,
\qquad\quad
(c^{\vec{m}})_{\vec{n}} \equiv\prod_{r=1}^{\|\vec{m}\|} (c^{m_r})_{n_r},\qquad\;
\mathcal{\upsilon}^{\vec{m}}_{\vec{n}} \equiv \prod_{r=1}^{s} \mathcal{\upsilon}^{m_r}_{n_r},
.
\label{def:vectorfuns}
\end{eqnarray}
The last two notations are used for special coefficients in this paper.

\item
For any vector index $\vec{a}=\{a_1,...,a_N\}$ we introduce
\begin{equation}\cev{a}=\{a_N,...,a_1\}\end{equation} 
with reverse order of components.

\item To simplify expressions we will use the following notation for products of tensors or vectors of the same type:
\begin{eqnarray}
T_{(i)}^{\otimes r_i} &\equiv& T_{\alpha_{a_1}}T_{\alpha_{a_2}}...T_{\alpha_{a_{r_i}}},
\nonumber\\
T_{(ii)}^{\otimes n_i} &\equiv& T_{\alpha_{a_1}\alpha_{a_2}}T_{\alpha_{a_3}\alpha_{a_4}}...T_{\alpha_{a_{n_i-1}}\alpha_{a_{n_i}}},
\nonumber\\
T_{(i)(j)}^{\otimes k_{ij}} &\equiv& T_{\alpha_{a_1}\beta_{b_1}}T_{\alpha_{a_2}\beta_{b_2}}...T_{\alpha_{a_{k_{ij}}}\beta_{b_{k_{ij}}}},\nonumber
\end{eqnarray}
and so on. For more complicated expressions like $T_{(ii)}^{\otimes n_i} q_{(i)}^{\otimes r_i}$ it is 
assumed that all indexes in different structures are different, i.e.
$$
T_{(ii)}^{\otimes n_i} q_{(i)}^{\otimes r_i} \equiv T_{\alpha_1\alpha_2}...T_{\alpha_{2n_i-1}\alpha_{2n_i}}q_{\alpha_{2n_i+1}}...q_{\alpha_{2n_i+r_i}}.
$$

\item $(...)$ denotes symmetrization of any structure inside parentheses in each group 
of indexes. For example,
\begin{equation}
\left( T_{(1)(2')}^{\otimes k} \right) \equiv \sum_{\{s_1,...,s_k\}_p\atop \{s'_1,...,s'_k\}_p} T_{\alpha_{s_1}\beta_{s'_1}}...T_{\alpha_{s_k}\beta_{s'_k}},
\label{def:symmetrization}
\end{equation}
where $\{...\}_p$ means only permutations, which lead to the sum of unique 
terms. This definition is slightly different from the usual 
symmetrization, where the sum is taken over all permutations and divided 
by the total number of terms.

\item $[...]$ and $[...]_{sym}$ denotes tensor constructions, which are traceless and symmetric (TS), i.e. 
like previous ones, but also traceless.

\item For traces we will use the following notations:
\begin{equation}
Sp_i T \equiv Sp_i T_{(\alpha)_{J_i}}\equiv g_{\alpha_s\alpha_{s'}} T_{\left(\alpha_1...\alpha_s...\alpha_{s'}...\alpha_{J_i}\right)}.
\end{equation}

\item For contractions we use the notation like in examples below
\begin{eqnarray}
A_{(1\mathbf{1})}\otimes B_{(\mathbf{1})} &\equiv& A_{\mu_1\mu_2}B_{\mu_2},\quad
A_{(r\mathbf{r})}\otimes B_{(\mathbf{r})} \equiv A_{\alpha_1\alpha_2}B_{\alpha_1},
\nonumber\\
A_{(\mathbf{1})(2)}\otimes B_{(\mathbf{1})} &\equiv& A_{\mu\nu}B_{\mu}, \quad
A_{(\mathbf{r})(s)}\otimes B_{(\mathbf{r})} \equiv A_{\mu\nu}B_{\mu},
\nonumber\\
A_{(\mu)_{r_1},(\nu)_{r_2}}\otimes B_{(\mu)_{r_1}} &\equiv& A_{(\mu_1...\mu_{r_1}),(\nu_1...\nu_{r_2})}B_{(\mu_1...\mu_{r_1})}
\nonumber
\end{eqnarray}
for some tensors $A$ and $B$.
Here we contract all the identical indexes using the usual Einstein summation rule. In some cases, when all indexes are contracted, we
will omit indexes like in
$$
\mathcal{V}^{J_1}\otimes\mathcal{W}^{\{J_1,J_{1'}\}}\otimes \mathcal{V}^{J_{1'}}.
$$
And also indexes are omitted, if all the indexes are in the same group of indexes, for example
$$
\left(\mathcal{V}^{J_1-r_1}q^{\otimes r_1}\right)\equiv \left(\mathcal{V}^{J_1-r_1}_{\mu_1...\mu_{J_1-r_1}}q_{\mu_{J_1-r_1+1}}...q_{\mu_{J_1}}\right).
$$

\item 
$\vec{k},\, \vec{k}',\, \vec{k}'',\dots$ ("$k$-like") denote vector indexes which we use to indicate multiple 
products of tensors like $g_{(i)(j)},\, G_{(i)(j)},\dots,i\neq j$, i.e. they denote number of "connections
between two different tensor legs", and
\begin{equation}
 \|\vec{k}\|=\|\vec{k}'\|=\dots=\mathbb{C}_{N_L}^2=N_L(N_L-1)/2, 
\end{equation}
where $N_L$ is the number of tensor legs.

$\vec{J}$, $\vec{\kappa}$, $\vec{n}$, $\vec{r}$ ("$J$-like") are vector indexes to indicate products
of tensors and vectors like $g_{(ii)},\, G_{(ii)},\, p_{i(i)},\,\dots$, and 
$$
\|\vec{J}\|=\|\vec{\kappa}\|=\|\vec{n}\|=\|\vec{r}\|=N_L.
$$

For different amplitudes (see section~\ref{section:ampset}) and their basic symmetric structures we have different "k-like" and "J-like" vector indices.
For the amplitude with four tensor legs ($H^{\vec{J}}$, $\mathcal{H}^{\vec{J}}$) and its structures $S^{H\, \vec{J}}_{\vec{k}\;\vec{n}}$ the vector indexes
are
\begin{eqnarray}
 \mathrm{"k-like"}: \vec{k}&\equiv&\{k_{11'},k_{22'},k_{12},k_{1'2'},k_{1'2},k_{12'}\},\quad k_{rs}\equiv k_{sr},\nonumber\\
\mathrm{"J-like"}: \vec{J} &\equiv& \{ J_1,J_{1'},J_2,J_{2'}\}, \nonumber\\
\vec{\kappa} &\equiv& \{ \sum_{s\neq 1} k_{1s},\sum_{s\neq 1'} k_{1's},\sum_{s\neq 1} k_{2s},\sum_{s\neq 2'} k_{2's}\}, \nonumber\\
\vec{n} &\equiv& \{ n_1, n_{1'}, n_2, n_{2'}\},\nonumber\\
\vec{r} &\equiv&\{ r_1,r_{1'},r_2,r_{2'}\}. \label{eq:notations4legs}
\end{eqnarray}

For the amplitude with three tensor legs ($Y^{\vec{J}}$, $\mathcal{Y}^{\vec{J}}$) and its structures $S^{Y\, \vec{J}}_{\vec{k}\;\vec{n}}$ the vector 
indexes are
\begin{eqnarray}
 \mathrm{"k-like"}: \vec{k}&\equiv&\{k_{12},k_{13},k_{23}\},\quad k_{rs}\equiv k_{sr},\nonumber\\
 \mathrm{"J-like"}: \vec{J} &\equiv& \{ J_1,J_2,J_3\}, \nonumber\\
\vec{\kappa} &\equiv& \{ \sum_{s\neq 1} k_{1s},\sum_{s\neq 2} k_{2s},\sum_{s\neq 3} k_{3s}\}, \nonumber\\
\vec{n} &\equiv& \{ n_1,n_2,n_3\},\nonumber\\
\vec{r} &\equiv&\{ r_1,r_2,r_3\}.\label{eq:notations3legs}
\end{eqnarray}

For the amplitudes with two tensor legs ($W^{\vec{J}}$, $\mathcal{W}^{\vec{J}}$, $F^{\vec{J}}$, $\mathcal{F}^{\vec{J}}$) and their structures 
$S^{W\, \vec{J}}_{k\;\vec{n}}$, $S^{F\, \vec{J}}_{k\;\vec{n}}$ we have the vector indexes
\begin{eqnarray}
\mathrm{"k-like"}: k\, && \mathrm{(single\; index)},\nonumber\\
\mathrm{"J-like"}: \vec{J} &\equiv& \{ J_i,J_{i'}\}\;(W,\mathcal{W})\quad\mathrm{or}\quad \{ J_i,J_j \}\;(F,\mathcal{F}),  \nonumber\\
\vec{n} &\equiv& \{ n_i,n_{i'}\}\quad\mathrm{or}\quad \{ n_i,n_j\}, \nonumber\\
\vec{r} &\equiv& \{ r_i,r_{i'}\}\quad\mathrm{or}\quad \{ r_i,r_j\},\nonumber\\
\vec{\kappa} &\equiv& \{k,k\} = k\cdot \vec{1},\nonumber\\
&& i,j \in \{1,2\},\quad j\neq i,i'\label{eq:notations2legs}
\end{eqnarray}
Symmetric structures $S^{T\; \vec{J}}_{\vec{k}\;\vec{n}}$  with single or several totally symmetric groups of Lorentz indexes are defined in Appendix~\ref{appendixB}.

\item
With vector indexes we use the following notations
\begin{eqnarray}
&& \mathrm{"J-like"}\; \vec{e}_i = \{0,...,0,1(i\mathrm{-th\; place}),0,...,0\}\nonumber\\
&& \mathrm{"k-like"}\; \vec{e}_{ij} = \{0,...,0,1(k_{ij}\mathrm{-th}\;\mathrm{place}),0,...,0 \},\label{def:eieij}
\nonumber\\
&& \vec{0}=\{0,0,\dots,0\},\qquad \vec{1}=\{1,1,\dots,1\}.
\end{eqnarray}
Here are some examples of index shifts:
\begin{equation}
h^{\vec{k}'}_{\vec{n}-r\,\vec{e}_{1'}} = h^{\vec{k}'}_{\{n_1,n_{1'}-r,n_2,n_{2'}\}},\quad
h^{\vec{k}'-r\,\vec{e}_{12}}_{\vec{n}} = h^{\{k'_{11'},k'_{22'},k'_{12}-r,k'_{1'2'},k'_{1'2},k'_{12'}\}}_{\vec{n}}
\label{def:OshiftExamples}.
\end{equation}

\item
For any differences of vectors with the same length we have the notation like
\begin{equation}
 \vec{J}^* \equiv \vec{J}-2\vec{n} -\vec{\kappa} \equiv \{ J_1-2n_1-\kappa_1,...\}
\label{def:Jbar}
\end{equation}

\item
Number of terms in different basic symmetric structures:
\begin{equation}
\mathcal{N}(S^{T\;\vec{J}}_{\vec{k}\;\vec{n}})=
\mathcal{N}^{T\;\vec{J}}_{\vec{k}\;\vec{n}}=
\frac{\vec{J}!}{2^{|\vec{n}|}\vec{n}!\,\vec{k}!\,\vec{J}^*!}.
\label{eq:NumberOfTerms}
\end{equation}

\item In the recurrent equations 
we often omit $(\vec{k};\vec{J})$ in coefficients for convenience, i.e.
\begin{equation}
h^{\vec{k}'(\vec{k};\vec{J})}_{\vec{n}}\equiv h^{\vec{k}'}_{\vec{n}};\qquad
y^{\vec{k}'(\vec{k};\vec{J})}_{\vec{n}}\equiv y^{\vec{k}'}_{\vec{n}}.
\label{eq:omitindex}
\end{equation}


\item
The generation function for any tensor amplitude 
$T^{\vec{J}}$ with $\|\vec{J}\|$ legs and its basic symmetryc structure $S^{T\;\vec{J}}_{\vec{k}\;\vec{n}}$ 
(symmetric groups of Lorentz indices) look as follows
\begin{equation}
\varPhi^T(\vec{X}_T) = \vec{\omega}^{\otimes \vec{J}} \otimes T^{\vec{J}},\quad 
{\varPhi'}^T_{\vec{k}\;\vec{n}}(\vec{X}_T) =   \vec{\omega}^{\otimes \vec{J}} \otimes S^{T\;\vec{J}}_{\vec{k}\;\vec{n}},\quad
\vec{\omega}^{\otimes \vec{J}} = \prod_{s\in\Omega_T} \omega_s^{\otimes J_s},
\label{def:genfunnotation}
\end{equation}
where $\omega_s\in \mathbb{M}_D$, $\vec{X}_T$ is the vector of all possible variables (invariant contractions of $\omega_s$ with different 
basic structures of $T^{\vec{J}}$). 

${\varPhi'}^T_{\vec{k}\;\vec{n}}$ is the simple monomial of variables contained in $\vec{X}_T$,  which 
is multiplied by number of unique terms in $S^{T\;\vec{J}}_{\vec{k}\;\vec{n}}$. All these functions, sets of variables
 and basic tensors for concrete $T$ can be found
in Appendix~\ref{appendixB}.

And for derivatives we have
\begin{equation}
\partial_{\vec{\omega}}\equiv \prod_{s\in\Omega_T} \partial_{\omega_{s(s)}},\;
\partial^{\otimes\vec{J}}_{\vec{\omega}}\equiv \prod_{s\in\Omega_T} \partial^{\otimes J_s}_{\omega_{s(s)}}.
\label{def:omegaderivatives}
\end{equation}
$\Omega_T$ can be found in~(\ref{def:qgpownonconserv}).

\end{enumerate}
Any other notations are explained in the text.

\section{Basics of the covariant reggeization framework}
\label{section1}

Basic elements of the covariant reggeization approach are vertex functions (irreducible
tensor representations of the Poincare group with multiple groups of Lorentz indices). As examples
we can consider the vertex ``scalar-scalar-spin-$J$ tensor'' in momentum space:
${\mathscr{V}}^{\mu_1\cdots\mu_{J}}(k,q)$, where
\begin{equation}
 \label{eq:TvertexDef}
\mathscr{V}^{\mu_1\dots\mu_J}(p,q)=<p-q|\mathscr{I}^{\mu_1\dots\mu_J}|p>,
\end{equation}
and the propagator
\begin{eqnarray}
&& \mathscr{P}^{\mu_1...\mu_J\nu_1...\nu_J}(q)=\nonumber\\
&&\int d^4x\; e^{iq(x-y)}
\left< 0|\;  \Phi^{\mu_1...\mu_J}(x) \Phi^{\nu_1...\nu_J}(y)\; |0 \right>\quad{,}\nonumber\\
&& = \Pi^{\mu_1...\mu_J, \nu_1...\nu_J}(q)/(m^2(J)-q^2),
\label{eq:WtensorDef}
\end{eqnarray}
which have the poles at
\begin{equation}
\label{eq:polesJ}
m^2(J)-q^2=0,\; \mbox{i.e.}\; J=\alpha_{{\mathbb R}}(q^2)\;,
\end{equation}
after an appropriate analytic continuation of the signatured amplitudes 
in $J$, where $\alpha_{\mathbb R}$ is the reggeon
trajectory.

The reggeization prescription for the amplitude after all contractions of Lorentz indexes is
\begin{equation}
\label{reggeizationpr}
\sum_J\frac{F^J}{(q^2-m^2)}\to 
\frac{\alpha_{{\mathbb R}}^{\prime}}{2}\eta_{{\mathbb R}}(q^2)\Gamma(-\alpha_{{\mathbb R}}(q^2))F^{\alpha_{{\mathbb R}}(q^2)}.
\end{equation}

$\mathscr{I}^{\mu_1\dots\mu_{J}}$ is the current operator related to the hadro\-nic spin-$J$ Heisenberg field operator,
\begin{equation}
\label{eq:KGJ}
\left( \square + m_J^2\right) \Phi^{\mu_1\dots\mu_J}(x)=\mathscr{I}^{\mu_1\dots\mu_J}(x)\;,
\end{equation}
and
\begin{eqnarray}
&&\partial_{\mu} \mathscr{I}^{\mu_1...\mu...\mu_J}=0\;{;}\label{eq:Icond1}\\
&&\mathscr{I}^{\mu_1...\mu_J}=\mathscr{I}^{\left(\mu_1...\mu_J\right)}\;{;}\;
\label{eq:Icond2}\\
&&g_{\mu_i\mu_k}\mathscr{I}^{\mu_1...\mu_i...\mu_k...\mu_J}=0.\label{eq:Icond3}
\end{eqnarray}
Here $\left(\mu_1...\mu_J\right)$ denotes symmetrization in all indices. (\ref{eq:Icond1})-(\ref{eq:Icond3})
are Rarita-Schwinger conditions for irreducible representations of the Poincar\'e algebra (transverse-sym\-met\-ric-traceless, TST). In 
momentum space these conditions look like
\begin{eqnarray}
&& q_{\lambda}\mathscr{V}^{\mu_1...\lambda ...\mu_J}=0\;{;}\label{eq:Vcond1}\\
&& \mathscr{V}^{\mu_1...\mu_J}=
\mathscr{V}^{\left(\mu_1...\mu_J\right)}\;{;}
\label{eq:Vcond2}\\
&&g_{\mu_i\mu_k}\mathscr{V}^{\mu_1...\mu_i...\mu_k...\mu_J}=0{.}\label{eq:Vcond3}
\end{eqnarray}
Similar conditions are imposed on all other irreducible tensors considered in this paper. 

\section{Set of amplitudes and hadronic form-factors for diffractive processes}
\label{section:ampset}

In this section we consider basic tensors, which we can use to construct amplitudes and cross-sections for all diffractive processes. Let us
at first introduce irreducible (TST) tensors:
\begin{figure}[hbt!]
	\centering
	\includegraphics[width=\textwidth]{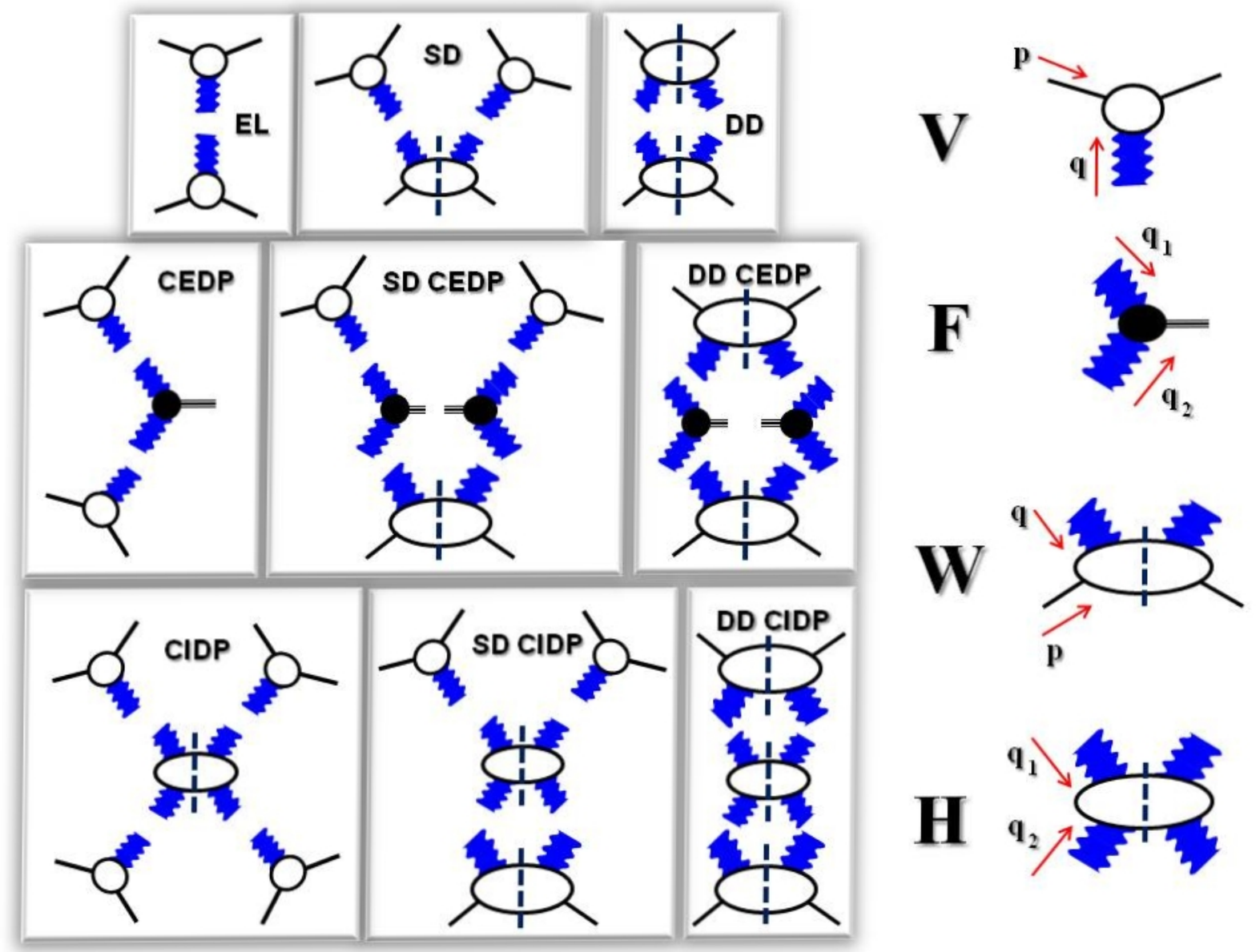}
	\caption{\label{fig:diffractiveLego} Elements of the Lego constructor for diffractive processes and final tensors.}
\end{figure}
\begin{itemize}
\renewcommand\labelitemi{$\bullet$}
\item The tensor with one symmetrized group of Lorentz indexes, which corresponds to the scalar-scalar-tensor vertex (Fig.~\ref{fig:diffractiveLego}): $$\mathcal{V}^{J_i}\equiv \mathcal{V}^{J_i}_{(\alpha)_{J_i}}(p_i,q_i)$$
\item The tensor with two symmetrized groups of Lorentz indexes, which corresponds to the scalar-tensor forward scattering amplitude (Fig.~\ref{fig:diffractiveLego}):
$$\mathcal{W}^{\vec{J}}\equiv \mathcal{W}^{\{J_i,J_{i'}\}}_{(\alpha)_{J_i},(\alpha')_{J_{i'}}}(p_i,q_i)$$
\item  The tensor with two symmetrized groups of Lorentz indexes, which corresponds to the tensor-tensor to scalar fusion amplitude (Fig.~\ref{fig:diffractiveLego}):
\begin{equation}
 \mathcal{F}^{\vec{J}}\equiv
\mathcal{F}^{\{J_i,J_j\}}_{(\alpha)_{J_i},(\beta)_{J_j}}(q_1,q_2),\qquad i,j\in\{1,1',2,2'\},\quad j\neq i,i'. 
\nonumber
\end{equation}
\item  The tensor with three symmetrized groups of Lorentz indexes, which corresponds to the tensor-tensor-tensor vertex (Fig.~\ref{fig3:3IPtensors}).
\begin{equation}
\mathcal{Y}^{\vec{J}}\equiv
\mathcal{Y}^{\{J_1,J_2,J_3\}}_{(\mu)_{J_1},(\nu)_{J_2},(\rho)_{J_3}}(q_1,q_2)
\nonumber
\end{equation}
\item The tensor with four symmetrized groups of Lorentz indexes, which corresponds to the tensor-tensor forward scattering amplitude (Fig.~\ref{fig:diffractiveLego}):
\begin{equation}
  \mathcal{H}^{\vec{J}}\equiv
 \mathcal{H}^{\{J_1,J_{1'},J_2,J_{2'}\}}_{(\mu)_{J_1},(\mu')_{J_{1'}},(\nu)_{J_2},(\nu')_{J_{2'}}}(q_1,q_2)\nonumber
\end{equation}
\end{itemize}

\begin{figure}[t!]
	\centering
	\includegraphics[width=0.45\textwidth]{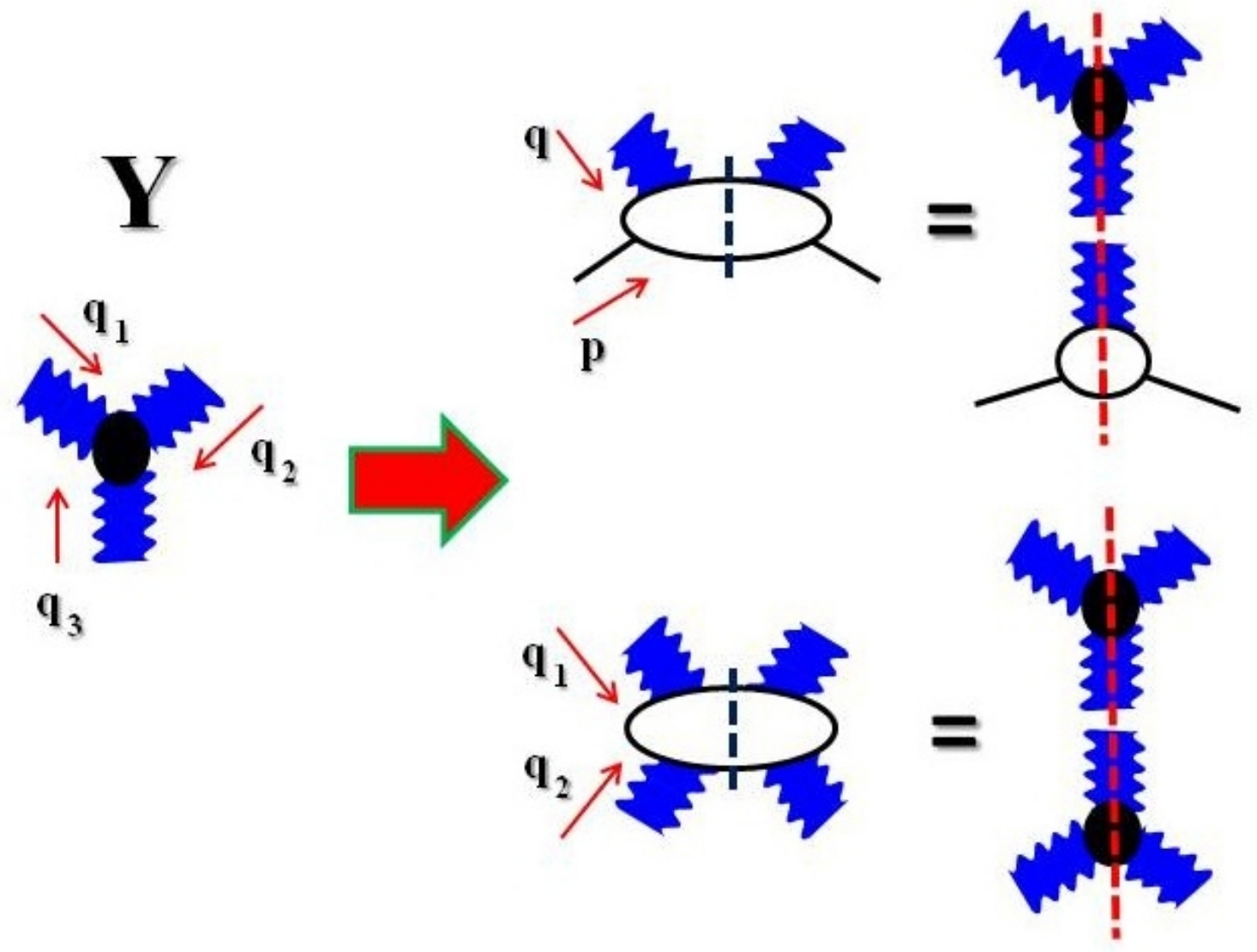}
	\caption{\label{fig3:3IPtensors}
		Amplitudes for single dissociation and
		reggeon-reggeon fusion processes in 3-reggeon approach,
		$\mathcal{W}_{3\mathbb{R}}^{\vec{J}}$ and
		$\mathcal{H}_{3\mathbb{R}}^{\vec{J}}$.
	}
\end{figure}

We can use contractions of three-reggeon vertexes $\mathcal{Y}^{\vec{J}}$ in the forward limit~(\ref{eq:YfwdLimit}) 
to construct amplitudes for single dissociation and
reggeon-reggeon fusion processes in 3-reggeon 
approach (see Fig.~\ref{fig3:3IPtensors}):
\begin{eqnarray}
\!\!\!\!\!\!\!
\mathcal{W}^{\vec{J}}_{3\mathbb{R}}(p,q)&=&
\left<
\mathcal{Y}_{fwd\,(\mu)_{J},(\mu')_{J'},(\rho)_{J_3}}^{\{J,J',J_3\}}(q,n)\otimes
\mathcal{V}_{(\rho)_{J_3}}^{J_3}(p,n)
\right>_{p,q},
\label{W3R}
\\
\!\!\!\!\!\!\!
\mathcal{H}^{\vec{J}}_{3\mathbb{R}}(q_1,q_2)&=&
\left<
\mathcal{Y}_{fwd\,(\mu)_{J_1},(\mu')_{J_{1'}},(\rho)_{J_3}}^{\{J_1,J_{1'},J_3\}}(q_1,n)\otimes
\mathcal{Y}_{fwd\,(\nu)_{J_2},(\nu')_{J_{2'}},(\rho)_{J_3}}^{\{J_2,J_{2'},J_3\}}(q_2,n)
\right>_{q_1,q_2},
\label{H3R}
\end{eqnarray}
\begin{eqnarray}
\left<n_{\alpha_1}...n_{\alpha_{2N}}\right>_{p,q}&=&
\frac{
	\int d^D n \;\delta(n^2+1)\;\delta(pn)\;\delta(qn)\;
	n_{\alpha_1}...n_{\alpha_{2N}}
}{
\int d^D n \;\delta(n^2+1)\;\delta(pn)\;\delta(qn)
}=
\nonumber\\
&=& 
\frac{
	(-1)^{N}
}{
2^{N} 
\left(\frac{D-2}{2}\right)_N
} 
\left(
\mathcal{G}_{(rs)}(p,q)
^{\otimes N}\right)_{all\;sym.},\qquad 
N=|\vec{J}|/2,
\label{eq:naverage}
\end{eqnarray}
where "$all\;sym.$" means the symmetrization of
all indices (even from different groups) in the sense 
of~(\ref{def:symmetrization}) and $r$, $s$ can be from 
any group of indices defined by $\Omega_{W,H}$~(\ref{def:qgpownonconserv}).

\section{Expansion in terms of irreducible tensors}
\label{section4}
It is important to note, that in some models~\cite{CloseIn}-\cite{CloseOut} reggeon currents may be nonconserved, so, we can introduce
general tensors in this case as expansions in terms of the above irreducible tensors.

Let us denote any irreducible multileg 
amplitude as
$\mathcal{T}^{\vec{J}}$  and
 amplitude, which is not transverse in
 momentum transfers as
$T^{\vec{J}}$, with the 
corresponding coefficients $\tau$:
\begin{equation}
\mathcal{T}=\{\mathcal{V}, \mathcal{W}, \mathcal{F}, \mathcal{Y}, \mathcal{H}\},\quad
T=\{V,W,F,Y,H\},\quad
\tau=\{v, w, f, y, h\}.
\label{def:alltensors}
\end{equation}
 Then we can 
write the expansion:
\begin{equation}
T^{\vec{J}}\equiv
 \displaystyle\sum_{\vec{r}=\vec{0}}^{\vec{J}}
\overline{\tau}_{\vec{r}}
 \left[ \mathcal{T}^{\vec{J}-\vec{r}}
q_.^{\otimes \vec{r}} \right],
\label{def:Tnonconserv}
\end{equation}
where
\begin{equation}
  \left[ \mathcal{T}^{\vec{J}-\vec{r}}
q_.^{\otimes \vec{r}} \right]
 = \sum_{\vec{a}=\vec{0}}^{[\vec{r}/2]}
 \tilde{\tau}_{\vec{a}}^{\vec{J}} 
q_.^{2\vec{a}}
 \left( \mathcal{T}^{\vec{J}-\vec{r}} 
q_.^{\otimes (\vec{r}-2\vec{a})} 
g_{..}^{\otimes \vec{a}} \right),
\label{def:TSTnonconserv}
\end{equation}
and
\begin{eqnarray}
q_.^{\otimes \vec{r}} &\equiv& 
\prod_{s\in \Omega_T} q_{s(s)}^{\otimes r_s},
\quad
g_{..}^{\otimes \vec{a}} \equiv
\prod_{s\in \Omega_T} g_{(ss)}^{\otimes a_s},
\quad
q_.^{2\vec{a}}=
\prod_{s\in \Omega_T} \left(q_s^2\right)^{a_s},\nonumber\\
\Omega_V&=&\{i\} ,\, \Omega_W=\{i,i'\},\, 
\Omega_F=\{i,j\},\; i,j\in\{1,1',2,2'\},\; j\neq i,i'.
\nonumber\\
\Omega_Y&=&\{1,2,3\},\,
\Omega_H=\{1,1',2,2'\},
\label{def:qgpownonconserv}
\end{eqnarray}

The above decompositions are possible, since
 the following orthogonal relations hold
\begin{equation}
\left[ \mathcal{T}^{\vec{J}-\vec{r}} 
q_.^{\otimes \vec{r}}
 \right]\otimes  
\left[ \mathcal{T}^{\vec{J}-\vec{r}\,'} 
q_.^{\otimes \vec{r}\,'}
 \right] \sim \delta_{\vec{r}\vec{r}\,'}.
\label{eq:sqTorthogonal} 
\end{equation}

All the coefficients in the above
 structures~(\ref{def:Tnonconserv})-(\ref{def:TSTnonconserv})
 are defined in Appendix~C.


Now let us repeat the list of processes in the introduction, but with general formula for their amplitudes and cross-sections (see also Fig.~\ref{fig:diffractiveLego}):
\begin{itemize}
\renewcommand\labelitemi{$\bullet$}

\item \textbf{EL}: 
\begin{equation}
\label{eq:finalgenEL}
A_{\mathrm{EL}}=V^J(p_1,q)\otimes V^J(p_2,q);
\end{equation}

\item \textbf{CEDP}: 
\begin{equation}
A_{\mathrm{CEDP}}=
F^{\vec{J}}(q_1,q_2) \otimes
 V^{J_1}(p_1,q_1)\otimes V^{J_2}(p_2,q_2);
\label{eq:finalgenCEDP}
\end{equation}

\item \textbf{CIDP}: 
\begin{eqnarray}
\sigma_{\mathrm{CIDP}}&\sim& H^{\vec{J}}(q_1,q_2)\otimes
 V^{J_1}(p_1,q_1) \otimes V^{J_{1'}}(p_1,q_1)\otimes
 \nonumber\\
&& \otimes\, 
V^{J_2}(p_2,q_2) \otimes V^{J_{2'}}(p_2,q_2);
\label{eq:finalgenCIDP}
\end{eqnarray}

\item \textbf{SD}: 
\begin{equation}
\sigma_{\mathrm{SD}}\sim
W^{\vec{J}}(p_2,q) \otimes 
V^{J_1}(p_1,q)\otimes V^{J_{1'}}(p_1,q) + (p_1\leftrightarrow p_2);
\label{eq:finalgenSD}
\end{equation}

\item \textbf{SD CEDP}:
\begin{eqnarray}
 \sigma_{\mathrm{SD\; CEDP}}&\sim&
F^{\{J_1,J_2\}}(q_1,q_2) \otimes F^{\{J_{1'},J_{2'}\}}(q_1,q_2) \otimes\nonumber\\ 
&& \otimes\,\left[ V^{J_1}(p_1,q_1)\otimes V^{J_{1'}}(p_1,q_1) \otimes  W^{\{J_2,J_{2'}\}}(p_2,q_2) + \right.\nonumber\\
&& \left.\phantom{ W^{\{J_2,J_{2'}\}}}\hspace*{-1cm} + (p_1,q_1\leftrightarrow p_2,q_2) \right];
\label{eq:finalgenSDCEDP}
\end{eqnarray}

\item \textbf{SD CIDP}: 
\begin{eqnarray}
 \sigma_{\mathrm{SD \;CIDP}}&\sim&
H^{\vec{J}}(q_1,q_2) \otimes
\left[ V^{J_1}(p_1,q_1)\otimes 
V^{J_{1'}}(p_1,q_1)\otimes W^{\{J_2,J_{2'}\}}(p_2,q_2) +\right.
\nonumber\\ 
&& \left.\phantom{W^{\{J_2,J_{2'}\}}} \hspace*{0.65cm}
 + (p_1,q_1\leftrightarrow p_2,q_2) \right];
\label{eq:finalgenSDCIDP}
\end{eqnarray}

\item \textbf{DD}:  
\begin{equation}
\sigma_{\mathrm{DD}}\sim
W^{\vec{J}}(p_1,q) \otimes W^{\vec{J}}(p_2,q);
\label{eq:finalgenDD}
\end{equation}

\item \textbf{DD CEDP}: 
\begin{eqnarray}
\sigma_{\mathrm{DD \;CEDP}}&\sim&
F^{\{J_1,J_2\}}(q_1,q_2) \otimes 
F^{\{J_{1'},J_{2'}\}}(q_1,q_2) \otimes\nonumber\\ 
&&\otimes W^{\{J_1,J_{1'}\}}(p_1,q_1) \otimes 
W^{\{J_2,J_{2'}\}}(p_2,q_2);
\label{eq:finalgenDDCEDP}
\end{eqnarray}

\item \textbf{DD CIDP}: 
\begin{equation}
\sigma_{\mathrm{DD \;CIDP}}\sim
H^{\vec{J}}(q_1,q_2) \otimes
 W^{\{J_1,J_{1'}\}}(p_1,q_1) \otimes 
W^{\{J_2,J_{2'}\}}(p_2,q_2).
\label{eq:finalgenDDCIDP}
\end{equation}

\end{itemize}
Here $\sigma_{...}\sim$ means that we should multiply the right hand side by 
an appropriate coefficient and take the imaginary part of
the corresponding forward amplitude due to the optical theorem
to obtain the exact cross-section. For conserved
 reggeon currents we simply replace 
$V$, $W$, $F$, $Y$, $H$ by 
 $\mathcal{V}$, $\mathcal{W}$, 
 $\mathcal{F}$, $\mathcal{Y}$, $\mathcal{H}$. In 
this paper we will obtain all tensors, and
give some examples for definite spins. And 
in the next paper we will consider 
cross-sections for 
any spins and the transition to
 the complex J-plane.

\section{Irreducible tensor representations: recurrent equations for coefficients and their solutions}
\label{section:recurrences}

Below in this section we show structures, equations and solutions for tensors 
$\mathcal{V}$, $\mathcal{W}$, $\mathcal{F}$, $\mathcal{Y}$, $\mathcal{H}$
and discuss their properties. To avoid unnecessary complication in the main part of the work, we 
consider basic tensor structures in Appendix~\ref{appendixA} and all the ways to obtain the equations in Appendix~\ref{appendixB}.

Contractions of different basic tensors and momenta are considered in Appendix~\ref{appendixA}.

We can use any basic "raw" tensors ($S^{T;\vec{J}}_{\vec{k}\,\vec{0}}$) as a basis, 
if we want to obtain irreducible tensors by the standart harmonization method, which removes
all traces from basic tensors. 

Any irreducible tensor can be represented as follows
\setlength{\fboxrule}{1pt}
\begin{empheq}[box=\fcolorbox{red}{white}]{align}
\mathcal{T}^{\vec{J}}=\sum_{\bar{\Omega}^T_{\vec{k}}} 
\hat{\tau}^{\vec{J}}_{\vec{k}} \mathcal{T}^{*\,\vec{J}}_{\vec{k}},\qquad
\mathcal{T}^{*\,\vec{J}}_{\vec{k}} =\sum_{\tilde{\Omega}^T_{\vec{k}'\,\vec{n}}}
\tau^{\vec{k}'(\vec{k};\vec{J})}_{\vec{n}} S^{T\,\vec{J}}_{\vec{k}'\,\vec{n}},
\label{def:anyirrep}
\end{empheq}
and regions of summation are
\begin{eqnarray}
&&\tilde{\Omega}^T_{\vec{k}'\,\vec{n}}: J^*_r\ge 0,\quad \kappa_r\ge\kappa'_r,\quad 
n_r\ge 0,\; k'_{rs}\ge 0,\quad
 r,s\in\Omega_T,
\; r\neq s, \nonumber\\
&& \bar{\Omega}^T_{\vec{k}} = \tilde{\Omega}^T_{\vec{k}\,\vec{0}}.
\label{def:regionsirrep}
\end{eqnarray}
$\mathcal{T}^{*\,\vec{J}}_{\vec{k}}$ is the TST basis (in this work we consider two different basises: standard and 
harmonical). Structures $S^{T\, \vec{J}}_{\vec{k}\,\vec{n}}$ are defined in~(\ref{def:SV})-(\ref{def:SH}).

In terms of generating functions
\begin{empheq}[box=\fcolorbox{red}{white}]{align}
{\varPhi^{\star}}^{\mathcal{T}}_{\vec{k}}(\vec{X}_T) &= \sum_{\tilde{\Omega}^T_{\vec{k}',\vec{n}}}  
\tau^{\vec{k}'(\vec{k};\vec{J})}_{\vec{n}} {\varPhi'}^T_{\vec{k}'\;\vec{n}}(\vec{X}_T) 
 = 
\frac{1}{\Pi^{\vec{J}\, \vec{k}}_0} 
 \hat{\Pi}(\square_{\vec{\omega}})\;\; {\varPhi'}^T_{\vec{k}\;\vec{0}}(\vec{X}_T) 
\label{eq:irrepHarmProjT}
\end{empheq}
\begin{eqnarray}
  \hat{\Pi}(\square_{\vec{\omega}}) &=& \prod_{s\in\Omega_T} \hat{\Pi}(\square_{\omega_s}),\qquad
 \square_{\omega_s} =g_{\alpha\beta}\; \partial_{\omega_{s\,\alpha}}\! \partial_{\omega_{s\,\beta}},
 \nonumber\\
\partial_{\omega_{s\,\alpha}} &=& \sum_{s'=1}^{\|\vec{X}_T\|} \frac{\partial (\vec{X}_T)_{s'}}{\partial_{\omega_{s\,\alpha}}} \partial_{(\vec{X}_T)_{s'}}.
\label{eq:irrepHarmProj}
\end{eqnarray}
${\varPhi'}^T_{\vec{k}'\;\vec{n}}(\vec{X}_T)$ 
is defined in~(\ref{def:genfunnotation}) and
it is simply the monomial multiplied 
by number of terms in $S^{T\, \vec{J}}_{\vec{k}\,\vec{n}}$.
\begin{empheq}[box=\fcolorbox{red}{white}]{align}
 \hat{\Pi}(\square_{\omega_s}) {\varPhi'}(\vec{X}_T) =
\sum_{m_s=0}^{[J_s/2]} \frac{y_s^{m_s}}{4^{m_s} m_s! (c^{J_s})_{m_s}} 
\left(\square_{\omega_s}\right)^{m_s} {\varPhi'}(\vec{X}_T) |_{\omega_s\to 0}
\label{eq:irrepHarmProjTsingle}
\end{empheq}
\begin{equation}
\Pi^{J_i\, \vec{k}}_0 ={}_2F_1 \left(-\frac{J_i-\kappa_i}{2},\frac{1-J_i+\kappa_i}{2};
c^{J_i}; 1\right),
\qquad
\Pi^{\vec{J}\, \vec{k}}_0 = \prod_{i\in\Omega_T} \Pi^{J_i\, \vec{k}}_0.
\label{HARMnormalization}
\end{equation}
\begin{equation}
 c^{J} =-\left( J+\frac{D-5}{2}\right).
\label{eq:irrepcJ}
\end{equation}
\begin{equation}
y_s=G_{(ss)}\otimes\omega_s\otimes\omega_s= \omega_s^2-\frac{(q_s\omega_s)^2}{q_s^2},
\label{eq:irrepHarmProjs}
\end{equation}
and in harmonical basis you have to make a replacement $y_s\to u_s+x_s^2$  
in the final result, since $y_s$ is the variable of the standard basis. $\Pi^{J\, k}_0$ is introduced
to normalize the coefficient of $S^{T\,\vec{J}}_{\vec{k}\,\vec{0}}$ in the tensor 
$\mathcal{T}^{*\,\vec{J}}_{\vec{k}}$ to 
unity ($\Pi^{J\, k}_0=1$ in the standard basis).

Now we have for irreducible tensors
\begin{empheq}[box=\fcolorbox{red}{white}]{align}
 \mathcal{T}^{\vec{J}} =\sum_{\bar{\Omega}^T_{\vec{k}}} \hat{\tau}_{\vec{k}}\;
\frac{1}{\vec{J}!}\; \partial^{\otimes\vec{J}}_{\vec{\omega}}\; {\varPhi^{\star}}^{\mathcal{T}}_{\vec{k}}(\vec{X}_T)|_{\vec{\omega}\to\vec{0}}
\label{eq:irrepTensorTHarm}
\end{empheq}

Tensors $G_{(ss)}$ are defined in Appendix~\ref{appendixA}, and 
$\Omega_T$, $\tilde{\Omega}^T_{\vec{k}'\,\vec{n}}$, $\bar{\Omega}^T_{\vec{k}}$ can be found in~(\ref{def:qgpownonconserv}) and~(\ref{def:regionsirrep}). 

Functions $\hat{\tau}_{\vec{k}}\equiv \hat{\tau}_{\vec{k}}(\{t_i\})$ are form-factors depending on the corresponding 
squares of momentum transfers.

From two different definitions of ${\varPhi^{\star}}^{\mathcal{T}}_{\vec{k}}(\vec{X}_T)$ in~(\ref{eq:irrepHarmProjT}) 
we can obtain $\tau^{\vec{k}'(\vec{k};\vec{J})}_{\vec{n}}$. It is very convenient, when we calculate tensors numerically, but
in analytical calculations it is rather inconvenient to obtain this coefficients from~(\ref{eq:irrepHarmProjT}) for further complicated 
manipulations. That is why we use recurrent equations below to get appropriate analytical expressions for the above coefficients. 

Let us consider recurrent equations and coefficients in two different basises defined in Appendix~\ref{appendixB}.
	
\subsection{\bf Scalar-scalar-tensor irreducible vertex}
\label{subsection:V}

For the vertex $\mathcal{V}$ we have the following general expression:
\begin{equation}
\label{eq:irrVexpression}
\mathcal{V}^J(p,q) = \hat{\mathcal{\upsilon}}_0(t) \sum_{n=0}^{\left[ J/2\right]} \mathcal{\upsilon}^J_n S^{V; J}_n
\end{equation}
where $S^{V; J}_n$ is defined in~(\ref{def:SV}) and coefficients obey the recurrent equation (obtained from the tracelessness 
condition~(\ref{eq:Vcond3}) in Appendix~\ref{subsection:trickFW}):
\begin{equation}
\begin{cases}
 &\mathcal{\upsilon}^J_{n-1}-2\left( c^J+n-1\right) \mathcal{\upsilon}^J_n  = 0,\; n>0,\\
 & \mathcal{\upsilon}^J_{0}=1,  \label{eq:irrVcoefs}
\end{cases}
\end{equation}
$c^J$ is defined in~(\ref{eq:irrepcJ}).

The solution of~(\ref{eq:irrVcoefs}) can be expressed as
\begin{equation}
\mathcal{\upsilon}^J_n=\frac{1}{2^n (c^J)_n},
\label{eq:irrVcoefsSolution}
\end{equation}
where $(...)_n$ is the Pochhammer symbol (see Appendix~\ref{appendixD}).

$\hat{\mathcal{\upsilon}}_0(t)$ is the scalar form-factor which is usually 
taken as an exponential or a dipole one in $t$.


\subsection{\bf Hadronic tensor for single dissociation (forward reggeon-hadron amplitude) and fusion 
reggeon-reggeon vertex of the exclusive central production}
\label{subsection:WF}

Using the general definition of irreducible tensors~(\ref{def:anyirrep}), where we substitute $\mathcal{W}$ and $\mathcal{F}$ instead 
of $\mathcal{T}$ and $w$, $f$ instead of $\tau$, expansion~(\ref{eq:Ftensorout}) developed in Appendix~\ref{subsection:trickFW} and definitions~(\ref{def:SF}),(\ref{def:SW}), we can write recurrent equations:
\begin{equation}
\begin{cases}
& f^{k'}_{\vec{n}-\vec{e}_i} -2(c^{J_i}+n_i-1)f^{k'}_{\vec{n}} +2\chi_m J^*_{i^{\star}} f^{k'+1}_{\vec{n}-\vec{e}_i}+\\
& \phantom{\frac{X^{X^{X\atop Y}}}{X^{X^{X\atop Y}}}\hspace*{-9mm}}
+ 2n_{i^{\star}} f^{k'+2}_{\vec{n}-\vec{e}_i-\vec{e}_{i^{\star}}} - \lambda_m J^*_{i^{\star}}\left( J^*_{i^{\star}}-1\right) f^{k'+2}_{\vec{n}-\vec{e}_i} = 0,\\
&\phantom{.}\\
& f^{k}_{\vec{n}}=\mathcal{\upsilon}^{\vec{J}}_{\vec{n}},\quad f^{k'}_{\vec{0}}=\delta_{k'k},\quad
 i=1,2,\; i^{\star}=2,1
\end{cases}
\label{eq:irrFcoefs}
\end{equation}
\begin{equation}
f^{k'}_{\vec{n}}\equiv f^{k'\,(k;\vec{J})}_{\vec{n}}|_{\tilde{\Omega}^F_{k'\,\vec{n}}},
\label{eq:fthetas}
\end{equation}

For $\mathcal{W}$ we can obtain equations from~(\ref{eq:irrFcoefs}), if we take the limit $\chi_m\to 1$ and $i^{\star}\to i'$:
\begin{equation}
\begin{cases}
& w^{k'}_{\vec{n}-\vec{e}_i} -2(c^{J_i}+n_i-1)w^{k'}_{\vec{n}} +2 J^*_{i'} w^{k'+1}_{\vec{n}-\vec{e}_i}+
 2n_{i'} w^{k'+2}_{\vec{n}-\vec{e}_i-\vec{e}_{i'}} = 0,\\
&\phantom{.}\\
& w^{k}_{\vec{n}}=\mathcal{\upsilon}^{\vec{J}}_{\vec{n}},\quad w^{k'}_{\vec{0}}=\delta_{k'k},\quad
i=1,1',\; i'=1',1
\end{cases}
\label{eq:irrWcoefs}
\end{equation}
\begin{equation}
 w^{k'}_{\vec{n}}\equiv w^{k'\,(k;\,\vec{J})}_{\vec{n}}|_{\tilde{\Omega}^W_{k'\,\vec{n}}}.
\label{eq:wthetas}
\end{equation} 
$c^{J_i}$ and $\mathcal{\upsilon}^J_n$ are defined by~(\ref{eq:irrepcJ}) and~(\ref{eq:irrVcoefsSolution}), 
$\mathcal{\upsilon}^{\vec{J}}_{\vec{n}}$ is calculated according to the rules~(\ref{def:vectorfuns}).

The detailed derivation and solution of the above equations are discussed in 
Appendices~\ref{subsection:trickFW} and~\ref{subsection:FWsolve}. In this section we present only 
final formula for coefficients:
\begin{eqnarray}
&&\hspace*{-5mm} f^{k'\,(k;\,\vec{J})}_{\vec{n}} = 2^{k-k'-n_1-n_2-1}\chi_m^{k-k'}\times\nonumber\\
&&\phantom{.}\nonumber\\
&&\hspace*{-5mm} \times \left\{
\frac{n_2!\left( J_2-2n_2-k'\right)!}{(c^{J_1})_{n_1}\chi_m^{2n_2}}\Lambda^{J_1J_2k}_{R\; k-k'\; n_1 n_2}(\hat{x})+
\frac{n_1!\left( J_1-2n_1-k'\right)!}{(c^{J_2})_{n_2}\chi_m^{2n_1}}\Lambda^{J_2J_1k}_{R\; k-k'\; n_2 n_1}(\hat{x})
\right\}\!,
\label{eq:irrFcoefsSolution}
\end{eqnarray}
where $\Lambda^{J_1J_2k}_{R\; k-k'\; n_1 n_2}(\hat{x})$ is the polynomial on $$\hat{x}=-(1-\chi_m^2)/\chi_m^2$$ 
defined in~(\ref{def:LambdaR}) (see Appendix~\ref{subsection:FWsolve}).

And for $w^{k'}_{\vec{n}}$ we have
\begin{eqnarray}
&&\hspace*{-5mm} w^{k'\,(k;\,\vec{J})}_{\vec{n}}  = 2^{k-k'-n_1-n_{1'}-1}\times\nonumber\\
&&\phantom{.}\nonumber\\
&&\hspace*{-5mm} \times\left\{
\frac{n_{1'}!\left( J_{1'}-2n_{1'}-k'\right)!}{(c^{J_1})_{n_1}}\Lambda^{J_1J_{1'}k}_{\mathbb{C}\; k-k'\; n_1 n_{1'}}+
\frac{n_1!\left( J_1-2n_1-k'\right)!}{(c^{J_{1'}})_{n_{1'}}}\Lambda^{J_{1'}J_1k}_{\mathbb{C}\; k-k'\; n_{1'} n_1}
\right\},
\label{eq:irrWcoefsSolution}
\end{eqnarray}
where the coefficient $\Lambda^{J_1J_{1'}k}_{\mathbb{C}\; k-k'\; n_1 n_{1'}}$ is defined
in~(\ref{def:LambdaC}) (see Appendix~\ref{subsection:FWsolve}). Examples of solutions for $\mathcal{F}$ and $\mathcal{W}$ for
fixed spins $J_{1,1',2}=2$ can
be found in Appendix~\ref{appendixE}.

The generating function can be expressed in terms of Srivastava-Daoust functions~(\ref{eq:genfunFSTANDstarFinal}).


Using the tricks of Appendix~\ref{appendixB} and trace operator~(\ref{eq:genWFtraceDiffHARM}) in the harmonical basis,
we obtain more simple equations for 
\begin{equation}
f^{k'}_{\vec{n}}=f^{k-2\ell\,(k;\vec{J})}_{\vec{n}}
\equiv \mathbf{f}^{\ell}_{\vec{n}}|_{\tilde{\Omega}^F_{k-2\ell\,\vec{n}}}
\,:
\label{eq:fthetasHARM}
\end{equation}
\begin{equation}
\begin{cases}
&  2(\lambda_k-2\ell+n_i-1)\cdot\mathbf{f}^{\ell}_{\vec{n}} +
\mathbf{f}^{\ell}_{\vec{n}-\vec{e}_i} + 
2n_{i^{\star}}\cdot\mathbf{f}^{\ell-1}_{\vec{n}-\vec{e}_i-\vec{e}_{i^{\star}}} = 0,\\
&\phantom{.}\\
& \lambda_k=\frac{D-2}{2}+k,\quad
\mathbf{f}^{\ell}_{\vec{0}} =B^{k\,\vec{J}}_{\ell},\quad
i\in\Omega_F,
\end{cases}
\label{eq:irrFcoefsHARM}
\end{equation}
and their solutions are obtained in Appendix~\ref{subsection:FWsolveHARM}:
\begin{eqnarray}
\mathbf{f}^{*\ell\, k}_{\vec{n}} &=&
\frac{\vec{n}!\,2^{2\ell-|\vec{n}|}(-1)^{|\vec{n}|}
}{
\ell!\,(\vec{n}-\vec{\ell})!\,(\lambda_k)_{\vec{n}-\vec{\ell}}\,(2-\lambda_k)_{\ell}
}	
\label{eq:basicsolutionHARM}
\\
\mathbf{f}^{\ell}_{\vec{n}} &=&\!\!\!
\sum_{m'=\max(0,\vec{\ell}-\vec{n})}^{\ell}
B_{m'}^{k\,\vec{J}}\; \mathbf{f}^{*\ell-m'\, k-2m'}_{\vec{n}}
,\qquad
\vec{\ell}=\{\ell,\ell\}.
\label{eq:fBsolutionHARM}
\end{eqnarray}

For $w^{k'}_{\vec{n}}$ we have the same 
equations, but $i^{\star}\to i'$, $f\to w$, $\mathbf{f}\to \mathbf{w}$, $F\to W$. 

To obtain the initial condition $B^{k\,\vec{J}}_{\ell}$, we have to use~(\ref{eq:irrepHarmProjT}) to extract coefficients from two different
representations for the generating function
 (see Appendix~\ref{subsection:FWsolveHARM}):
\begin{eqnarray}
B^{k\,\vec{J}}_{\ell}&=& 
\frac{(\lambda_k-2\ell)_{\ell}\,(\vec{J}-\vec{k}+2\vec{\ell})!}{4^{\ell}\ell!\,(c^{\vec{J}})_{\vec{\ell}}\,(\vec{J}-\vec{k})!}
\cdot\frac{\mathcal{B}^{k\,\vec{J}}_{\ell}}{\mathcal{B}^{k\,\vec{J}}_{0}},
\qquad
\mathcal{B}^{k\,\vec{J}}_{\ell}=\prod_{i\in\Omega_F} \mathcal{B}^{k\,J_i}_{\ell},
\nonumber\\
\mathcal{B}^{k\,J}_{\ell} &=&
 \sum_{m'=0}^{[(J-k)/2]}
\frac{1}{4^{m'}m'!\, (J-k-2m')!\, (c^{J-\ell})_{m'}}=
\nonumber\\
&=&\frac{1}{(J-k)!} \phantom{.}_2F_1\left(
-\frac{J-k}{2},\,
\frac{1-J+k}{2};\,
c^{J-\ell};\,
1
\right) =\nonumber\\
&=&
\frac{(c^{J-\ell-r^*_k})_{r_k}}{(J-k)! (c^{J-\ell})_{r_k}}, \nonumber\\
r_k&=&\left[\frac{J-k}{2}\right],
\qquad 
r^*_k=r_k+(-1)^{J-k+1}/2,
\qquad
c^{J-\ell}=c^J+\ell
.
\label{eq:BellHARM0}
\end{eqnarray}
After some simplifications we have
\begin{eqnarray}
B^{k\,\vec{J}}_{\ell}&=&
\frac{(\lambda_k-2\ell)_{\ell}\,(\vec{J}-\vec{k}+2\vec{\ell})!}{4^{\ell}\ell!\,(\vec{J}-\vec{k})!}
\frac{(c^{k+1/2})_{\vec{\ell}}}{(c^{\vec{J}-\vec{r}_k})_{\vec{\ell}}\,
		(c^{\vec{J}-\vec{r}_k^{\,*}})_{\vec{\ell}}
		}=\nonumber\\
&=&
\frac{(-1)^{\ell}(1-\lambda_k)_{2\ell}\,
(1-\lambda_k)_{\ell}\,
(\vec{J}-\vec{k}+2\vec{\ell})!
}{
4^{\ell}\ell!\, (\vec{J}-\vec{k})!\,
(c^{\frac{\vec{J}+\vec{k}}{2}})_{\vec{\ell}}\,
(c^{\frac{\vec{J}+\vec{k}+1}{2}})_{\vec{\ell}}
}, 
\label{eq:BellHARM}		
\end{eqnarray}
\begin{equation}
\vec{r}_k=\{r_{k\,1},r_{k\,2}\},\; \vec{r}_k^{\,*}=\{r_{k\,1}^*,r_{k\,2}^*\},\qquad r_{k\,i}=r_k|_{J\to J_i}.
\label{def:vectorsrk}
\end{equation}
  This solution is simpler than~(\ref{eq:irrFcoefsSolution}), that is why the generating function can be expressed in terms of hypergeometric/Gegenbauer functions~(\ref{eq:genfunFstarFinal}).

\subsection{\bf Spin-J propagator tensor structure}
\label{subsection:Propagator}

For the propagator of spin-J particle we have the following tensor structure, which can be obtained
from $\mathcal{W}^{\{J,J\}}$ for zero vector $P$.

\begin{equation}
\mathcal{P}^J(q)=\sum_{n=0}^{[J/2]} \frac{n!}{(c^J)_n}S^{P\; J}_n,
\label{eq:propagatorJ}
\end{equation}
where $S^{P\; J}_n$ is defined in~(\ref{def:SP}) and $c^J$ can be found in~(\ref{eq:irrepcJ}).

\subsection{\bf 3-reggeon vertex}
\label{subsection:Y}

\subsubsection{Standard basis}
\label{subsection:Ystand}
Using again basic notations, the general definitions in Appendix~\ref{appendixA} and~(\ref{def:anyirrep}), where we substitute $\mathcal{Y}$ and $y$, expansion~(\ref{eq:genYtrace}) and the method of generating functions~\cite{genfun0}-\cite{genfun5} developed 
in Appendix~\ref{subsection:genfun} with definitions~(\ref{def:SY}), we can write recurrent equations
for $y^{\vec{k}'(\vec{k};\,\vec{J})}_{\vec{n}}$:
\begin{eqnarray}
&&\hspace*{-0.2cm} 
y^{\vec{k}'}_{\vec{n}} =\frac{1}{2(c^{J_i}+n_i-1)} 
  \left( 
P_i^2 y^{\vec{k}'}_{\vec{n}-\vec{e}_i}
 + \left[
 2k'_{rs} y^{\vec{k}'+\vec{e}_{ir}+\vec{e}_{is}-\vec{e}_{rs}}_{\vec{n}-\vec{e}_i}
 +
2\hat{\lambda}^i_{rs} J^*_r J^*_s y^{\vec{k}'+\vec{e}_{ir}+\vec{e}_{is}}_{\vec{n}-\vec{e}_i}
\right]+
\phantom{  \sum_{j\neq i\atop j\in \Omega_Y} \hat{O}^{+2}_{k'_{ij}}}
\right.\nonumber\\
&& \phantom{\hspace*{-0.2cm} \hat{\mathcal{A}}^Y_i = \hat{O}^{-1}_{n_i}
	\left( 
	\bar{\kappa}_i\right.}\left. +
\sum_{j\neq i\atop j\in \Omega_Y}
\left[ 
2\bar{\chi}_{ij} J^*_j y^{\vec{k}'+\vec{e}_{ij}}_{\vec{n}-\vec{e}_i}
+
2n_j y^{\vec{k}'+2\vec{e}_{ij}}_{\vec{n}-\vec{e}_i-\vec{e}_j}
\!-\!\tilde{\lambda}^j_{ij} J^*_j \left( J^*_j\!-\!1\right) y^{\vec{k}'+2\vec{e}_{ij}}_{\vec{n}-\vec{e}_i}
\right]  \hspace*{-0.12cm}
\right)\nonumber\\
&&\hspace*{-0.2cm}  r,s,i,j\in\Omega_Y;\quad i\neq r\neq s.\,
\label{eq:irrYcoefs}
\end{eqnarray}
$J^*_i$ and $\Omega_Y$ are defined in~(\ref{def:Jbar}),(\ref{def:qgpownonconserv}).

And the general solution is
\begin{eqnarray}
&&\mathbf{y}^{\vec{k}'}_{\vec{n}} 
= \frac{1}{3!}
\sum\limits_{ \{i,r,s\}\atop\in \{1,2,3\}_P}
\mathbf{A}_i\,\,
\!\textcolor{red}{
	\hat{\mathcal{S}}
	\left[ 
	\phantom{\frac{X^{X^{X^{X^\frac{X^X}{X^X}}}}}{X^{X^{X^{X^\frac{X^X}{X^X}}}}}}\hspace*{-1.7cm}
	\right.
}
\left(\frac{\mathbf{A}_r}{\mathbf{A}_i}\right)^{ n_i=0}_{ n_{r,s}\neq 0}
\textcolor{blue}{
	\hat{\mathcal{S}}_{a_{is}}
	\left[ 
	\phantom{\frac{X^{X^{X^\frac{X^X}{X^X}}}}{X^{X^{X^\frac{X^X}{X^X}}}}}\hspace*{-1.5cm}
	\right.
}
\left(\frac{\mathbf{A}_s}{\mathbf{A}_r}\right)^{ n_{i,r}=0}_{ n_s\neq 0}
\textcolor{green}{
	\hat{\mathcal{S}}_{4,5}
	\left[ 
	\phantom{\frac{X^{X^{X^X}}}{X^{X^{X^X}}}}\hspace*{-1.1cm}
	\right.
}
\frac{\mathbf{\delta}_{\vec{k}'\vec{k}}}{\mathbf{A}^{\vec{k}}_{s; \vec{0}}}
\textcolor{green}{
	\left. 
	\phantom{\frac{X^{X^{X^X}}}{X^{X^{X^X}}}}\hspace*{-1.1cm}
	\right]
}
\textcolor{blue}{
	\left. 
	\phantom{\frac{X^{X^{X^\frac{X^X}{X^X}}}}{X^{X^{X^\frac{X^X}{X^X}}}}}\hspace*{-1.5cm}
	\right]
}
\textcolor{red}{
	\left. 
	\phantom{\frac{X^{X^{X^{X^\frac{X^X}{X^X}}}}}{X^{X^{X^{X^\frac{X^X}{X^X}}}}}}\hspace*{-1.74cm}
	\right]
}
\nonumber	\\
&& \hat{\mathcal{S}}_{...} \equiv \hat{\mathcal{S}}^{\;\vec{\zeta}^{\,Y}_i,\; \mathcal{M}^Y_{\Delta}}_{...} ,\quad
\mathbf{A}_i \equiv \mathbf{A}^{\vec{k}'}_{i;\; \vec{n}}
,\quad  a_{is}\equiv
\begin{cases} 4 & i<s\\ 5 & i>s \end{cases}
\end{eqnarray}
$\{1,2,3\}_P$ are all possible permutations of $\{1,2,3\}$. Notations 
for $\mathbf{A}_i$ and operators $\hat{\mathcal{S}}_{...}$, the detailed 
derivation and solution of the above equations are discussed in 
Appendices~\ref{subsection:genfun} and~\ref{subsection:Ysolve} in~(\ref{eq:irrYcoefsChVars})-(\ref{inih:step4aY}).  


\subsubsection{Harmonical basis}
\label{subsection:Yharm}

As in the previous case we use~(\ref{def:genYvarsHARM}),(\ref{eq:genYtraceDiffHARM}) plus 
general definitions and obtain the following equations:
\begin{eqnarray}
&&\hspace*{-0.2cm} 
\mathbf{y}^{\vec{\ell}}_{\vec{n}} = -\frac{1}{2(\lambda_i-2\ell_i+n_i-1)} 
\left( 
P_i^2 \mathbf{y}^{\vec{\ell}}_{\vec{n}-\vec{e}_i}
+ 
2\sum_{r\neq i} n_r \mathbf{y}^{\vec{\ell}-\vec{e}_i-\vec{e}_r}_{\vec{n}-\vec{e}_i-\vec{e}_r}
+
2 \mathbf{y}^{\vec{\ell}-\vec{e}_i}_{\vec{n}-\vec{e}_i}
\right)\nonumber\\
&& y^{\vec{k}'}_{\vec{n}}=\mathbf{y}^{(\vec{\kappa}-\vec{\kappa}')/2}_{\vec{n}},\qquad
\mathbf{y}^{\vec{0}}_{\vec{0}}=1,\qquad
\mathbf{y}^{\vec{\ell}}_{\vec{0}}=B_{\vec{\ell}}, \nonumber\\
&& \vec{l}\in\left[\vec{0},\left[ \frac{\vec{\kappa}}{2}\right]\right],\qquad
\vec{n}\in \left[\vec{0},\left[ \frac{\vec{J}-\vec{\kappa}}{2}\right]+\vec{\ell}\,\right],\qquad
\lambda_i=(D-2)/2+\kappa_i,
\nonumber\\
&&  r,i\in\Omega_Y;\quad i\neq r\,
\label{eq:irrYcoefsHARM}
\end{eqnarray}
This equation is similar to~(\ref{eq:irrFcoefsHARM}), and we could solve 
it by the same method. But the most convenient way is to use
the expansion~(\ref{eq:irrepHarmProjT}) and a closed expression for 
$\square^{\vec{m}}_{\vec{\omega}}$~(\ref{squaremYHARM}). We
can expand all the summations written in compressed form and
get a hypergeometric like function~(\ref{YHARMgenfun}).






\subsection{\bf Inclusive reggeon-reggeon hadronic tensor (forward amplitude) and form-factors}
\label{subsection:H}

Here we consider a solution for $\mathcal{H}$, similar to the method presented in subsection~\ref{subsection:Y}. 

\subsubsection{Standard basis}
\label{subsection:Hstand}
Here 
we substitute $\mathcal{H}$ and $h$ to~(\ref{def:anyirrep}), and use the 
expansion~(\ref{eq:genHtrace}) and the method of Appendix~\ref{subsection:genfun} 
with definitions~(\ref{def:SH}). Then we write recurrent equations
for $h^{\vec{k}'(\vec{k};\,\vec{J})}_{\vec{n}}$:

\begin{eqnarray}
&&\hspace*{-2mm} h^{\vec{k}'}_{\vec{n}} =\frac{1}{2(c^{J_i}+n_i-1)} \times\nonumber\\
&&\hspace*{-2mm} \left(
\phantom{ \sum_{r\neq i,i'\atop r\in \Omega_H} 
\left[ 
2\chi_m J^*_r \hat{O}^{+1}_{k'_{ir}} \right]
\hspace*{-2.9cm}}
h^{\vec{k}'}_{\vec{n}-\vec{e}_i} 
+
2J^*_i h^{\vec{k}'+\vec{e}_{ii'}}_{\vec{n}-\vec{e}_i} 
+
2n_{i'} h^{\vec{k}'+2\vec{e}_{ii'}}_{\vec{n}-\vec{e}_i-\vec{e}_{i'}} 
+
2k'_{jj'} 
h^{\vec{k}'+\vec{e}_{ij}+\vec{e}_{ij'}-\vec{e}_{jj'}}_{\vec{n}-\vec{e}_i} 
-
2\lambda_m  J^*_{j} J^*_{j'}
h^{\vec{k}'+\vec{e}_{ij}+\vec{e}_{ij'}}_{\vec{n}-\vec{e}_i} 
+\right.\nonumber\\
&&\hspace*{-2mm} \left. + \sum_{r\neq i,i'\atop r\in \Omega_H} 
\left[ 
2\chi_m J^*_r h^{\vec{k}'+\vec{e}_{ir}}_{\vec{n}-\vec{e}_i} 
+
2k'_{i'r}
h^{\vec{k}'+\vec{e}_{ii'}+\vec{e}_{ir}-\vec{e}_{i'r}}_{\vec{n}-\vec{e}_i} 
+
2n_r 
h^{\vec{k}'+2\vec{e}_{ir}}_{\vec{n}-\vec{e}_i-\vec{e}_r} 
-
\lambda_m J^*_r \left( J^*_r-1\right) 
h^{\vec{k}'+2\vec{e}_{ir}}_{\vec{n}-\vec{e}_i} 
\right]
\right),\nonumber\\
&&\hspace*{-2mm}  r,i,i',j,j'\in\Omega_H,\qquad \{j,j'\} =\{i^{\star},{i^{\star}}'\}\neq \{i,i'\}.
\label{eq:irrHcoefs}
\end{eqnarray}
$J^*_i$ and $\Omega_H$ are defined in~(\ref{def:Jbar}),(\ref{def:qgpownonconserv}).


And the general solution is
\begin{eqnarray}
 \mathbf{h}^{\vec{k}'}_{\vec{n}} 
&=& \frac{1}{4!}\!\!\!\!\!\!
\sum\limits_{ \{i,j,r,s\}\atop\in \{1,1',2,2'\}_P}\!\!\!
\!\!\!\!\!\!\!\mathbf{A}_i\;
\textcolor{red}{
	\hat{\mathcal{S}}\!\!
	\left[ 
	\phantom{\frac{X^{X^{X^{X^\frac{X^X}{X^X}}}}}{X^{X^{X^{X^\frac{X^X}{X^X}}}}}}\hspace*{-1.8cm}
	\right.
}
\left(\!\frac{\mathbf{A}_j}{\mathbf{A}_i}\!\right)^{\!\! n_i=0}_{\!\! n_{j,r,s}\neq 0}
\!\textcolor{magenta}{
	\hat{\mathcal{S}}_{a_{rsj}}\!\!
	\left[ 
	\phantom{\frac{X^{X^{X^\frac{X^{\frac{X}{X}}}{X^X}}}}{X^{X^{X^\frac{X^{\frac{X}{X}}}{X^X}}}}}\hspace*{-1.7cm}
	\right.
}
\left(\!\frac{\mathbf{A}_s}{\mathbf{A}_j}\!\right)^{\!\! n_{i,j}=0}_{\!\! n_{r,s}\neq 0}
\times
\nonumber\\
&& \hspace*{1.4cm}\times\,
\,\textcolor{blue}{
	\hat{\mathcal{S}}_{\{a_1,a_2\}_{rs}}\!\!
	\left[ 
	\phantom{\frac{X^{X^{X^\frac{X}{X^X}}}}{X^{X^{X^\frac{X}{X^X}}}}}\hspace*{-1.6cm}
	\right.
}
\left(\!\frac{\mathbf{A}_r}{\mathbf{A}_s}\!\right)^{\!\! n_{i,j,s}=0}_{\!\! n_r\neq 0}\!\!\!
\textcolor{green}{
	\;\hat{\mathcal{S}}_{10,11,12}\!\!
	\left[ 
	\phantom{\frac{X^{X^{X^X}}}{X^{X^{X^X}}}}\hspace*{-1.2cm}
	\right.
}
\frac{\mathbf{\delta}_{\vec{k}'\vec{k}}}{\mathbf{A}^{\vec{k}}_{r; \vec{0}}}
\textcolor{green}{
	\left. 
	\phantom{\frac{X^{X^{X^X}}}{X^{X^{X^X}}}}\hspace*{-1.2cm}
	\right]
}
\textcolor{blue}{
	\left. 
	\phantom{\frac{X^{X^{X^\frac{X}{X^X}}}}{X^{X^{X^\frac{X}{X^X}}}}}\hspace*{-1.6cm}
	\right]
}
\textcolor{magenta}{
	\left. 
	\phantom{\frac{X^{X^{X^\frac{X^{\frac{X}{X}}}{X^X}}}}{X^{X^{X^\frac{X^{\frac{X}{X}}}{X^X}}}}}\hspace*{-1.5cm}
	\right]
}
\textcolor{red}{
	\left. 
	\phantom{\frac{X^{X^{X^{X^\frac{X^X}{X^X}}}}}{X^{X^{X^{X^\frac{X^X}{X^X}}}}}}\hspace*{-1.6cm}
	\right]
}
\nonumber\\
 \hat{\mathcal{S}}_{...} &\equiv& \hat{\mathcal{S}}^{\;\vec{\zeta}^{\,H},\; \mathcal{M}^H_{\Delta}}_{...} ,\quad
\mathbf{A}_i \equiv \mathbf{A}^{\vec{k}'}_{i;\; \vec{n}} \phantom{\frac{X^X}{X^X}}
\nonumber\\
  a_{rsj}&\equiv&
\begin{cases} 10 & r=j^*, s={j^*}'\\ 11 & r=j', s={j^*}' \\ 12 & r=j', s=j^* \end{cases}\quad,\quad
 \{a_1,a_2\}_{rs}\equiv
\begin{cases} \{10,11\} & r={s^*}'\\ \{10,12\} & r=s^* \\ \{11,12\} & r=s' \end{cases}
\nonumber\\
 1^*&\!=&\!2,\quad 2^*=1,\quad (1')'=1,\quad (2')'=2\phantom{\frac{X^X}{X^X}}
\end{eqnarray}
$\{1,1',2,2'\}_P$ are all possible permutations of $\{1,1',2,2'\}$. Notations 
for $\mathbf{A}_i$ and operators $\hat{\mathcal{S}}_{...}$,
the detailed derivation and solution of the above equations are discussed in 
Appendix~\ref{subsection:genfun} and~\ref{subsection:Hsolve} in~(\ref{eq:irrHcoefsChVars})-(\ref{inih:step4a}). 

\subsubsection{Harmonical basis}
\label{subsection:Hharm}
Using definitions~(\ref{def:genHallvars})-(\ref{def:genfunH}) and usual tricks of 
Appendix~\ref{appendixB} we obtain the equations
\begin{eqnarray}
h^{\vec{k}'}_{\vec{n}} 
&=&
-\frac{1}{2(\frac{D-2}{2}+\kappa_i'+n_i-1)}
\left\{
	h^{\vec{k}'}_{\vec{n}-\vec{e}_i}+
	2\sum_{r\neq i} n_r h^{\vec{k}'+2\vec{e}_{ir}}_{\vec{n}-\vec{e}_i-\vec{e}_r}+
	2 \sum_{s,r\neq i\atop s>r} h^{\vec{k}'+\vec{e}_{ir}+\vec{e}_{is}-\vec{e}_{rs}}_{\vec{n}-\vec{e}_i}
	\right\}
\nonumber\\
&&i,r,s\in \Omega_H
\label{eq:irrHcoefsHARM}
\end{eqnarray}

As in the case~(\ref{eq:irrYcoefsHARM}) it is more 
convenient to use direct expansion~(\ref{eq:irrepHarmProjT}) and 
a closed expression for 
$\square^{\vec{m}}_{\vec{\omega}}$~(\ref{squaremHHARM}). 
And then expand all the summations to
obtain a hypergeometric like function~(\ref{HHARMgenfun}).




\section{Discussions and conclusions}
\label{section:disconc}

Based on the picture of reggeon as a “reggeized” tensor quantum field of arbitrary integer spin in 
a $D$-dimensional Minkowski space, this work derives the hadronic irreducible 
tensors that represent the amplitudes of all the important diffractive processes. The 
corresponding cross sections can be obtained by various contractions of these tensors.

A generalization to the case of non‑conserved currents is also considered, in which the final 
tensors can be decomposed into irreducible ones and momenta transfers.

It should be noted that the “non‑conservation” of currents may be an artifact 
of higher-dimensional space-time. For this reason we have considered here a 
“toy” model of Minkowski space with an arbitrary number of dimensions, in order to understand 
how it affects the behavior of diffractive cross 
sections. For example, if the currents are conserved, then the differential cross 
section for single dissociation as a function of the momentum transfer squared 
must exhibit a minimum 
at very small values of  the momentum transfer squared, as was shown in Refs.~\cite{mySDandPomCS,myVisualization}. Those 
works also demonstrated that unitarization can somewhat 
soften this behavior, but the trend of decreasing cross section at small momentum transfer 
squared remains. Consequently, some authors~\cite{CloseIn},\cite{CloseOut} have studied a case
with a non‑conserved current (which, however, may be conserved in a space with extra dimensions, so 
that its projection onto 
our four‑-dimensional space-time appears effectively “non‑conserved.”).  Of course, in future 
work one should consider more realistic scenarios with curved extra dimensions, for example an anti-de Sitter bulk.

As already mentioned in the Introduction, we apply a tensorial formalism here for transparency. In 
principle one could obtain similar results using approaches based on “continuous” spin and Wigner’s 
equations~\cite{CSPin}-\cite{CSPout}, on higher spin 
field theories, hyperfields and string methods, or by using the 
helicity formalism~\cite{helicityIn}-\cite{helicityOut} to 
compute amplitudes (which will 
be done later to generalize this approach to vertices with half‑-integer spins and their reggeized analogs). The 
helicity formalism becomes especially important for multi‑-Reggeon vertices, since the number of recurrent equations 
grows with the number of Reggeon legs in amplitudes and the solutions for coefficient functions become cumbersome.

Particular attention must be paid to the possible behavior of the form--factors at small momentum transfer, which is
one of characteristics for diffractive processes. As will be shown in the second part of this study, the properties of 
form factors at low momentum transfer (finiteness of the hadronic tensors) impose additional constraints 
on the hadronic tensors, allowing one to simplify the cross-‑section expressions significantly and 
to extract these form-factors by the use of existing experimental data.

The solutions for irreducible tensors obtained in this work may be applied not only to diffractive processes 
but also to the amplitudes in effective models for higher spin particles, since the 
approach presented here is mathematically quite general.

\section*{Appendices}
\appendix 

\section{Basic notations for momenta and simple transverse tensor structures}
\label{appendixA}

Notations for momenta of different processes and vertexes can be found in Figs.~\ref{fig:alldiffprocs}, \ref{fig:diffractiveLego}.

Let us introduce basic variables and transverse structures, which we use to construct irreducible tensor representations.

\subsection{\bf Momenta which are orthogonal to momenta transfers}
\label{appendixA:momenta}

\begin{itemize}
\renewcommand\labelitemi{$\bullet$}
\item In $\mathcal{V}^J(p,q)$, $\mathcal{W}^{\vec{J}}(p,q)$:

\vspace*{0.3cm}
Momentum which is orthogonal to $q$:
\begin{eqnarray}
 P_{\alpha}&\equiv& \frac{\left( p_{\alpha}-\frac{pq}{q^2}q_{\alpha}\right)}{\sqrt{p^2-(pq)^2/q^2}}\equiv 
\frac{\left( p_{\alpha}-\frac{q_{\alpha}}{2}\right)}{\sqrt{m^2+Q^2/4}},\nonumber\\
 Q^2&=&-q^2\equiv -t,\quad
P_{\alpha}q_{\alpha} 
=
0.
\phantom{X^{X^X}}\label{eq:VWPtransversity}
\end{eqnarray}
\vspace*{0.1cm}

\item In $\mathcal{F}^{\vec{J}}(q_1,q_2)$:

\vspace*{0.3cm}
Momenta which are orthogonal to $q_1,q_2$:
\begin{eqnarray}
 p_c&=&q_1+q_2,\; M_c^2=p_c^2,\nonumber\\
 P_{i\;\alpha}&\equiv& \frac{\left( p_{c\;\alpha}-\frac{p_c q_i}{q_i^2}q_{i\;\alpha}\right)}{\sqrt{M_c^2-(p_cq_i)^2/q_i^2}}\equiv \nonumber\\
&\equiv& \frac{2Q_i\left( p_{c\;\alpha}-\frac{p_c q_i}{q_i^2}q_{i\;\alpha}\right)}{\lambda^{1/2}(M_c^2,q_1^2,q_2^2)}\equiv
\frac{2Q_i\left( q_{i^{\star}\;\alpha}-\frac{q_1q_2}{q_i^2}q_{i\;\alpha}\right)}{\lambda^{1/2}(M_c^2,q_1^2,q_2^2)},\nonumber\\ 
 Q_i^2&=&-q_i^2\equiv -t_i,\quad i\in \{1,2\},\quad 
 P_{i\;\alpha}q_{i\;\alpha}=0 \phantom{\frac{X}{X}\hspace*{-9mm}}
 \label{eq:FPtransversity}
\end{eqnarray}

In this case it is convenient to introduce auxiliary momenta:
\begin{eqnarray}
 \tilde{q}_i&=&q_i/Q_i,\quad \tilde{q}_i^2=-1,\quad  \tilde{q}_i\tilde{q}_{i^{\star}}\equiv\tilde{q}_1\tilde{q}_2=1/\chi_m,\;\label{eq:tildeqi}\\
 \chi_m&=&\frac{Q_1 Q_2}{q_1q_2},\quad \lambda_m=1-\chi_m^2,\label{eq:chimandlambdam}\\
 P_i&=&\frac{1}{\sqrt{\lambda_m}}\left( \tilde{q}_i+\chi_m \tilde{q}_{i^{\star}}\right)\label{eq:Pibytildeqi}.
\end{eqnarray}

To obtain reggeon-hadron forward amplitude $\mathcal{W}^{\vec{J}}(p,q)$ from $\mathcal{F}^{\vec{J}}(q_1,q_2)$
we have to change $2\to 1'$, $i^{\star}\to  i'$ and then
\begin{equation}
 P_1\to P, \quad P_{1'}\to P,\quad q_1\to q,\quad q_{1'}\to -q\quad\Rightarrow\quad
\chi_m\to 1,\quad \lambda_m\to 0.
\label{eq:FtoWrelation}
\end{equation}
This is useful relation to obtain solutions for the 
reggeon-hadron forward amplitude from solutions for the reggeon-reggeon fusion process.
\vspace*{0.7cm}

\item In $\mathcal{Y}^{\vec{J}}(q_1,q_2)$:

\vspace*{0.3cm}
Momenta which are orthogonal to $q_1,q_2,q_3$:
\begin{eqnarray}
&& \hspace*{2cm}q_3+q_1+q_2=0,  \phantom{\frac{X}{X}\hspace*{-9mm}}
\nonumber\\
&& P_{i\;\alpha}\equiv 
\frac{ 
(q_iq_s)q_{r\;\alpha}-(q_iq_r)q_{s\;\alpha}
}{Q_i\sqrt{|\mathcal{K}|}},\, r<s;\qquad
 P_{i\;\alpha}q_{i\;\alpha}=0.\label{eq:YPtransversity}
\end{eqnarray}

Let us introduce also auxiliary quantities:
\begin{eqnarray}
\hspace*{-5mm}\mathcal{K} &\equiv& (q_iq_j)^2-q_i^2 q_j^2 = \frac{1}{4} \lambda (q_1^2,q_2^2,q_3^2)
,\nonumber\\
\hspace*{-5mm} Q_i^2&=&-q_i^2\equiv -t_i,\quad P_i^2 = \mathrm{sign}[-q_i^2 \mathcal{K}],\quad 
i,j\in \{1,2,3\},\, i\neq j,
  \phantom{\frac{X}{X}\hspace*{-9mm}}\nonumber\\
\hspace*{-5mm} \chi_{ij} &=& \frac{Q_iQ_j}{|q_iq_j|},\quad
\lambda_{ij}=1-\mathrm{sign}[q_i^2 q_j^2]\cdot\chi_{ij}^2
,\phantom{\frac{X}{X}\hspace*{-9mm}}\nonumber\\
\hspace*{-5mm}
\bar{\eta}_{ij}&\equiv&
\begin{cases} \mathrm{sign}[s-i] & j=r\\ \mathrm{sign}[i-r] & j=s\\ r<s, & r,s\neq i \end{cases}
,\quad r,s\in \{1,2,3\}
,\phantom{\frac{X}{X}\hspace*{-9mm}}\nonumber\\
\hspace*{-5mm} \hat{\lambda}^i_{rs} &\equiv& \hspace*{-0.4cm}
\prod_{a\neq b\atop a,b\in\{1,2,3\}}\hspace*{-0.4cm}
\mathrm{sign}[q_aq_b]\cdot 
\mathrm{sign}[i-s] \cdot\mathrm{sign}[i-r]\cdot
\chi_{rs}\sqrt{|\lambda_{ir}||\lambda_{is}|},\nonumber\\
\tilde{\lambda}^r_{ir}&\equiv& \mathrm{sign}[-q_r^2]\cdot |\lambda_{ir}|,\quad 
\bar{\chi}_{ij} = \mathrm{sign}[-q_i^2 \cdot q_iq_j]\cdot  \bar{\eta}_{ij} \chi_{ij}.
\phantom{\frac{X}{X}\hspace*{-9mm}}\label{eq:auxYquantities}
\end{eqnarray}

Some relations between different momenta and quantities are valid:
\begin{equation}
Q_1 P_1 +Q_3 P_3 = Q_2 P_2,
\quad Q_s\frac{\sqrt{|\lambda_{is}|}}{\chi_{is}} = 
 Q_r\frac{\sqrt{|\lambda_{ir}|}}{\chi_{ir}}.
\label{eq:Ymomentarelations}
\end{equation}
\vspace*{0.2cm}

\item In the forward (space-like) limit 
$
\mathcal{Y}_{fwd}^{\vec{J}}(q,n)=\mathcal{Y}^{\vec{J}}(q_1,q_2)|_{\Omega_{fwd}}
$, where
\begin{empheq}[box=\fcolorbox{red}{white}]{align}
 \Omega_{fwd}:\quad
    q_1&\to q+\frac{\epsilon}{2}\cdot n,\quad
	q_2\to -q+\frac{\epsilon}{2}\cdot n, \quad 
	q_3=-\epsilon\cdot n\nonumber\\
n^2&=\tilde{q}^2=-1, \quad qn=\tilde{q}n=0, \quad
\epsilon\to 0\label{eq:YfwdLimit}
\end{empheq}

\begin{equation}
P_{1,2}=\tilde{q},\quad P_3=n, \qquad
Q^2=-q^2, \quad \tilde{q}=q/Q,\qquad
P_i^2=-1.
\label{eq:fwdYmomenta}
\end{equation}	
\vspace*{0.2cm}

\item In $\mathcal{H}^{\vec{J}}(q_1,q_2)$:

\vspace*{0.3cm}
Notations for momenta which are orthogonal to $q_1,q_2$ are similar
to notations for $\mathcal{F}^{\vec{J}}(q_1,q_2)$. Only the difference is that
we have additional groups of indexes $(1')$ and $(2')$, for which momenta are
the same as for $(1)$ and $(2)$. Below we always assume that 
$$
P_{1'}=P_{1},\quad P_{2'}=P_{2},
$$
i.e. pairs of indexes $(1,1')$, $(2,2')$ are related to the forward scattering amplitude 
with identical initial and final momenta, and other pairs with different indexes and momenta ($(1,2)$, $(1'2')$, $(1',2)$, $(1,2')$) are related to the fusion 
channel or cross-channel. 
\end{itemize}

Let us remind you that $P_{1'}=P_1$ defines the four-vector $P_{1'}$ from the vector $P_1$, but
$$P_{(1)}\equiv P_{\mu_i}\neq P_{(1')}\equiv P_{\mu'_i}$$ defines the same vector $P$, but with different Lorentz indices.

\subsection{\bf Transverse tensors of the second rank}
\label{appendixA:G}

In the standard basis we use tensors orthogonal to one of the momenta transfer  
\begin{eqnarray}
&& G_{(rr)}\equiv g_{(rr)}-\frac{q_{r(r)}q_{r(r)}}{q_r^2}=g_{(rr)}+\tilde{q}_{r(r)}\tilde{q}_{r(r)}, \label{def:Grr}\\
&& G_{(rr')}\equiv g_{(rr')}-\frac{q_{r(r)}q_{r(r')}}{q_r^2}=g_{(rr')}+\tilde{q}_{r(r)}\tilde{q}_{r(r')},\label{def:Grrprime}\\
&& \hat{G}_{(r)(s)}\equiv g_{(rs)}-\frac{q_{s(r)}q_{r(s)}}{q_rq_s}=g_{(rs)}-\chi_m\, \tilde{q}_{s(r)}\tilde{q}_{r(s)},\quad 
s\neq r,r',\label{def:Ghatrs}\\
&& \phantom{...}\nonumber\\
&& G_{(r\mathbf{r})}\otimes q_{r(\mathbf{r})} = G_{(r\mathbf{r}')}\otimes q_{r(\mathbf{r}')} = 0,\quad 
\hat{G}_{(\mathbf{r})(s)}\otimes q_{r(\mathbf{r})} = \hat{G}_{(r)(\mathbf{s})}\otimes
q_{s(\mathbf{s})} = 0. 
\label{def:Gqrule}
\end{eqnarray}
In the harmonical basis we use symmetric tensors orthogonal to two different momenta
\begin{equation}
\mathcal{G}_{\alpha\beta} (v_1,v_2) = g_{\alpha\beta} + \frac{v_2^2 v_{1\alpha} v_{1\beta} +
v_1^2 v_{2\alpha} v_{2\beta} -(v_1v_2) (
v_{1\alpha} v_{2\beta} + v_{2\alpha} v_{1\beta}
)	
}{
(v_1v_2)^2-v_1^2 v_2^2
}
\label{def:GHarmrs}
\end{equation}
\begin{equation}
\mathcal{G}_{\alpha\beta}v_{1\alpha} = \mathcal{G}_{\alpha\beta}v_{2\alpha} = 0,\qquad v_{1,2}\in\mathbb{M}_D.
\end{equation}

\vspace*{0.1cm}

\begin{itemize}
\renewcommand\labelitemi{$\bullet$}
\item In $\mathcal{V}^J(p,q)$ we have only one structure $G_{(rr)}$ like in~(\ref{def:Grr}) and
use only one basis, since the vertex has a single tensor leg. 

\vspace*{0.3cm}

\item In $\mathcal{W}^{\{J_1,J_{1'}\}}(p,q)$ and $\mathcal{P}^J(q)$ (propagator)
we have only structures $G_{(11)}$, $G_{(11')}$ and  $G_{(1'1')}$ like 
in~(\ref{def:Grr}),(\ref{def:Grrprime}) and in harmonical basis we use
\begin{equation}
\mathcal{G}_{(rs)} =\mathcal{G}_{(rs)} (p,q) = G_{(rs)} - P_{(r)} P_{(s)},\qquad r,s=\{1,1'\}
\label{eq:GharmWeqs}
\end{equation}

\item In $\mathcal{F}^{\{J_1,J_2\}}(q_1,q_2)$ we have 
 $G_{(11)}$, $\hat{G}_{(1)(2)}$ and  $G_{(22)}$
 like in~(\ref{def:Grr}),(\ref{def:Ghatrs}) and in harmonical basis we use
\begin{equation}
\mathcal{G}_{(rs)}\!\! \!=\!\! \mathcal{G}_{(rs)} (\tilde{q}_1,\tilde{q}_2)\! = \!
 G_{(rs)} - P_{r(r)} P_{r(s)} = \hat{G}_{(r)(s)} - \chi_m P_{r(r)} P_{s(s)}
\label{eq:GharmFeqs}
\end{equation} 

\vspace*{0.3cm}

\item In $\mathcal{Y}^{\vec{J}}(q_1,q_2)$ we have 
 $G_{(rr)}$, $\hat{G}_{(r)(s)}$, $r,s\in\{ 1,2,3\}$, 
like  in~(\ref{def:Grr}),(\ref{def:Ghatrs}) and in harmonical basis
$\mathcal{G}_{(rs)}(\tilde{q}_r,\tilde{q}_s)$

\vspace*{0.3cm}

\item In $\mathcal{Y}_{fwd}^{\vec{J}}(\tilde{q},n)$ we use
harmonical basis, since 
$\mathcal{G}_{(rs)}(\tilde{q},n)$ is well defined,
but $\hat{G}_{(i)(3)}$ are not defined in this limit.

\vspace*{0.3cm}

\item In $\mathcal{H}^{\vec{J}}(q_1,q_2)$ we have 
 $G_{(rr)}$, $G_{(rr')}$ and $\hat{G}_{(r)(s)}$, 
 like in~(\ref{def:Grr})-(\ref{def:Ghatrs}), $r,s\in\{1,1',2,2'\}$. And in 
 harmonical basis the tensor is like~(\ref{eq:GharmFeqs}).
\end{itemize}

\subsection{\bf Contractions of basic structures}
\label{appendixA:contractions}


\begin{itemize}
	\renewcommand\labelitemi{$\bullet$}
	\item in all the structures we have
	the following relations:
	\begin{eqnarray}
	&& G_{(i\mathbf{i})} \otimes P_{i\;(\mathbf{i})} = P_{i\;(i)},\nonumber\\
	&& G_{(\mathbf{i}\mathbf{i})} = D-1,\nonumber\\
	&& G_{(i\mathbf{i})} \otimes G_{(i\mathbf{i})} = G_{(ii)},\nonumber\\
	&& G_{(\mathbf{i}i)} \otimes \hat{G}_{(\mathbf{i}j)} = \hat{G}_{(ij)}. 
	\label{eq:allcGGP}
	\end{eqnarray}
	
	\item For $\mathcal{V}^J(p,q)$ and $\mathcal{W}^{\{J_1,J_{1'}\}}(p,q)$ we have also
	\begin{eqnarray}
	&& G_{(\mathbf{i}i')} \otimes P_{(\mathbf{i})} = P_{(i')},\nonumber\\ 
	&& G_{(i\mathbf{i}')} \otimes P_{(\mathbf{i}')} = P_{(i)}.
	\label{eq:VWcGP}
	\end{eqnarray}
	and
	\begin{eqnarray} 
	&& G_{(i\mathbf{i})} \otimes G_{(\mathbf{i}i')} = G_{(ii')},\nonumber\\ 
	&& G_{(i\mathbf{i}')} \otimes G_{(i\mathbf{i}')} = G_{(ii)},\nonumber\\
	&& G_{(\mathbf{i}i')} \otimes G_{(\mathbf{i}i')} = G_{(i'i')},
	\label{eq:VWcGG}
	\end{eqnarray}
	
	\item For $\mathcal{F}^{\{J_1,J_2\}}(q_1,q_2)$ we have
	\begin{eqnarray}
	&& \hat{G}_{(\mathbf{i})(j)} \otimes P_{i\;(\mathbf{i})} = \chi_m P_{j\;(j)},\nonumber\\
	&& \hat{G}_{(i)(\mathbf{j})} \otimes P_{j\;(\mathbf{j})} = \chi_m P_{i\;(i)}.
	\label{eq:FcGP}
	\end{eqnarray}
	and
	\begin{eqnarray}
	&& \hat{G}_{(i)(\mathbf{j})} \otimes \hat{G}_{(i)(\mathbf{j})} = G_{(ii)}-\lambda_m P_{i\;(i)} P_{i\;(i)},\nonumber\\
	&& \hat{G}_{(\mathbf{i})(j)} \otimes \hat{G}_{(\mathbf{i})(j)} = G_{(jj)}-\lambda_m P_{j\;(j)} P_{j\;(j)}.
	\label{eq:FcGG}
	\end{eqnarray}
	
	\item For $\mathcal{Y}^{\vec{J}}(q_1,q_2)$ we have
	\begin{eqnarray}
	&& \hat{G}_{(\mathbf{i})(j)} \otimes P_{i\;(\mathbf{i})} = \bar{\chi}_{ij} P_{j\;(j)},\nonumber\\
	&& \hat{G}_{(i)(\mathbf{j})} \otimes P_{j\;(\mathbf{j})} = \bar{\chi}_{ji} P_{i\;(i)}.
	\label{eq:YcGP}
	\end{eqnarray}
	and
	\begin{eqnarray}
	&& \hat{G}_{(i)(\mathbf{j})} \otimes \hat{G}_{(i)(\mathbf{j})} = G_{(ii)}-\tilde{\lambda}^i_{ij} P_{i\;(i)} P_{i\;(i)},\nonumber\\
	&& \hat{G}_{(\mathbf{i})(j)} \otimes \hat{G}_{(\mathbf{i})(j)} = G_{(jj)}-\tilde{\lambda}^j_{ij} P_{j\;(j)} P_{j\;(j)},
	\nonumber\\
	&& \hat{G}_{(\mathbf{i})(r)} \otimes \hat{G}_{(\mathbf{i})(s)} = \hat{G}_{(r)(s)}+\hat{\lambda}^i_{rs} P_{r\;(r)} P_{s\;(s)},\nonumber\\
	&& i,r,s\in\{1,2,3\},\quad i\neq r\neq s\phantom{\frac{X}{X}\hspace*{-9mm}}.
	\label{eq:YcGG}
	\end{eqnarray}
	
	\item For  $\mathcal{H}^{\vec{J}}(q_1,q_2)$ we have~(\ref{eq:allcGGP})-(\ref{eq:FcGG}), and in~(\ref{eq:FcGP}), (\ref{eq:FcGG}) $i$ and $j$ are from different groups of indices, for example $i\in \{1,1'\}$ and $j\in\{ 2,2'\}$ and vice versa.
	
\end{itemize}

Contractions of harmonical tensor $\mathcal{G}_{(rs)}$ with different tensors and momenta can be
obtained from its representations~(\ref{eq:GharmWeqs}),(\ref{eq:GharmFeqs}). Since $\mathcal{G}_{(rs)}$
is orthogonal to $P_{r,s}$, there are some simplifications in every equation.

\section{Tricks to obtain basic equations and their solutions}
\label{appendixB}

In this Appendix we consider all possible ways to obtain recurrent equations and their solutions
for coefficients of different TST structures from the tracelessness condition in each group of
indexes, i.e. for a tensor with several groups of indexes it looks as
\begin{equation}
Sp_i T^{\vec{J}} = 0, \qquad i\in\Omega_T
\label{notationsB}
\end{equation}

Let us first introduce some short notations for transverse and symmetric tensors in two different basises 
(standard and harmonical).

For $V$ we use only one basis, since it has only one group of indices:
\begin{equation}
 S_{n_1}^{V\, J_1} \equiv \left( P_{(1)}^{\otimes J_1-2n_1} G_{(11)}^{\otimes n_1}\right),\label{def:SV}\\
 \end{equation}
For $W$,$F$ we have two versions: 
 \begin{eqnarray}
 S_{k\,\vec{n}}^{F\,\vec{J}} &\equiv& \left(
 \hat{G}_{(1)(2)}^{\otimes k} \prod_{i\in\Omega_F}P_{i(i)}^{\otimes J^*_i}G_{(ii)}^{\otimes n_i}
\right)\quad\mathrm{and}\quad
 \left(
\mathcal{G}_{(12)}^{\otimes k} \prod_{i\in\Omega_F}P_{i(i)}^{\otimes J^*_i}\mathcal{G}_{(ii)}^{\otimes n_i}
\right),
 \label{def:SF}\\
 S_{k\,\vec{n}}^{W\, \vec{J}} &\equiv&\left(
G_{(11')}^{\otimes k} \prod_{i\in\Omega_W} P_{(i)}^{\otimes J^*_i}G_{(ii)}^{\otimes n_i}
\right)
\quad\mathrm{and}\quad
\left(
\mathcal{G}_{(11')}^{\otimes k} \prod_{i\in\Omega_W} P_{(i)}^{\otimes J^*_i}\mathcal{G}_{(ii)}^{\otimes n_i}
\right)
,\label{def:SW}
\end{eqnarray}
For the structures in the propagator of spin-$J$ particle we can obtain
\begin{equation}
S_{n}^{P\, J} =  S_{J-2n\; \{n,n\}}^{W\; \{J,J\}}
\equiv\left(
G_{(11')}^{\otimes J-2n} G_{(11)}^{\otimes n} G_{(1'1')}^{\otimes n}
\right). \label{def:SP}
\end{equation}
For $Y$ vertex we get structures
\begin{equation}
 S_{\vec{k}\,\vec{n}}^{Y\; \vec{J}} \equiv
 \left(
\prod_{i\in\Omega_Y}P_{i(i)}^{\otimes J^*_i}G_{(ii)}^{\otimes n_i}
\prod_{\forall r\neq s \atop r,s\in\Omega_Y} \hat{G}_{(r)(s)}^{\otimes k_{rs}}
\right)
\quad\mathrm{and}\quad
\left(
\prod_{i\in\Omega_Y}P_{i(i)}^{\otimes J^*_i}\mathcal{G}_{(ii)}^{\otimes n_i}
\prod_{\forall r\neq s \atop r,s\in\Omega_Y} \mathcal{G}_{(rs)}^{\otimes k_{rs}}
\right).
\label{def:SY}
\end{equation}

In forward case $Y_{fwd}$ we have (in harmonical basis):
\begin{equation}
S_{\vec{k}\,\vec{n}}^{Y_{fwd}\; \vec{J}} \equiv
\left(
n_{(1)}^{\otimes J^*_1}n_{(2)}^{\otimes J^*_2} \tilde{q}_{(3)}^{\otimes J^*_3}
\prod_{i\in\Omega_Y}\mathcal{G}_{(ii)}^{\otimes n_i}
\prod_{\forall r\neq s \atop r,s\in\Omega_Y} \mathcal{G}_{(rs)}^{\otimes k_{rs}}
\right).
\label{def:SYfwd}
\end{equation}

And for $H$ we have structures
\vspace*{-3mm}
\begin{equation}
S_{\vec{k}\,\vec{n}}^{H\; \vec{J}} \equiv 
\left(
\!\prod_{i=1,2} \!\!\!G_{(ii')}^{\otimes k_{ii'}}
\!\!\prod_{j\in\Omega_H}\!\!P_{j(j)}^{\otimes J^*_j}G_{(jj)}^{\otimes n_j}
\!\!\!\prod_{\forall r\neq r',s \atop r,s\in\Omega_H} \!\!\!\!
\hat{G}_{(r)(s)}^{\otimes k_{rs}}\!
\right)
\,\mathrm{and}\,
\left(\!
\prod_{i\in\Omega_H}\!\!P_{i(i)}^{\otimes J^*_i}\mathcal{G}_{(ii)}^{\otimes n_i}
\!\!\!\prod_{\forall r\neq s \atop r,s\in\Omega_H} \!\!\!\mathcal{G}_{(rs)}^{\otimes k_{rs}}
\!\right),
\label{def:SH}
\end{equation}
\vspace*{-2mm}
where $J^*_i=J_i-2n_i-\kappa_i$ and all indexes are defined in the notations.

\subsection{\bf Tensors extraction trick}
\label{subsection:trickFW}

This trick consists in extracting the basic simplest tensor structures from the tensor and contracting the metric with these structures. 

For the tensor $\mathcal{V}^J$ this method can be presented in the following form:
\vspace*{-3mm}
\begin{eqnarray}
  Sp_1  S_{n_1}^{V\, J_1} &=& g_{\mu_1\mu_2} 
\left[\phantom{\sum_{ij}  G_{\mu_1\mu_i} G_{\mu_2\mu_j} S_{n_1-2}^{V\, J_1-4}} \hspace*{-3.5cm}
P_{\mu_1}P_{\mu_2} S_{n_1}^{V\, J_1-2} 
+ \sum_i \left( P_{\mu_1}  G_{\mu_2\mu_i} +
P_{\mu_2}  G_{\mu_1\mu_i}
\right) S_{n_1-1}^{V\, J_1-3}+
\right. \nonumber\\
&+&\left. \sum_{i<j}  G_{\mu_1\mu_i} G_{\mu_2\mu_j} S_{n_1-2}^{V\, J_1-4}  + 
G_{\mu_1\mu_2} S_{n_1-1}^{V\, J_1-2}
\right]=\nonumber\\
&=& S_{n_1}^{V\, J_1-2}+2 
\frac{(J_1-2)\mathcal{N}^{J_1-3}_{n_1-1}}{\mathcal{N}^{J_1-2}_{n_1-1}} S_{n_1-1}^{V\, J_1-2}+\nonumber\\
&+& \frac{(J_1-2)(J_1-3)\mathcal{N}^{J_1-4}_{n_1-2}}{\mathcal{N}^{J_1-2}_{n_1-1}} S_{n_1-1}^{V\, J_1-2}+ 
(D-1) S_{n_1-1}^{V\, J_1-2}=\nonumber\\
&=& S_{n_1}^{V\, J_1-2} +(2(J_1-2n_1)+2(n_1-1)+D-1) S_{n_1-1}^{V\, J_1-2},
\label{eq:Spur1V}
\end{eqnarray}
\begin{eqnarray}
  Sp_1 \mathcal{V}^{J_1} &=&  
\hat{\mathcal{\upsilon}}_0(t) \sum_{n_1=0}^{\left[ J_1/2\right]} \, \mathcal{\upsilon}^{J_1}_{n_1} Sp_1  S_{n_1}^{V\, J_1} = 
\nonumber\\
\!\!&=&\hat{\mathcal{\upsilon}}_0(t)\,\left[ 
\sum_{n_1=1}^{\left[ J_1/2\right]}  S_{n_1-1}^{V\, J_1-2}
\left( 
\mathcal{\upsilon}^{J_1}_{n_1-1}  
- 2\mathcal{\upsilon}^{J_1}_{n_1} (c^{J_1}+n_1-1)
 \right)
\right]=0,
\label{eq:Vtensorout}
\end{eqnarray}
where
\begin{equation}
\label{eq:NumTermsSV}
\mathcal{N}^{J}_{n}=\frac{J!}{2^n n! (J-2n)!}
\end{equation}
is the number of terms in the tensor structure $S_{n}^{V\, J}$ and $c^J$ is defined in~(\ref{eq:irrepcJ}). From the above 
equation we obtain recurrent equations for $\mathcal{\upsilon}^{J_1}_{n_1}$, which have the simple solution~(\ref{eq:irrVcoefsSolution}).


For the tensor $\mathcal{F}^{\vec{J}}$ this method can be presented in the following form:\vspace*{-4mm}
\begin{eqnarray}
&Sp_1&  \!\!S_{k\,\vec{n}}^{F\, \vec{J}} = g_{\mu_1\mu_2} 
\left[\phantom{\sum_{ij}  G_{\mu_1\mu_i} G_{\mu_2\mu_j} S_{k,\,\{n_1-2,n_2\}}^{F\, \{J_1-4,J_2\}}} \hspace*{-4.3cm}
P_{1\mu_1}P_{1\mu_2} S_{k\, \vec{n}}^{F\, \vec{J}-2\vec{e}_1} +
 \sum_i \left( 
P_{1\mu_1}  G_{\mu_2\mu_i} +
P_{1\mu_2}  G_{\mu_1\mu_i} 
\right) S_{k\,\vec{n}-\vec{e}_1}^{F\, \vec{J}-3\vec{e}_1}+
\right. \nonumber\\
&+&\left. \hspace*{-0.1cm}
\sum_{i<j}  G_{\mu_1\mu_i} G_{\mu_2\mu_j} S_{k\,\vec{n}-2\vec{e}_1}^{F\, \vec{J}-4\vec{e}_1}  + 
\sum_{i,j} \left( 
G_{\mu_1\mu_i}  \hat{G}_{\mu_2\nu_j} +
G_{\mu_2\mu_i}  \hat{G}_{\mu_1\nu_j}  
\right) S_{k-1,\,\vec{n}-\vec{e}_1}^{F\, \vec{J}-3\vec{e}_1-\vec{e}_2}+
\right.\nonumber\\
&+& \left.\hspace*{-0.1cm}
G_{\mu_1\mu_2} S_{k,\,\vec{n}-\vec{e}_1}^{F\, \vec{J}-2\vec{e}_1}+
\sum_i \left( 
P_{1\mu_1}  \hat{G}_{\mu_2\nu_i} +
P_{1\mu_2}  \hat{G}_{\mu_1\nu_i}  
\right) S_{k-1,\,\vec{n}-\vec{e}_1}^{F\, \vec{J}-2\vec{e}_1-\vec{e}_2}+
\right.\nonumber\\
&+&\left.
\phantom{\sum_{ij}  G_{\mu_1\mu_i} G_{\mu_2\mu_j} S_{k,\,\vec{n}}^{F\, \vec{J}-4\vec{e}_1}} \hspace*{-3.8cm}
\sum_{i<j} 
\hat{G}_{\mu_1\nu_i}  \hat{G}_{\mu_2\nu_j}  S_{k-2\,\vec{n}}^{F\, \vec{J}-2\vec{e}_1-2\vec{e}_2}
\right]=\nonumber\\
 &=& S_{k\,\vec{n}}^{F\, \vec{J}-2\vec{e}_1}+ 2 
\frac{(J_1-2)
\mathcal{N}^{F\,\vec{J}-3\vec{e}_1}_{k\,\vec{n}-\vec{e}_1}}{\mathcal{N}^{F\,\vec{J}-2\vec{e}_1}_{k\,\vec{n}-\vec{e}_1}} 
S_{k\,\vec{n}-\vec{e}_1}^{F\, \vec{J}-2\vec{e}_1}+\nonumber\\
 &+& \frac{(J_1-2)(J_1-3)\mathcal{N}^{F\, \vec{J}-4\vec{e}_1}_{k\,\vec{n}-2\vec{e}_1}}{\mathcal{N}^{F\, \vec{J}-2\vec{e}_1}_{k\,
\vec{n}-\vec{e}_1}} S_{k\,\vec{n}-\vec{e}_1}^{F\, \vec{J}-2\vec{e}_1}
 +2 
\frac{(J_1-2)\mathcal{N}^{F\,\vec{J}-3\vec{e}_1-\vec{e}_2}_{k-1\,\vec{n}-\vec{e}_1}}{\mathcal{N}^{F\, \vec{J}-2\vec{e}_1}_{k\,\vec{n}-\vec{e}_1}} 
S_{k\,\vec{n}-\vec{e}_1}^{F\, \vec{J}-2\vec{e}_1}+\nonumber\\
&&\!\! \phantom{.}\nonumber\\
&+& (D-1) S_{k\,\vec{n}-\vec{e}_1}^{F\,\vec{J}-2\vec{e}_1}+
2 \chi_m
\frac{\mathcal{N}^{F\,\vec{J}-2\vec{e}_1-\vec{e}_2}_{k-1\,\vec{n}}}{\mathcal{N}^{F\,\vec{J}-2\vec{e}_1}_{k-1\,\vec{n}}} 
S_{k-1\,\vec{n}}^{F\,\vec{J}-2\vec{e}_1}+\nonumber\\
&+&\left(
\frac{\mathcal{N}^{F\,\vec{J}-2\vec{e}_1-2\vec{e}_2}_{k-2\,\vec{n}}}{\mathcal{N}^{F\,\vec{J}-2\vec{e}_1}_{k-2\,\vec{n}+\vec{e}_2}} 
S_{k-2\,\vec{n}+\vec{e}_2}^{F\, \vec{J}-2\vec{e}_1}
 -\lambda_m \frac{\mathcal{N}^{F\, \vec{J}-2\vec{e}_1-2\vec{e}_2}_{k-2\,\vec{n}}}{\mathcal{N}^{F\,\vec{J}-2\vec{e}_1}_{k-2\,\vec{n}}} S_{k-2\,
\vec{n}}^{F\, \vec{J}-2\vec{e}_1}
\right)
=\label{eq:Ftensorout}\\
&&\!\! \phantom{X}\nonumber\\
&=&S_{k\,\vec{n}}^{F\, \vec{J}-2\vec{e}_1} +
\left[\phantom{2^{2^2}}\hspace*{-0.5cm} 
2(J_1-2n_1-k)+2(n_1-1) 
+(D-1)+2k\right] S_{k\,\vec{n}-\vec{e}_1}^{F\, \vec{J}-2\vec{e}_1}+\nonumber\\
&+& 2\chi_m(J_2-2n_2-k+1) S_{k-1\,\vec{n}}^{F\, \vec{J}-2\vec{e}_1}
+ 2(n_2+1) S_{k-2\,\vec{n}+\vec{e}_2}^{F\, \vec{J}-2\vec{e}_1}- \phantom{\frac{X^X}{X^X}}\nonumber\\
&-&\lambda_m(J_2-2n_2-k+2) (J_2-2n_2-k+1) S_{k-2\,\vec{n}}^{F\, \vec{J}-2\vec{e}_1},\phantom{\frac{X^X}{X^X}}\nonumber
\end{eqnarray}
where
\begin{equation}
\label{eq:NumTermsSF}
\mathcal{N}^{F\, \vec{J}}_{k\,\vec{n}}=\frac{\vec{J}!}{2^{|\vec{n}|} \vec{n}! \vec{J}^*! k!}
\end{equation}
is the number of terms in the tensor structure $S_{k\,\vec{n}}^{F\,\vec{J}}$. 

Next, using the trick as in the previous example we can get equations~(\ref{eq:irrFcoefs}). It is easy to show 
that the trace expansion like~(\ref{eq:Ftensorout}) and 
equations for coefficients $w^{k'\,(k\, \vec{J})}_{\vec{n}}$ can be obtained from~(\ref{eq:irrFcoefs}) in the limit
\begin{equation}
\label{eq:SDlimitAll}
q_1=-q_2=q,\quad P_1=P_2=P,\quad \chi_m\to 1.
\end{equation} 

We could apply the same method to the tensor $\mathcal{H}^{\vec{J}}$ to obtain recurrent equations~(\ref{eq:irrHcoefs}) for 
$h^{\vec{k}'\,(\vec{k};\vec{J})}_{\vec{n}}$, but 
this requires more cumbersome calculations. That is why 
we apply the method of generating functions in this case.

\subsection{\bf Generating functions method to obtain equations}
\label{subsection:genfun}

Here we obtain recurrent equations for $\mathcal{Y}^{\vec{J}}$ and $\mathcal{H}^{\vec{J}}$ and
give trace operators, variables and generating functions for other cases, since 
the algorithm to obtain equations is the same.

\subsubsection*{$\mathcal{Y}^{\vec{J}}$ functions, variables and operators}
\label{subsubsection:genfunY}

In terms of generating functions~(\ref{def:genfunnotation}),(\ref{def:omegaderivatives}) we can write:
\begin{eqnarray}
 \vec{X}_Y&=&\{ x_1,x_2,x_3,y_1,y_2,y_3,z_{12},z_{13},z_{23}\},\nonumber\\
 \vec{x}&=&\{x_1,x_2,x_3\},\;
\vec{y}=\{y_1,y_2,y_3\},\;
\vec{z}=\{z_{12},z_{13},z_{23}\},\nonumber\\
&& \phantom{.}\nonumber\\
 x_r&=&P_r\omega_r,\qquad y_r=\omega_r^2-\frac{\left(\omega_rq_r\right)^2}{q_r^2},\nonumber\\ 
 z_{rs}&=&\omega_r\omega_s-\frac{\left(\omega_sq_r\right)\left(\omega_rq_s\right)}{q_rq_s},\; 
z_{rs}\equiv z_{sr},\, r\neq s.
\label{def:genYvars}
\end{eqnarray}

\begin{equation}
 {\varPhi'}^{\mathcal{Y}}_{\vec{k}\;\vec{n}}(\vec{X}_Y) =   \vec{\omega}^{\otimes \vec{J}} \otimes 
S^{Y\,\vec{J}}_{\vec{k}\,\vec{n}} =  \mathcal{N}^{Y\, \vec{J}}_{\vec{k}\,\vec{n}}\cdot
\vec{x}^{\vec{J}^*} \vec{y}^{\,\vec{n}} \vec{z}^{\,\vec{k}},
\label{def:monomY}
\end{equation}
\begin{equation}
 \varPhi^{\mathcal{Y}}(\vec{X}_Y) =
\sum_{\bar{\Omega}^Y_{\vec{k}}} \hat{y}^{\vec{J}}_{\vec{k}} \varPhi^{*\mathcal{Y}}_{\vec{k}}(\vec{X}_Y),
\qquad
 \varPhi^{*\mathcal{Y}}_{\vec{k}}(\vec{X}_Y) =
\sum_{\tilde{\Omega}^Y_{\vec{k}'\,\vec{n}}} y^{\vec{k}'\,(\vec{J};\,\vec{k})}_{\vec{n}} 
\mathcal{N}^{Y\, \vec{J}}_{\vec{k}\,\vec{n}}\cdot \vec{x}^{\vec{J}^*} \vec{y}^{\,\vec{n}} \vec{z}^{\,\vec{k}}.
\label{def:genfunY}
\end{equation}

The trace differential operator for any $i=1,2,3$  in terms of the generating function variables looks as follows
\begin{eqnarray}
 && \hspace*{-6mm}\square_{\omega_i}=
  P_i^2 \partial^2_{x_i x_i}+4x_i \partial^2_{x_i y_i}+
4y_i \partial^2_{y_i y_i}
+2 \sum_{r\neq i} \bar{\chi}_{ir} x_r\partial^2_{x_i z_{ir}}+
4\sum_{r\neq i}z_{ir}\partial^2_{y_i z_{ir}}+\nonumber\\
&& 
+ \sum_{r\neq i}(y_r-\tilde{\lambda}^r_{ir} x_r^2)\partial^2_{z_{ir}z_{ir}}
 + 2 (z_{rs}+\hat{\lambda}^i_{rs} x_r x_s)_{r\neq s\neq i}\partial^2_{z_{ir}z_{is}}
+2(D-1)\partial_{y_i}, \nonumber\\
&& i,r,s\in\Omega_Y.
\label{eq:genYtraceDiff}
\end{eqnarray}

When we apply this operator to $ \varPhi^{*\mathcal{Y}}_{\vec{k}}(\vec{X}_Y)$, we get
\begin{eqnarray}
 y^{\vec{k}'\,(\vec{k};\, \vec{J})}_{\vec{n}}\mathcal{N}^{Y\, \vec{J}}_{\vec{k}'\,\vec{n}}\, 
 &P_i^2& \hspace*{-7.5mm}\left[\hspace*{4mm}J^*_i \left( J^*_i-1\right) x_i^{J^*_i-2}... +
4 J^*_i n_i y_i^{n_i-1}...+4n_i(n_i-1) y_i^{n_i-1}\!\!\!\!...\!+
\phantom{\frac{X^X}{X^X}}\hspace*{-0.5cm} \right. \nonumber\\
 &+&\left.
2 J^*_i \sum_{r\neq i} \bar{\chi}_{ir} k'_{ir}
 x_i^{ J^*_i-1}x_{r}^{ J^*_{r}+1}z_{ir}^{k'_{ir}-1}\!\!\!\!...+
 4n_i\sum_{r\neq i}k'_{ir}y_i^{n_i-1}...+
  \right. \nonumber\\
 &+&\left. \sum_{r\neq i}k'_{ir}(k'_{ir}-1)
\left(
y_r^{n_r+1}-\tilde{\lambda}^r_{ir} x_r^{J^*_r+2} 
\right)z_{ir}^{k'_{ir}-2}...+ \right. \nonumber\\
&+& \left. 2k'_{ir}k'_{is}\left(
z_{rs}^{k'_{rs}+1}+\hat{\lambda}^i_{rs} x_r^{J^*_{r}+1}x_{s}^{J^*_{s}+1}
\right)_{r\neq s\neq i}
z_{ir}^{k'_{ir}-1}z_{is}^{k'_{is}-1}
...+\right. \nonumber\\
&+& \left. 
2(D-1)n_i y_i^{n_i-1}...
\phantom{\frac{X^X}{X^X}}\hspace*{-0.5cm} 
\right]=0\label{eq:genYtrace},
\end{eqnarray}
where $...$ means product of other variables with unchanged powers. From the above equation
we can get recurrent equations~(\ref{eq:irrYcoefs}) by the use of indexes shifts like in the
method developed in the previous subsection.

In the harmonical basis we have
\begin{eqnarray}
 \vec{X}_Y&=&\{ x_1,x_2,x_3,u_1,u_2,u_3,u_{12},u_{13},u_{23}\},\nonumber\\
 \vec{x}&=&\{x_1,x_2,x_3\},\;
\vec{u}=\{u_1,u_2,u_3\},\;
\dvec{u}=\{u_{12},u_{13},u_{23}\},\nonumber\\
 x_r&=&P_r\omega_r,\nonumber\\
 u_r&=&\mathcal{G}_{(rr)}\otimes\omega_{r(r)}\otimes\omega_{r(r)}=y_r-x_r^2,\nonumber\\ 
 u_{rs}&=&\mathcal{G}_{(rs)}\otimes\omega_{r(r)}\otimes\omega_{s(s)}=z_{rs}-\bar{\chi}_{rs} x_r x_s,\quad 
u_{rs}\equiv u_{sr},\, r\neq s.
\label{def:genYvarsHARM}
\end{eqnarray}
The trace differential operator in this case is more simple
\begin{eqnarray}
\square_{\omega_i}&=&
P_i^2\partial^2_{x_ix_i}+
\square_{\omega_i}^*,\nonumber\\
\square_{\omega_i}^*&=&
\sum_{r\neq i} u_r \partial^2_{u_{ir}u_{ir}} +
2 u_{rs}|_{s>r \atop r\neq i} \partial^2_{u_{ir}u_{is}}+\nonumber\\
&+&
4 \left[
\frac{(D-2)}{2} +
u_i\partial_{u_i}+
\sum_{r\neq i}
u_{ir}\partial_{u_{ir}}
\right]
\partial_{u_i}
\label{eq:genYtraceDiffHARM}
\end{eqnarray}
And the generating function looks as
\begin{eqnarray}
\varPhi^{*\mathcal{Y}}_{\vec{k}}(\vec{X}_Y)&=&
\frac{1}{\Pi_0^{\vec{J}\,\vec{k}}}
\frac{\vec{J}!}{(\vec{J}-\vec{\kappa})!\vec{k}!}
\sum_{\vec{m}}^{[\vec{J}/2]}
\frac{1}{
	4^{|\vec{m}|}\vec{m}!
	\left(c^{\vec{J}}\right)_{\vec{m}}
}
\sum_{\vec{m}'=\vec{0}}^{\vec{m}}
\mathbb{C}_{\vec{m}}^{\vec{m}'}
\vec{x}^{2(\vec{m}-\vec{m}')}
\vec{u}^{\,\vec{m}'}\times
\nonumber\\
&\times&\sum_{\vec{r}=\vec{0}}^{\vec{m}}
\mathbb{C}_{\vec{m}}^{\vec{r}}
\vec{x}^{\vec{J}-\vec{\kappa}-2(\vec{m}-\vec{r})}
\prod_{i\in\Omega_Y}
\left(P_i^2\right)^{m_i-r_i}
\square^{*\vec{r}}_{\vec{\omega}}
\dvec{u}^{\,\vec{k}},
\label{YHARMgenfun}
\end{eqnarray}
with $\Pi_0^{\vec{J}\,\vec{k}}$ from~(\ref{HARMnormalization}) and definitions 
\begin{eqnarray}
\square^{*r_i}_{\omega_i}
\vec{u}^{\,\vec{n}}
\dvec{u}^{\,\vec{k}}
&=&
\sum_{\vec{a}^{(i)}}
\mathbb{A}_{\vec{a}^{(i)}\,r_i}^{\vec{k}\;\vec{n}}
\vec{u}^{\,
	\vec{n}+\vec{\delta}_i
}
\dvec{u}^{\,
	\vec{k}+\ddvec{\delta}_i
},
\\
&\phantom{x}&
\nonumber\\
\mathbb{A}_{\vec{a}^{(i)}\,r_i}^{\vec{k}\;\vec{n}}&=&
\frac{r_i!\, a^{\smash[b]{(i)}}_i!\,
	(\lambda_i+n_i-r_i)_{r_i-a^{\smash[b]{(i)}}_i}
}{
(r_i-a^{\smash[b]{(i)}}_i)!\,
(2a^{\smash[b]{(i)}}_i-|\vec{a}^{(i)}|)!\,
\vec{a}^{(i)}!
}
\frac{
	n_i!\, 2^{2r_i-|\vec{a}^{(i)}|}
}{
(n_i-r_i+a^{\smash[b]{(i)}}_i)!
}\times
\nonumber\\
&\times&
\prod_{j\neq i}
\frac{k_{ij}!}{
	(k_{ij}+2(\cev{a}^{\!(i)}
	\vec{e}_{ij})-|\vec{a}^{\smash[b]{(i)}}|)!
},
\nonumber\\
&\phantom{x}&
\nonumber\\
\vec{a}^{(i)}&=&
\{ a^{\smash[b]{(i)}}_1,a^{\smash[b]{(i)}}_2,
a^{\smash[b]{(i)}}_3\},
\quad
\cev{a}^{\!(i)}=
\{
a^{\smash[b]{(i)}}_3,
a^{\smash[b]{(i)}}_2,
a^{\smash[b]{(i)}}_1
\},\nonumber\\
\lambda_i &=&(D-2)/2+\kappa_i.
\end{eqnarray}

\begin{eqnarray}
\square^{*\vec{r}}_{\vec{\omega}}
\dvec{u}^{\,\vec{k}}
&=&
\sum_{\{\vec{a}^{\smash[b]{(i)}}\}
	\atop i\in\Omega_Y
}
\delta_{r_1\,a^{\smash[b]{(1)}}_1}\;
\prod_{j\in\Omega_Y}
\mathbb{A}_{
	\vec{a}^{(j)}\,r_j
}^{
\vec{k}+\ddvec{\Delta}_j
\;\;\vec{\Delta}_j
}\;
\cdot
\vec{u}^{\,
	\vec{\Delta}
}\;
\dvec{u}^{\,
	\vec{k}+\ddvec{\Delta}
},\nonumber\\
\dvec{\delta}_s &=& 
 \left(2\cev{a}^{\!(s)}-
|\vec{a}^{(s)}|\cdot\vec{1}\right),\quad
\vec{\delta}_s = 
 \left(\vec{a}^{(s)}-\vec{e}_s r_s\right),
\quad j,s\in\Omega_Y,\nonumber\\
\vec{\Delta}_1&=&\vec{0},\quad 
\vec{\Delta}_j =
\sum_{s<j} \vec{\delta}_s,\quad
\vec{\Delta} =
\sum_{s\in\Omega_Y} \vec{\delta}_s,\nonumber\\
\dvec{\Delta}_1&=&\vec{0},\quad
\dvec{\Delta}_j =
\sum_{s<j} \dvec{\delta}_s,\quad
\dvec{\Delta} =
\sum_{s\in\Omega_Y} \dvec{\delta}_s.
\label{squaremYHARM}
\end{eqnarray}

\subsubsection*{$\mathcal{H}^{\vec{J}}$ functions, variables and operators}
\label{subsubsection:genfunH}

\begin{eqnarray}
 \vec{X}_H&=&\{ x_1,x_{1'},x_2,x_{2'},y_1,y_{1'},y_2,y_{2'},z_{11'},z_{22'},z_{12},z_{1'2'},z_{12'},z_{1'2}\},\nonumber\\
 \vec{x}&=&\{x_1,x_{1'},x_2,x_{2'}\},\;
\vec{y}=\{y_1,y_{1'},y_2,y_{2'}\},\;
\vec{z}=\{z_{11'},z_{22'},z_{12},z_{1'2'},z_{12'},z_{1'2}\},\nonumber\\
&&\phantom{.}\nonumber\\
 x_r&=&P_r\omega_r,\qquad y_r=\omega_r^2-\frac{\left(\omega_rq_r\right)^2}{q_r^2},\nonumber\\ 
 z_{rr'} &=& \omega_r\omega_{r'}-\frac{\left(\omega_rq_r\right)\left(\omega_{r'}q_r\right)}{q_r^2},\nonumber\\
 z_{rs}&=&\omega_r\omega_s-\frac{\left(\omega_sq_r\right)\left(\omega_rq_s\right)}{q_rq_s},\; 
z_{rs}\equiv z_{sr},\, r\neq s,s'.
\label{def:genHallvars}
\end{eqnarray}

\begin{equation}
{\varPhi'}^{\mathcal{H}}_{\vec{k}\;\vec{n}}(\vec{X}_H) =   \vec{\omega}^{\otimes \vec{J}} \otimes 
S^{H\,\vec{J}}_{\vec{k}\,\vec{n}} =  \mathcal{N}^{H\, \vec{J}}_{\vec{k}\,\vec{n}}\cdot
\vec{x}^{\vec{J}^*} \vec{y}^{\,\vec{n}} \vec{z}^{\,\vec{k}},
\label{def:monomH}
\end{equation}

\begin{equation}
\varPhi^{\mathcal{H}}(\vec{X}_H) =
\sum_{\bar{\Omega}^H_{\vec{k}}} \hat{h}^{\vec{J}}_{\vec{k}} \varPhi^{*\mathcal{H}}_{\vec{k}}(\vec{X}_H),
\qquad
\varPhi^{*\mathcal{H}}_{\vec{k}}(\vec{X}_H) =
\sum_{\tilde{\Omega}^H_{\vec{k}'\,\vec{n}}} h^{\vec{k}'\,(\vec{J};\,\vec{k})}_{\vec{n}} 
\mathcal{N}^{H\, \vec{J}}_{\vec{k}\,\vec{n}}\cdot \vec{x}^{\vec{J}^*} \vec{y}^{\,\vec{n}} \vec{z}^{\,\vec{k}}.
\label{def:genfunH}
\end{equation}

The trace differential operator for any $i=1,1',2,2'$  in terms of the generating function variables looks as follows
\begin{eqnarray}
\square_{\omega_i}&=&
\partial^2_{x_i x_i}+4x_i \partial^2_{x_i y_i}+
4y_i \partial^2_{y_i y_i}+2x_{i'} \partial^2_{x_i z_{ii'}}+
2\chi_m \sum_{r\neq i,i'}x_r\partial^2_{x_i z_{ir}}+\nonumber\\
&+&
4\sum_{r\neq i}z_{ir}\partial^2_{y_i z_{ir}}+y_{i'}\partial^2_{z_{ii'}z_{ii'}}+ 
\sum_{r\neq i,i'}(y_r-\lambda_m x_r^2)\partial^2_{z_{ir}z_{ir}}+
\nonumber\\
&+&
2(z_{jj'}-\lambda_m x_jx_{j'})\partial^2_{z_{ij}z_{ij'}}+ 
\sum_{r\neq i,i'}2y_{i'r}\partial^2_{z_{ii'}z_{ir}}+2(D-1)\partial_{y_i},\nonumber\\
\{j,j'\}&\neq&\{i,i'\},\, j\equiv i^{\star},\, j'\equiv{i^{\star}}', \qquad i,j,r\in\Omega_H.
\label{eq:genHtraceDiff}
\end{eqnarray}

When we apply this operator to $ \varPhi^{*\mathcal{H}}_{\vec{k}}(\vec{X}_H)$, we get
\begin{eqnarray}
h^{\vec{k}'\,(\vec{k};\,\vec{J})}_{\vec{n}} \mathcal{N}^{H\, \vec{J}}_{\vec{k}\,\vec{n}}\phantom{x} 
&J^*_i& 
\hspace*{-7mm}\left[\hspace*{3mm} 
\left( J^*_i-1\right) x_i^{J^*_i-2}... + 
4 J^*_i n_i y_i^{n_i-1}...+4n_i(n_i-1) y_i^{n_i-1}...+
\phantom{\frac{\frac{\frac{X}{X}}{\frac{X}{X}}}{\frac{X}{X}}}\hspace*{-0.5cm} 
\right. \nonumber\\
&+&\hspace*{-2mm}  \left.  2J^*_ik'_{ii'} x_i^{ J^*_i-1}x_{i'}^{ J^*_{i'}+1}z_{ii'}^{k'_{ii'}-1}...+
2\chi_m  J^*_i \sum_{r\neq i,i'} k'_{ir}
x_i^{ J^*_i-1}x_{r}^{ J^*_{r}+1}z_{ir}^{k'_{ir}-1}...+
\right. \nonumber\\
&+&\hspace*{-2mm}  \left. 4n_i\sum_{r\neq i}k'_{ir}y_i^{n_i-1}...+k'_{ii'}(k'_{ii'}-1)y_{i'}^{n_{i'}+1}z_{ii'}^{k'_{ii'}-2}...+ 
\right. \nonumber\\
&+&\hspace*{-2mm}  \left. \sum_{r\neq i,i'}k'_{ir}(k'_{ir}-1)
\left(
y_r^{n_r+1}-\lambda_m x_r^{J^*_r+2} 
\right)z_{ir}^{k'_{ir}-2}...+ \right. \nonumber\\
&+&\hspace*{-2mm} \left. 2k'_{ij}k'_{ij'}\left(
z_{jj'}^{k'_{jj'}+1}-\lambda_m x_j^{J^*_{j}+1}x_{j'}^{J^*_{j'}+1}
\right)z_{ij}^{k'_{ij}-1}z_{ij'}^{k'_{ij'}-1}
...+\right. \nonumber\\
&+&\hspace*{-2mm}  \left. 
\sum_{r\neq i,i'}2k'_{ii'}k'_{ir}
z_{ii'}^{k'_{ii'}-1}z_{ir}^{k'_{ir}-1}z_{i'r}^{k'_{i'r}+1}...+
2(D-1)n_i y_i^{n_i-1}...
\phantom{\frac{\frac{\frac{X}{X}}{\frac{X}{X}}}{\frac{X}{X}}}\hspace*{-0.4cm} 
\right]\!\!=\!0\label{eq:genHtrace},
\end{eqnarray}
where $...$ means product of other variables with unchanged powers. From the above equation
we can get recurrent equations~(\ref{eq:irrHcoefs}).

In the harmonical basis we have
\begin{eqnarray}
 \vec{X}_H&=&\{ x_1,x_{1'},x_2,x_{2'},u_1,u_{1'},u_2,u_{2'},u_{11'},u_{22'},u_{12},u_{1'2'},u_{12'},u_{1'2}\},\nonumber\\
 \vec{x}&=&\{x_1,x_{1'},x_2,x_{2'}\},\;
\vec{u}=\{u_1,u_{1'},u_2,u_{2'}\},\;
\dvec{u}=\{u_{11'},u_{22'},u_{12},u_{1'2'},u_{12'},u_{1'2}\},\nonumber\\
&&\phantom{.}\nonumber\\
 x_r&=&P_r\omega_r,\qquad u_r=\mathcal{G}_{(rr)}\otimes\omega_{r(r)}\otimes\omega_{r(r)}=y_r-x_r^2,\nonumber\\ 
 u_{rs}&=&\mathcal{G}_{(rs)}\otimes\omega_{r(r)}'\otimes\omega_{s(s)},\quad 
u_{rs}\equiv u_{sr},\, r\neq s,\nonumber\\
 u_{rr'}&=&z_{rr'}-x_r x_{r'},\qquad
u_{rs}=z_{rs}-\chi_m x_r x_s,\; s\neq r,r'. 
\label{def:genHvarsHARM}
\end{eqnarray}
The trace differential operator in this case is
\begin{eqnarray}
\square_{\omega_i}&=&
\partial^2_{x_ix_i}+
\square_{\omega_i}^*,\nonumber\\
\square_{\omega_i}^*&=&
\sum_{r\neq i} u_r \partial^2_{u_{ir}u_{ir}} +
\sum_{s,r\neq i
	\atop s>r}
2 u_{rs} \partial^2_{u_{ir}u_{is}}+\nonumber\\
&+&
4 \left[
\frac{(D-2)}{2} +
u_i\partial_{u_i}+
\sum_{r\neq i}
u_{ir}\partial_{u_{ir}}
\right]
\partial_{u_i},\nonumber\\
i,r,s&\in&\Omega_H.
\label{eq:genHtraceDiffHARM}
\end{eqnarray}
The generating function can be expressed as
\begin{eqnarray}
\varPhi^{*\mathcal{H}}_{\vec{k}}(\vec{X}_H)&=&
\frac{1}{\Pi_0^{\vec{J}\,\vec{k}}}
\frac{\vec{J}!}{(\vec{J}-\vec{\kappa})!\vec{k}!}
\sum_{\vec{m}}^{[\vec{J}/2]}
\frac{1}{
	4^{|\vec{m}|}\vec{m}!
	\left(c^{\vec{J}}\right)_{\vec{m}}
}
\sum_{\vec{m}'=\vec{0}}^{\vec{m}}
\mathbb{C}_{\vec{m}}^{\vec{m}'}
\vec{x}^{\,2(\vec{m}-\vec{m}')}
\vec{u}^{\,\vec{m}'}\times
\nonumber\\
&\times&
\sum_{\vec{r}=\vec{0}}^{\vec{m}}
\mathbb{C}_{\vec{m}}^{\vec{r}}\;\;
\vec{x}^{\,\vec{J}-\vec{\kappa}-2(\vec{m}-\vec{r})}
\square^{*\vec{r}}_{\vec{\omega}}
\dvec{u}^{\,\vec{k}}\;,
\label{HHARMgenfun}
\end{eqnarray}
with  $\Pi_0^{\vec{J}\,\vec{k}}$ from~(\ref{HARMnormalization}) and 
definitions
\begin{eqnarray}
\square^{*r_i}_{\omega_i}
\vec{u}^{\,\vec{n}}
\dvec{u}^{\,\vec{k}}
&=&
\sum_{\vec{a}^{(i)},
	{\vec{a}}^{\,\prime(i)}
}
\mathbb{A}_{\vec{a}^{(i)}\;
	{\vec{a}}^{\,\prime(i)}
	\;r_i}^{\vec{k}\;\vec{n}}
\vec{u}^{\,
	\vec{n}+\vec{\delta}_i
}
\dvec{u}^{\,
	\vec{k}+
	\ddvec{\delta}_i
},
\\
&\phantom{x}&
\nonumber\\
\mathbb{A}_{\vec{a}^{(i)}\;
	{\vec{a}}^{\,\prime(i)}\;
	r_i}^{\vec{k}\;\vec{n}}
&=&
\frac{r_i!\;
	(\lambda_i+n_i-r_i)_{(r_i-
		|\vec{a}^{(i)}|-
		|{\vec{a}}^{\,\prime(i)}|)
	}
}{
(r_i-|\vec{a}^{(i)}|-|{\vec{a}}^{\,\prime(i)}|)!\,
\vec{a}^{(i)}!\,
{\vec{a}}^{\,\prime(i)}!
}
\times
\nonumber\\
&\times&
\frac{
	n_i!\, 2^{2(r_i-|\vec{a}^{(i)}|)-
		|{\vec{a}}^{\,\prime(i)}|
	}
}{
(n_i-r_i+
|\vec{a}^{(i)}|+
|{\vec{a}}^{\,\prime(i)}|
)!
}\,
\prod_{j\neq i}
\frac{k_{ij}!}{
	(
	k_{ij}-2a^{\smash[b]{(i)}}_j-
	\sum_{r\neq i,j}a^{\,\prime(i)}_{jr}
	)!
},
\nonumber\\
&\phantom{x}&
\nonumber\\
\vec{a}^{(i)}&=&\sum_{j\neq i}
\vec{e}_j a^{\smash[b]{(i)}}_{j},
\quad
{\vec{a}}^{\,\prime(i)}=
\sum_{j,s\neq i\atop s>j}
a^{\smash[b]{(i)}}_{js} \vec{e}_{js}
,\nonumber\\
\lambda_i &=&(D-2)/2+\kappa_i.
\end{eqnarray}
and
\begin{eqnarray}
\square^{*\vec{r}}_{\vec{\omega}}
\dvec{u}^{\,\vec{k}}
&=&
\sum_{\{\vec{a}^{(i)},
	{\vec{a}}^{\,\prime(i)}\}
	\atop i\in\Omega_H
	}
	\!\!\!
\delta_{r_1\,|\vec{a}^{(1)}|+
	|{\vec{a}}^{\,\prime(1)}|}\;
\prod_{j\in\Omega_H}	
\mathbb{A}_{
	\vec{a}^{(j)}\;
	{\vec{a}}^{\,\prime(j)}\;
	r_j
}^{
\vec{k}+\ddvec{\Delta}_j\;\;
\vec{\Delta}_j
}	
\;\;
\vec{u}^{
	\,\vec{\Delta}
}\;
\dvec{u}^{\,
	\vec{k}+\ddvec{\Delta}
},\nonumber\\
\vec{\delta}_s&=&[(-r_s+|\vec{a}^{(s)}|+
|{\vec{a}}^{\,\prime(s)}|)\vec{e}_s+
\vec{a}^{(s)}],
\nonumber\\
\dvec{\delta}_s&=&
\left(
{\vec{a}}^{\,\prime(s)} -
\sum_{r\neq s} \vec{e}_{sr}
\left(2 a^{\smash[b]{(s)}}_r + \sum_{s'\neq s,r}a^{\,\prime(s)}_{rs'}\right)
\right),\nonumber\\
\vec{\Delta}_1&=&\vec{0},
\quad
\vec{\Delta}_j = \sum_{s<j} \vec{\delta}_s,
\quad 
\vec{\Delta} = \sum_{s\in\Omega_H} \vec{\delta}_s,\nonumber\\
\dvec{\Delta}_1&=&\vec{0},
\quad
\dvec{\Delta}_j=\sum_{s<j} \dvec{\delta}_s,\quad
\dvec{\Delta}=\sum_{s\in\Omega_H} \dvec{\delta}_s,
\quad j,s,r\in\Omega_H.
\label{squaremHHARM}
\end{eqnarray}

\subsubsection*{$\mathcal{W}^{\vec{J}}$ and $\mathcal{F}^{\vec{J}}$ functions, variables and operators}
\label{subsubsection:genfunWF}

In the standard basis 
\begin{equation}
\vec{X}_{F,W}=\{ x_i,x_j,y_i,y_j,z_{ij}\},\quad
\vec{x}=\{x_i,x_j\},\quad \vec{y}=\{y_i,y_j\},\quad i,j\in \Omega_{F,W},
\label{def:genWFvars}
\end{equation}
and variables are defined as in~(\ref{def:genHallvars}).
The trace differential operator for $\mathcal{F}$ is
\begin{eqnarray}
\square_{\omega_i}&=&
\partial^2_{x_ix_i}+
4x_i \partial^2_{x_i y_i}+
4y_i \partial^2_{y_i y_i}+
4 z_{ii^{\star}}  \partial^2_{y_i z_{ii^{\star}}} +\nonumber\\
&+&
2\chi_m x_{i^{\star}} \partial^2_{x_i z_{ii^{\star}}}+
(y_{i^{\star}}-\lambda_m x_{i^{\star}}^2) \partial^2_{z_{ii^{\star}} z_{ii^{\star}}}+
2(D-1)\partial_{y_i}
\label{eq:genFtraceDiff}
\end{eqnarray}	
and for $\mathcal{W}$ we have
\begin{eqnarray}
\square_{\omega_i}&=&
\partial^2_{x_ix_i}+
4x_i \partial^2_{x_i y_i}+
4y_i \partial^2_{y_i y_i}+
4z_{ii'}  \partial^2_{y_i z_{ii'}} +\nonumber\\
&+&2x_{i'} \partial^2_{x_i z_{ii'}}+
y_{i'} \partial^2_{z_{ii'}z_{ii'}}+
2(D-1)\partial_{y_i} 
\label{eq:genWtraceDiff}
\end{eqnarray}	

In the harmonical basis we have the same sets both for $\mathcal{F}$ and $\mathcal{W}$:
\begin{eqnarray}
\vec{X}_{F,W}&=&\{ x_i,x_j,u_i,u_j,u_{ij}\},\quad
\vec{x}=\{x_i,x_j\},\quad \vec{u}=\{u_i,u_j\},\quad
i,j\in\Omega_{F,W}
\label{def:genWFvarsHARM}
\\
&&\phantom{.}\nonumber\\
\square_{\omega_i}&=&
\partial^2_{x_ix_i}+
4u_j\partial^2_{u_{ij}u_{ij}}+
4\left[(D-2)/2+
u_i \partial_{u_i}+
u_{ij}\partial_{u_{ij}}
\right]\partial_{u_i},\quad j\neq i,
\label{eq:genWFtraceDiffHARM}
\end{eqnarray}
with variables defined as in~(\ref{def:genHvarsHARM}). 

Generating function for $\mathcal{F}$ in this case is
\begin{equation}
\varPhi^{\mathcal{F}}(\vec{X}_F) =
\sum_{\bar{\Omega}^F_k} \hat{f}^{\vec{J}}_{k} \varPhi^{*\mathcal{F}}_{k}(\vec{X}_F),
\label{def:genfunF}
\end{equation}
and $\varPhi^{*\mathcal{F}}_k(\vec{X}_F)$ can be 
expressed in 
terms of Srivastava-Daoust and hypergeometric 
functions~(\ref{hypergeomseries}),(\ref{srivastavadaoust}). For the 
standard basis it looks like
\begin{equation}
\varPhi^{*\mathcal{F}}_k(\vec{X}_F)=
\frac{
	\vec{J}!\,
	\vec{x}^{\,\vec{J}-k\cdot\vec{1}}\,
	z^k
}{
(\vec{J}-k\cdot\vec{1})!\, k!
}\cdot
\frac{1}{2}\left(
F_{SD}^{k\,\{J_1,J_2\}} (\vec{x},\vec{y},z)
+
F_{SD}^{k\,\{J_2,J_1\}} (\cev{x},\cev{y},z)
\right),
\end{equation}
\begin{eqnarray}
&&
F_{SD}^{k\,\{J_1,J_2\}} (
\{x_1,x_2\},\{y_1,y_2\},z)=
\nonumber\\
&=&F_{4}^{2}
\left(
\begin{array}{c}
(-k): A_1 ,\;
(J_1-k+1): A_2; \\
(J_2-k+1): B_1,\;
(J_1-k+1): B_2,\;
c^{J_2}: B_3,\;
c^{J_1}: B_4;
\end{array}\;
\Bigg| z_1,\dots,z_7
\right),
\nonumber
\end{eqnarray}
\begin{eqnarray}
A_1&=&\{0,0,2,2,2,1,1\},
\qquad\quad\;\;
A_2=\{0,0,2,0,0,1,0\},
\nonumber\\
B_1&=& \{0,-2,0,0,0,-1,0\}, \qquad
B_2=\{-1,0,2,0,0,1,-1\},
\nonumber\\
B_3&=&\{0,1,1,0,0,1,0\},\qquad\quad\;\;
B_4=\{1,0,0,1,1,0,1\},
\nonumber\\
z_1&=& \frac{y_1}{4x_1^2},
\qquad\;
z_2= \frac{y_1}{4x_2^2\chi_m^2},
\qquad\quad\;
z_3=-\frac{\lambda_m}{4\chi_m^2}
\frac{y_2 x_1^2}{z^2},
\qquad
z_4=-\frac{\lambda_m}{4}
\frac{y_1 x_2^2}{z^2},\nonumber\\
z_5&=&\frac{y_1 y_2}{4z^2},
\qquad
z_6=-\frac{y_2 x_1}{2 x_2 z \chi_m},
\qquad
z_7=-\frac{y_1 x_2 \chi_m}{2 x_1 z}.
\label{eq:genfunFSTANDstarFinal}
\end{eqnarray} 
and for harmonical basis we have
\begin{eqnarray}
\varPhi^{*\mathcal{F}}_k(\vec{X}_F) &=&
\frac{\vec{J}!}{(\vec{J}-\vec{k})!}
\vec{x}^{\vec{J}}
\sum_{m=0}^{\lfloor k/2\rfloor}
\frac{(1-\lambda_{k})_{2m}\,(1-\lambda_k)_m}{4^m m! (k-2m)!}\,
\frac{(-1)^m\left(\frac{u_{12}}{x_1x_2}\right)^{k-2m}}{(c^{(\vec{J}+\vec{k})/2})_{\vec{m}}(c^{(\vec{J}+\vec{k}+\vec{1})/2})_{\vec{m}}}\times
\nonumber\\
&\times&
\,
{}_2F_1\!\left(
-\frac{k-2m}{2},\;\frac{1-(k-2m)}{2};\;
2-\lambda_{k-2m};\; 
\frac{u_1 u_2}{u_{12}^2}
\right)
\times\nonumber\\
&\times&
\prod_{i=1}^{2}
\Bigg[
{}_2F_1\!\left(
-\frac{J_i-k+2m}{2},\;\frac{1-(J_i-k+2m)}{2};\,
\lambda_{k-2m}
;\; -\frac{u_i}{x_i^2}
\right)
\Bigg],\nonumber\\
\lambda_{k-2m}&=&\lambda_k-2m=(D-2)/2+k-2m.
\label{eq:genfunFstarFinal}
\end{eqnarray}
The above function can be also expressed in terms of, for example, a single Srivastava-Daoust function. But this form is more convenient for further calculations.

\subsection{\bf Algorithm to solve basic recurrent equations in standard basis for $\cal{F}$ and $\cal{W}$}
\label{subsection:FWsolve}

Let us start to solve recurrent equations. At first we solve~(\ref{eq:irrFcoefs}). We can change
variables to simplify equatios:
\begin{eqnarray}
 f^{k-l}_{\{n_1, n_2\}} &=& \frac{2^{l-n_1-n_2}}{\left(c^{J_1}\right)_{n_1}\left(c^{J_2}\right)_{n_2}}\mathbf{f}^{l}_{n_1 n_2}=
2^l\mathcal{\upsilon}^{J_1}_{n_1}\mathcal{\upsilon}^{J_2}_{n_2}\mathbf{f}^{l}_{n_1 n_2} 
;\nonumber\\
 \mathbf{f}^{l}_{n_1 n_2} &=& \mathbf{f}^{l}_{i;\; n_1 n_2} J^*_i! n_i! \left(c^{J_i}\right)_{n_i} ,\,\, i=1,2;
\nonumber\\
 \mathbf{f}^{0}_{n_1 n_2}&=&1,\,\,
\mathbf{f}^{l}_{00}=\delta_{l0}
;\quad J^*_i=\left( J_i-2n_i-k+l\right),
\label{eq:irrFcoefsChVars}
\end{eqnarray}
and $c^J$,$\mathcal{\upsilon}^J_n$ can be found in~(\ref{eq:irrVcoefs}),(\ref{eq:irrVcoefsSolution}). After that we get new equations
\begin{equation}
\begin{cases}
& \mathbf{f}^{l}_{2;\; n_1-1\; n_2} -\mathbf{f}^{l}_{2;\; n_1 n_2} +
\chi_m \mathbf{f}^{l-1}_{2;\;n_1-1\; n_2}
+ \mathbf{f}^{l-2}_{2;\;n_1-1\; n_2-1} 
- \frac{\lambda_m}{4} \mathbf{f}^{l-2}_{2;\; n_1-1\; n_2} = 0,\\
&\phantom{.}\\
& \mathbf{f}^{l}_{1;\; n_1\; n_2-1} -\mathbf{f}^{l}_{1;\; n_1 n_2} +
\chi_m \mathbf{f}^{l-1}_{1;\;n_1\; n_2-1}
 + \mathbf{f}^{l-2}_{1;\;n_1-1\; n_2-1} 
- \frac{\lambda_m}{4} \mathbf{f}^{l-2}_{1;\; n_1\; n_2-1} = 0,\\
\end{cases}
\label{eq:irrFcoefsV2}
\end{equation}
with
\begin{equation}
\begin{cases}
& \mathbf{f}^{l}_{2;\; n_1\; 0} = \frac{\bar{\mathbf{f}}^l_{n_1}}{(J_2-k)!},\quad
\mathbf{f}^{l}_{1;\; 0\; n_2} = \frac{\bar{\mathbf{f}}^l_{n_2}}{(J_1-k)!},\\
&\phantom{.}\\
& \mathbf{f}^{l}_{2;\; 0\; n_2} = \alpha^{J_1J_2k}_{l\;n_2}\bar{\mathbf{f}}^l_{n_2},\quad
 \mathbf{f}^{l}_{1;\; n_1\; 0} = \alpha^{J_2J_1k}_{l\;n_1}\bar{\mathbf{f}}^l_{n_1},\\
&\phantom{.}\\
& \alpha^{J_1J_2k}_{l\;n} = \frac{(J_1-k+l)!}{(J_1-k)!}\frac{1}{(J_2-k+l-2n)!n!( c^{J_2})_n},
\end{cases}
\label{eq:irrFcoefsV2add}
\end{equation}
and the following recurrent equation for $\bar{\mathbf{f}}^{l}_{n}$
\begin{equation}
\begin{cases}
& \bar{\mathbf{f}}^{l}_{n}=\bar{\mathbf{f}}^{l}_{n-1}+\chi_m \bar{\mathbf{f}}^{l-1}_{n-1} 
-\frac{\lambda_m}{4}\bar{\mathbf{f}}^{l-2}_{n-1},\\
& \bar{\mathbf{f}}^{l}_0=\delta_{l0},\; \bar{\mathbf{f}}^{0}_{n} =1,\; 
\bar{\mathbf{f}}^{1}_{n}=\chi_m\;n,\\
\end{cases}
\label{eq:irrFcoefsV2add2}
\end{equation}

Let us introduce several generating functions
\begin{eqnarray}
&& \rho(t,z)=\sum_{n,l=0}^{\infty} \bar{\mathbf{f}}^{l}_{n} t^n z^l,
\label{eq:irrFcoefsV2rho}\\
&& \rho_2(y,z)=\sum_{n_2,l=0}^{\infty} \mathbf{f}^{l}_{2;\; 0\;n_2} y^{n_2} z^l,
\qquad \rho_1(x,z)=\sum_{n_1,l=0}^{\infty} \mathbf{f}^{l}_{1;\; n_1\;0} x^{n_1} z^l,
\label{eq:irrFcoefsV2rho12}\\
&& f_i(x,y,z)= \sum_{n_{1,2},l=0}^{\infty} \mathbf{f}^{l}_{i;\; n_1\;n_2} x^{n_1} y^{n_2} z^l.
\label{eq:irrFcoefsV2fi}
\end{eqnarray}
Then we have for $\rho(t,z)$
\begin{eqnarray}
&& \hspace*{-5mm}\left[ \rho(t,z)-\rho(0,z)\right]-
\left[ \rho(t,0)-\rho(0,0)\right]  -z\left[ \rho_z(t,0)-\rho_z(0,0)\right]=\nonumber\\
 &=&t\left\{ \phantom{\frac{\lambda_m}{4}}\hspace*{-0.55cm}
\rho(t,z) -\rho(t,0)-z \rho_z(t,0)
 + \chi_m z \left[ \rho(t,z)-\rho(t,0)\right]-
\frac{\lambda_m}{4} z^2 \rho(t,z) 
 \right\}.
\label{eq:irrFcoefsV2eqrho}
\end{eqnarray}
And the solution of this equation is
\begin{eqnarray}
\rho(t,z)&=&\frac{1}{\left( 
1-t \left( 
1+\chi_m z-\frac{\lambda_m}{4}z^2
\right) 
\right)}=\nonumber\\
&=& \sum_{s=0}^{\infty}\sum_{i=0}^s\sum_{j=0}^i \mathbb{C}_s^i \mathbb{C}_i^j
\chi_m^j \left( -\frac{\lambda_m}{4}\right)^{s-i}
t^s z^{2(s-i)+j}.
\label{eq:irrFcoefsV2solrho}
\end{eqnarray}
Then we simply compare this expression with~(\ref{eq:irrFcoefsV2rho}) and obtain
\begin{eqnarray}
&& \hspace*{-8mm}s\to n,\; 2(s-i)+j\to l \Rightarrow j=l-2n+2i,\nonumber\\
&& \phantom{.}\nonumber\\
&& \hspace*{-8mm}\bar{\mathbf{f}}^{l}_{n} = \!\!\sum_i \!
\mathbb{C}_n^i \mathbb{C}_i^{l-2n+2i}
\chi_m^{l-2n+2i} \left( -\frac{\lambda_m}{4}\right)^{n-i}\!\!\!\!\!\!\!\!=\!
 \chi_m^l \sum_i \!
\mathbb{C}_n^{n-i} \mathbb{C}_{n-i}^{l-2i}\left(\frac{\hat{x}}{4}\right)^i \!= \!\chi_m^l \mathcal{R}^l_n(\hat{x}),
\label{eq:irrFcoefsV2solbarfln}\\
&& \hspace*{-8mm}\mathcal{R}^L_N(\hat{x}) = \sum_{i=0}^N \mathbb{C}_N^{N-L+2i}\mathbb{C}_{N-L+2i}^i (\hat{x}/4)^i,\quad
\mathcal{R}^L_N(0)=\mathbb{C}_N^L,
\quad
\hat{x}= -\lambda_m/\chi_m^2, 
\label{eq:irrFcoefsV2x}
\end{eqnarray} 
$\mathbb{C}_n^i=\mathbb{C}_n^{n-i}$ is the binomial coefficient. For $\hat{x}\neq 0$ the function $\mathcal{R}^L_N(\hat{x})$ 
can be expressed in terms of hypergeometric functions~\cite{hypergeom}:
\begin{equation}
\mathcal{R}^L_N\!(\hat{x}) = \mathbb{C}_N^L \phantom{.}_2F_1
\left(
-\frac{L-1}{2},-\frac{L}{2},N-L+1;\; \hat{x}
\right)
 \label{eq:irrFcoefsV2RNLhyp}
\end{equation}

Let us introduce the differential operator, which extracts coefficient in the
expansion of any function of three variables $f=f(x,y,z)$:
\begin{equation}
\hat{\mathcal{D}}^l_{n_1 n_2}f\equiv \frac{1}{l! n_1!n_2!} \left[
\frac{\partial^{n_1}}{\partial x^{n_1}}
\frac{\partial^{n_2}}{\partial y^{n_2}}
\frac{\partial^{l}}{\partial z^{l}} f\right]_{
x,y,z\to 0
}
\label{def:Dhatln1n2}
\end{equation}
The usual property holds for this operator
\begin{equation}
\hat{\mathcal{D}}^l_{n_1 n_2} (u\cdot v) =
\sum_{i_1=0}^{n_1} \sum_{i_2=0}^{n_2} \sum_{j=0}^{l} 
\hat{\mathcal{D}}^j_{i_1\; i_2} (u)  \hat{\mathcal{D}}^{l-j}_{n_1-i_1\; n_2-i_2} (v), 
\label{eq:DhatContr}
\end{equation}
where $u=u(x,y,z)$, $v=v(x,y,z)$ are some functions.

From equations~(\ref{eq:irrFcoefsV2}),(\ref{eq:irrFcoefsV2add}) we can obtain equations like~(\ref{eq:irrFcoefsV2eqrho}) 
for $f_i(x,y,z)$:
\begin{eqnarray}
&& f_1(x,y,z)=f_0(y,x,z) \rho_1(x,z),\nonumber\\
&& f_2(x,y,z)=f_0(x,y,z) \rho_2(y,z),\label{eq:f12FW}\\
&& f_0(x,y,z) = \left(1-x\left[1+\chi_m z +y\left(x-\frac{\lambda_m}{4}\right)z^2\right]\right)^{-1}.
\label{eq:f0FW}
\end{eqnarray} 

We apply the operator $\hat{\mathcal{D}}^l_{n_1 n_2}$, for example, to\linebreak 
$f_2(x,y,z)$ (case for $f_1(x,y,z)$ can be obtained by simple 
change of variables and indexes $1\leftrightarrow 2$), use multinomial
expansion for $f_0$ and expressions~(\ref{eq:irrFcoefsV2rho12}). Finally
the chain of calculations looks as 
\begin{eqnarray}
 \mathcal{\beta}^l_{n_1 n_2}&=&\hat{\mathcal{D}}^l_{n_1 n_2} f_0(x,y,z) =  \nonumber\\
&=& \hat{\mathcal{D}}^l_{n_1 n_2}  \sum_{s=0}^{\infty}
\sum_{i,j,\hat{j}} \frac{s!}{i!j!\hat{j}! (s-i-j-\hat{j})!} 
 \chi_m^j \left( -\frac{\lambda_m}{4} \right)^{\hat{j}}  x^s y^i z^{2i+j+2\hat{j}}=
\nonumber\\
&=& \left/
\begin{array}{ll}
& s\to n_1,\, i\to n_2\\
& 2i+j+2\hat{j}\to l\\
& j\to n_1-n_2-j
\end{array}
 \right/ =
\nonumber\\
&=& \chi_m^{l-2n_2}\frac{n_1!}{n_2!}\sum_{j=\max [0,l-n_1-n_2]}^{\min [n_1-n_2,\left[ \frac{l-2n_2}{2}\right]]}
\left( \frac{\hat{x}}{4}\right)^j \cdot
\frac{1}{j!(l-2n_2-2j)!(n_1+n_2-l+j)!} =\nonumber\\
&& \phantom{...}\nonumber\\
&=& \mathbb{C}_{n_1}^{n_2} \chi_m^{l-2n_2} \mathcal{R}^{l-2n_2}_{n_1-n_2}(\hat{x})
\label{eq:obtainsolF1}
\end{eqnarray}
\begin{eqnarray}
 \mathcal{B}^l_{n_1 n_2} &=& \hat{\mathcal{D}}^l_{n_1 n_2} \rho_2(y,z) = \delta_{n_1 0} 
\chi_m^l \mathcal{R}^l_{n_2}(\hat{x})  \alpha^{J_1J_2k}_{l\;n_2}
\label{eq:obtainsolF2}\\ 
\hat{\mathcal{D}}^l_{n_1 n_2}f_2 &=& \!\!\!
\sum_{i_1=0}^{n_1} \sum_{i_2=0}^{n_2} \sum_{j=0}^{l} 
\mathcal{\beta}^j_{i_1\; i_2}  \mathcal{B}^{l-j}_{n_1-i_1\; n_2-i_2} = \nonumber\\
&=& \!\!\left/\!\!
\begin{array}{ll}
&  j\to l-j\\
& i_2\to n_2-i_2
\end{array}
 \right/ \!\!=
\chi_m^{l-2n_2} \Lambda^{J_1J_2k}_{R\; l\; n_1 n_2}(\hat{x}),\label{eq:obtainsolF3}
\end{eqnarray}
\begin{eqnarray}
 \Lambda^{J_1J_2k}_{R\; l\; n_1 n_2}(\hat{x}) &=& 
\sum_{i=0}^{n_2} 
\mathbb{C}_{n_1}^{n_2-i}
\sum_{j=0}^{l} \alpha^{J_1J_2k}_{j\; i} \mathcal{R}^j_i(\hat{x}) \mathcal{R}^{l-2n_2+2i-j}_{n_1-n_2+i}(\hat{x})
\label{def:LambdaR}\\
  \Lambda^{J_1J_2k}_{C\; l\; n_1 n_2} &=& \lim_{\hat{x}\to 0}
\Lambda^{J_1J_2k}_{R\; l\; n_1 n_2}(\hat{x}) 
 =
\sum_{i=0}^{n_2} 
\mathbb{C}_{n_1}^{n_2-i}
\sum_{j=0}^{l} \alpha^{J_1J_2k}_{j\; i} \mathbb{C}^j_i \mathbb{C}^{l-2n_2+2i-j}_{n_1-n_2+i}. 
\label{def:LambdaC}
\end{eqnarray}

\subsection{\bf Algorithm to solve basic recurrent equations in harmonical basis for $\cal{F}$ and $\cal{W}$}
\label{subsection:FWsolveHARM}

From the equation~(\ref{eq:irrFcoefsHARM}) for $\ell=0$ we have
\begin{equation}
2(\lambda_k+n_i-1)\mathbf{f}^{0\, k}_{\vec{n}} + \mathbf{f}^{0\, k}_{\vec{n}-\vec{e}_i} = 0.
\label{eq:WFHARMzeroell}
\end{equation}
Then, applying the shift in each component we can obtain
\begin{equation}
\mathbf{f}^{0\, k}_{\vec{n}} = -\frac{1}{2(\lambda_k+n_i-1)}\mathbf{f}^{0\, k}_{\vec{n}-\vec{e}_i} = \frac{(-1)^{n_i}}{2^{n_i}(\lambda_k)_{n_i}} \mathbf{f}^{0\, k}_{n_{i^{\star}}\vec{e}_{i^{\star}}} = \frac{(-1)^{|\vec{n}|}}{2^{|\vec{n}|}(\lambda_k)_{\vec{n}}},
\label{eq:WFHARMzeroellSol} 
\end{equation}
since $\mathbf{f}^{0\, k}_{\vec{0}} = 1$.

Let us write the solution as
\begin{equation}
\mathbf{f}^{\ell\, k}_{\vec{n}} = C_\ell \frac{(-1/2)^{|\vec{n}|-2\ell} \vec{n}!}{ (\vec{n}-\vec{\ell})! (\lambda_k)_{\vec{n}-\vec{\ell}} }.
\label{eq:WFHARMSol1}
\end{equation}
After that we obtain
\begin{equation}
-C_\ell \ell(1-\lambda_k+\ell) + C_{\ell-1} = 0 \implies C_\ell = \frac{C_{\ell-1}}{\ell(2-\lambda_k+\ell-1)} = \frac{1}{\ell! (2-\lambda_k)_\ell}.
\label{eq:WFHARMSol2}
\end{equation}
Finally we obtain the basic solution 
$\mathbf{f}^{*\ell\, k}_{\vec{n}}$~(\ref{eq:basicsolutionHARM}) and the general 
solution~(\ref{eq:fBsolutionHARM}).

To find $B^{k\,\vec{J}}_{\ell}$ we can use the general representation
\begin{eqnarray}
\varPhi^{*\mathcal{F}}_k(\vec{X}_F) &=&
\sum_{\tilde{\Omega}^F_{k'\,\vec{n}}} f^{k'\,(\vec{J};\,k)}_{\vec{n}} 
\mathcal{N}^{F\, \vec{J}}_{k\,\vec{n}}\cdot \vec{x}^{\vec{J}^*} \vec{u}^{\,\vec{n}} u_{12}^k =
\sum_{\ell=0}^{[k/2]} 
\sum_{\vec{n}=\vec{0}}^{[(\vec{J}-\vec{k})/2]+\vec{\ell}}
\!\!\!\!\!\!\!\sum_{m'=\atop =\max(0,\vec{\ell}-\vec{n})}^{\ell}
\hspace*{-4mm}B^{k\,\vec{J}}_{m'}
\times\nonumber\\
&\times& \frac{\vec{n}!\,2^{2(\ell-m')-|\vec{n}|}(-1)^{|\vec{n}|}
}{
(\ell-m')!\,(\vec{n}-\vec{\ell}+\vec{m}')!\,(\lambda_k-2m')_{\vec{n}-\vec{\ell}+\vec{m}'}\,(2-\lambda_k+2m')_{\ell-m'}
}\times
\nonumber\\
&\times& 
\frac{\vec{J}!}{
	2^{|\vec{n}|} \vec{n}! (k-2\ell)! (\vec{J}-2(\vec{n}-\vec{\ell})-\vec{k})!
}
\;\vec{x}^{\vec{J}-2(\vec{n}-\vec{\ell})-\vec{k}}
\vec{u}^{\,\vec{n}}
u_{12}^{k-2\ell}\label{def:genfunFstar}
\end{eqnarray}
and the closed
form for the harmonical representation~(\ref{eq:irrepHarmProjT}),(\ref{eq:irrepHarmProjTsingle}). Let us start from
\begin{eqnarray}
\varPhi^{*\mathcal{F}}_k(\vec{X}_F) &=&
\frac{1}{\Pi_0^{\vec{J}\,\vec{k}}}
\frac{\vec{J}!}{(\vec{J}-\vec{k})!k!}
\sum_{\vec{m}=\vec{0}}^{[\vec{J}/2]}\sum_{\vec{r}\,'=\vec{0}}^{\vec{m}} \mathbb{C}_{\vec{m}}^{\vec{r}\,'}
\frac{\vec{x}^{2(\vec{m}-\vec{r}\,')}\vec{u}^{\,\vec{r}\,'}}{4^{|\vec{m}|}\vec{m}!\,(c^{\vec{J}})_{\vec{m}}}
\square^{\vec{m}}_{\vec{\omega}} \vec{x}^{\vec{J}-\vec{k}}u_{12}^k,
\label{eq:genfunFstarHARM0}
\end{eqnarray}
($i=1,j=2$ in~(\ref{def:genWFvarsHARM}) and $\Pi_0^{\vec{J}\,\vec{k}}$~(\ref{HARMnormalization})), and then express the action of the operator $\square^{\vec{m}}_{\vec{\omega}}$.
\begin{eqnarray}
\square_{\omega_2}^{m_2} (x_2^{J_2-k}u_{12}^k) &=&
\sum_{r_2=0}^{m_2}\mathbb{C}_{m_2}^{r_2} 
\frac{(J_2-k)!k!\,x_2^{J_2-k-2(m_2-r_2)} u_1^{r_2} u_{12}^{k-2r_2}}{(J_2-k-2(m_2-r_2))!(k-2r_2)!},
\label{calc:squarem2}\\
\!\!\!\!\!\!\square_{\omega_1}^{m_1} (x_1^{J_1-k}u_1^{r_2}u_{12}^{k-2r_2}) &=&
\sum_{r_1=0}^{m_1}\mathbb{C}_{m_1}^{r_1}  
\frac{(J_1-k)!\,x_1^{J_1-k-2(m_1-r_1)}}{(J_1-k-2(m_1-r_1))!}
\square^{*r_1}_{\omega_1} (u_1^{r_2}u_{12}^{k-2r_2}),\nonumber\\
\square^*_{\omega_1}&=&
u_2 \partial^2_{u_{12}u_{12}}+
4\left[\frac{(D-2)}{2}+
u_1 \partial_{u_1}+
u_{12}\partial_{u_{12}}
\right] \partial_{u_1},\quad \vec{s}=\{s,s\},\nonumber\\
\square^{*r_1}_{\omega_1} (u_1^{r_2}u_{12}^{k-2r_2})&=&
\!\!\!\!\!\!\!\sum_{s=0}^{\min(r_1,r_2)}\!\!\!
\frac{4^s \vec{r}\,! (k-2r_2)! (\lambda_k-|\vec{r}\,|)_s}{s!(\vec{r}-\vec{s})!(k-2(|\vec{r}\,|-s))!}
u_1^{r_2-s}u_2^{r_1-s}u_{12}^{k-2(|\vec{r}\,|-s)}
\label{calc:squarem1}
\end{eqnarray}

\begin{eqnarray}
\varPhi^{*\mathcal{F}}_k(\vec{X}_F) &=&
\frac{\vec{J}!}{\Pi_0^{\vec{J}\,\vec{k}}} \!
\sum_{\vec{m}=\vec{0}}^{[\vec{J}/2]}
\sum_{\vec{r}\,'=\vec{0}}^{\vec{m}}
\sum_{\vec{r}=\vec{0}}^{\vec{m}}
\sum_{s=0}^{\min(r_1,r_2)}\!\!\!\!\!\!
\frac{
	4^{s-|\vec{m}|} (\lambda_k-|\vec{r}\,|)_s\vec{m}!
	}{
	\vec{r}\,'!(\vec{m}-\vec{r}\,')!(\vec{m}-\vec{r})!
	(\vec{J}-\vec{k}+2(\vec{m}-\vec{r}))!(\vec{r}-\vec{s})!
	}
\times\nonumber\\
&\times& 
\frac{1}{s!(k-2(|\vec{r}\,|-s))!(c^{\vec{J}})_{\vec{m}}}
\vec{x}^{\vec{J}-\vec{k}+2(\vec{r}-\vec{r}\,')}
u_1^{r_1'+r_2-s} u_2^{r_2'+r_1-s} u_{12}^{k-2(|\vec{r}\,|-s)}
\label{eq:genfunFstarHARM}
\end{eqnarray}
Let's equate the degrees of all the variables in~(\ref{eq:genfunFstarHARM}) and~(\ref{def:genfunFstar}) and set $\vec{n}=\vec{0}$, then
\begin{equation}
m'=\ell,\qquad 
\vec{r}-\vec{r}^{\,\prime}=\vec{\ell},\qquad
|\vec{r}\,|-s=\ell,
\label{eq:genfunFHARMpowers}
\end{equation}
and $\vec{r}-\vec{s}=-\vec{r}^{\,\prime}$. We have $\vec{r}^{\,\prime}!$ and $(-\vec{r}^{\,\prime})!$ in the denominator, so $\vec{r}^{\,\prime}=\vec{0}$. And now it is easy to obtain the expression for $B^{k\,\vec{J}}_{\ell}$~(\ref{eq:BellHARM}) from~(\ref{eq:genfunFstarHARM}) and~(\ref{def:genfunFstar}).

\subsection{\bf General linear recurrence with vector index and its solution.}
\label{subsection:GENERALsolve}

Let us consider the general problem of multiindex (vector index) linear recurrence 
solution (see, for example~\cite{recur1},\cite{recur2}). We have some coefficients $b^{\vec{\upsilon}'}_n$, where $$\vec{\upsilon}'=\{\upsilon'_1,\upsilon'_2,...,\upsilon'_N\}$$ is the vector 
of indexes. We can define also
the fixed vectors 
$$
\vec{\upsilon}=\{\upsilon_1,\upsilon_2,...,\upsilon_N\},\qquad 
\vec{\Delta}_i=\{\Delta_{i,1},...\Delta_{i,N}\} \, (\mathrm{shifts})
$$
which we will use in this task. 


At first, we have the following linear recurrent equation
\begin{equation}
b^{\vec{\upsilon}'}_n =
\gamma b^{\vec{\upsilon}'}_{n-1}+\sum_{i=1}^L \zeta_i b^{\vec{\upsilon}'+\vec{\Delta}_i}_{n-1},
\label{brecurrence}
\end{equation}
with the initial condition
\begin{equation}
b^{\vec{\upsilon}'}_0=B^{\vec{\upsilon}'},
\label{brecurrenceINI}
\end{equation}
where $\zeta_i$ are some constants (let us denote them as a vector $\vec{\zeta} =\{\zeta_1,...,\zeta_L\}\in \Omega_L$) and 
$\vec{\upsilon}' \in \Omega_N$, $\vec{\upsilon} \in \Omega^{\upsilon}_N$. $\Omega_L \in\mathbb{R}^L$. $\Omega_N, \Omega^{\upsilon}_N \in\mathbb{Z}_+^N$. For each component of the vector $\vec{\Delta}_i$ we have $\Delta_{i,j}\in\left[ -1,2\right]$.
Here
\begin{equation}
L=\left[\frac{(\|\vec{J}\|-1)(\|\vec{J}\|+5)}{2}\right]-\tilde{N},\qquad N=\frac{(\|\vec{J}\|-1)(\|\vec{J}\|+2)}{2},
\label{eq:LNvalues}
\end{equation}
$\tilde{N}=1$ for forward amplitudes $\mathcal{W}$,$\mathcal{H}$, otherwise it is zero, and $\|\vec{J}\|$
is the number of tensor legs in the amplitude.

To solve the recurrence we can introduce the indexed functions
\begin{equation}
 \beta^{\vec{\upsilon}'}(x)=\sum_{n=0}^{\infty}b^{\vec{\upsilon}'}_n x^n\nonumber\\
\label{defbeta}.
\end{equation}

In terms of $\beta^{\vec{\upsilon}'}(x)$ we have the following recurrent relation:
\begin{equation}
 \beta^{\vec{\upsilon}'}(x)=
 \frac{B^{\vec{\upsilon}'}}{1-\gamma x}+\frac{x}{1-\gamma x} \sum_{i=1}^L \zeta_i \beta^{\vec{\upsilon}'+\vec{\Delta}_i}(x)=
\frac{B^{\vec{\upsilon}'}}{1-\gamma x}+\frac{x}{1-\gamma x} \hat{T} \beta^{\vec{\upsilon}'}(x),
\label{betarecurrence}
\end{equation}
It can be solved by the inverse operator method
\begin{eqnarray}
 \beta^{\vec{\upsilon}'}(x)&=& 
\left(
1-\frac{x}{1-\gamma x} \hat{T}
\right)^{-1}
 \frac{B^{\vec{\upsilon}'}}{1-\gamma x} =
\sum_{m=0}^{\infty} \frac{x^m}{(1-\gamma x)^{m+1}} \hat{T}^m B^{\vec{\upsilon}'} =
\nonumber\\
&=& 
 \sum_{n=0}^{\infty} x^n \sum_{m=0}^n \mathbb{C}_n^m \gamma^{n-m} \hat{T}^m B^{\vec{\upsilon}'}.
\label{betarecurrencesol}
\end{eqnarray}
Now for $b^{\vec{\upsilon}'}_n$ we have from multinomial expansion of the $\hat{T}^m$:
\begin{eqnarray}
 b^{\vec{\upsilon}'}_n &=&  \sum_{m=0}^n \mathbb{C}_n^m \gamma^{n-m} \hat{T}^m B^{\vec{\upsilon}'} =
 \sum_{m=0}^n \mathbb{C}_n^m \gamma^{n-m}
\sum_{\sum\limits_{j=1}^L\!\!\! m_j =m\atop m_i\ge 0}\!\!\!\!
m! \prod_{i=1}^L \frac{\zeta_i^{m_i}}{m_i!}  
 B^{\vec{\upsilon}'+\sum\limits_{r=1}^L m_r \vec{\Delta}_r} =\nonumber\\
 &=&  \sum_{m=0}^n \frac{n!}{(n-m)!} \gamma^{n-m} \hspace*{-0.1cm} 
\sum_{\sum\limits_{j=1}^L\!\!\! m_j =m\atop m_i\ge 0}
\prod_{i=1}^L \frac{\zeta_i^{m_i}}{m_i!}  
 B^{\vec{\upsilon}'+\sum\limits_{r=1}^L m_r \vec{\Delta}_r} \equiv 
 \hat{\mathcal{S}}^{\;\vec{\zeta},\; \mathcal{M}_{\Delta}} \left[ B^{\vec{\upsilon}'} \right],
\label{brecurrencesol}\\
&& {\left( \mathcal{M}_{\Delta}\right)}_{ij} \equiv 
\Delta_{i,j}, 
\label{brecurrencesolM}
\end{eqnarray}
i.e. it is the final solution of~(\ref{brecurrence}) based on the initial value $B^{\vec{\upsilon}'}$, and 
$\hat{\mathcal{S}}^{\;\vec{\zeta},\; \mathcal{M}_{\Delta}}$
 is the operator of this solution. Also we
introduce generalization of this operator:
\begin{equation} 
\hat{\mathcal{S}}^{\;\vec{\zeta},\; \mathcal{M}_{\Delta}}_{a_1,...,a_{N'}} \left[ B^{\vec{\upsilon}'} \right] 
 = \hat{\mathcal{S}}^{\;\vec{\zeta},\; \mathcal{M}_{\Delta}} \left[ B^{\vec{\upsilon}'} \right]\Big|_{m_{a_1}=...=m_{a_{N'}}=0},
\label{def:SoperatorGen}
\end{equation}
where $m_{a_1},...,m_{a_{N'}}$ are some indexes in the multinomial sum, which are set to zero by hand.

\subsection{\bf Algorithm to solve recurrent equations for $\mathcal{H}$ in standard basis}
\label{subsection:Hsolve}

In this subsection we solve the recurrence~(\ref{eq:irrHcoefs}). As in the solution for $\mathcal{F}$, we begin
with changing variables:

\begin{eqnarray}
&& h^{\vec{k}'}_{\vec{n}} \equiv  h^{\vec{k}',(\vec{k};\vec{J})}_{\vec{n}}
 = \frac{2^{|\vec{k}|-|\vec{k}'|-|\vec{n}|}
 }{
(c^{\vec{J}})_{\vec{n}}} \mathbf{h}^{\vec{k}'}_{\vec{n}} =
2^{|\vec{k}|-|\vec{k}'|} 
 \mathcal{\upsilon}^{\vec{J}}_{\vec{n}} \,\cdot
\mathbf{h}^{\vec{k}'}_{\vec{n}}
,\; \label{eq:irrHcoefsChVars}\\
&&\mathbf{h}^{\vec{k}}_{\vec{n}}=1,\quad
\mathbf{h}^{\vec{k}'}_{\vec{0}}=\delta_{\vec{k}'\vec{k}}.
\label{eq:irrHcoefsChVarsInitial}\\
&&
\mathbf{h}^{\vec{k}'}_{\vec{n}} = \mathbf{h}^{\vec{k}'}_{i;\; \vec{n}}  \mathbf{A}^{\vec{k}'}_{i;\; \vec{n}},\quad
 \mathbf{A}^{\vec{k}'}_{i;\; \vec{n}} = 
\prod_{r\neq i\atop r\in\Omega_H}\!\!\!
J^*_r! n_r! \left(c^{J_r}\right)_{n_r}\!\!\!\!
\prod_{r,s\in\Omega_H\atop r\neq s\neq i} \!\!\! k'_{rs}!,
\quad i\in\Omega_H.
\label{eq:irrHcoefsChVars2}
\end{eqnarray}
Coefficients can be found in the text: $J^*_r$~(\ref{def:Jbar}), $\Omega_H$~(\ref{def:qgpownonconserv}), 
$c^J$~(\ref{eq:irrVcoefs}), $\mathcal{\upsilon}^J_n$~(\ref{eq:irrVcoefsSolution}).

After that we get new equations where $\gamma=1$:
\begin{eqnarray}
\!\!\! \mathbf{h}^{\vec{k}'}_{i;\; \vec{n}} &=& 
 \left(
\mathbf{h}^{\vec{k}'}_{i;\; \vec{n}-\vec{e}_i}
+
 \mathbf{h}^{\vec{k}'+\vec{e}_{ii'}}_{i;\; \vec{n}-\vec{e}_i}
 +
 \mathbf{h}^{\vec{k}'+2\vec{e}_{ii'}}_{i;\; \vec{n}-\vec{e}_i-\vec{e}_{i'}}
  +
 \mathbf{h}^{\vec{k}'+\vec{e}_{ij}+\vec{e}_{ij'}-\vec{e}_{jj'}}_{i;\; \vec{n}-\vec{e}_i}
 -
\frac{\lambda_m}{2} 
\mathbf{h}^{\vec{k}'+\vec{e}_{ij}+\vec{e}_{ij'}}_{i;\; \vec{n}-\vec{e}_i}
+
\phantom{ \sum_{r\neq i,i'\atop r\in \Omega_H} \frac{\lambda_m}{4} \hat{O}^{+2}_{k'_{ir}}}
\hspace*{-3cm}
\right.\nonumber\\
 &+& \left.  \sum_{r\neq i,i'\atop r\in \Omega_H} 
\left[ 
\chi_m \mathbf{h}^{\vec{k}'+\vec{e}_{ir}}_{i;\; \vec{n}-\vec{e}_i}
+
\mathbf{h}^{\vec{k}'+\vec{e}_{ii'}+\vec{e}_{ir}-\vec{e}_{i'r}}_{i;\; \vec{n}-\vec{e}_i}
+
\mathbf{h}^{\vec{k}'+2\vec{e}_{ir}}_{i;\; \vec{n}-\vec{e}_i\vec{e}_r}
-
\mathbf{h}^{\vec{k}'+2\vec{e}_{ir}}_{i;\; \vec{n}-\vec{e}_i}
\right]
\right),\nonumber\\
&& \hspace*{-10mm} r,i,i',j,j'\in\Omega_H;\quad \{j,j'\}=\{i^{\star},{i^{\star}}'\}\neq \{i,i'\}.
\label{eq:irrHBicoefs}
\end{eqnarray}

Let us introduce some auxiliary operators and vectors to use the algorithm of the previous subsection for the solution of~(\ref{eq:irrHBicoefs})
with $N=9$ and $L=12$.

\begin{equation}
\vec{\zeta}^{\,H} = \{1, 1, -\lambda_m/2, \chi_m, 1, -\lambda_m/4, \chi_m, 1, -\lambda_m/4, 1, 1, 1\},
\label{def:omegaH}
\end{equation}
\begin{equation}
\mathcal{M}^H_{\Delta} = 
\begin{pmatrix}
 1 & 0 & 0 & 0 & 0 & 0 & 0 & 0 & 0 \\
 0 & 1 & 1 & 0 & 0 & -1 & 0 & 0 & 0 \\
 0 & 1 & 1 & 0 & 0 &  0 & 0 & 0 & 0 \\
 0 & 1 & 0 & 0 & 0 &  0 & 0 & 0 & 0 \\
 1 & 1 & 0 & -1& 0 &  0 & 0 & 0 & 0 \\
 0 & 2 & 0 & 0 & 0 &  0 & 0 & 0 & 0 \\
 0 & 0 & 1 & 0 & 0 &  0 & 0 & 0 & 0 \\
 1 & 0 & 1 & 0 & -1&  0 & 0 & 0 & 0 \\
 0 & 0 & 2 & 0 & 0 &  0 & 0 & 0 & 0 \\
 2 & 0 & 0 & 0 & 0 &  0 & -1& 0 & 0 \\
 0 & 2 & 0 & 0 & 0 &  0 & 0 & -1& 0 \\
 0 & 0 & 2 & 0 & 0 &  0 & 0 & 0 & -1
\end{pmatrix}
\label{def:DeltaMatrixH}
\end{equation}

We define the transformation rule to obtain $\vec{\upsilon}_i$, $\vec{\upsilon}'_i$ from $\vec{k}$, $\vec{k}'$, $\vec{n}$:
\begin{eqnarray}
&& \vec{\upsilon}'^H_i = \{ k'_{ii'},k'_{ij},k'_{ij'},k'_{i'j},k'_{i'j'},k'_{jj'},n_{i'},n_j,n_{j'} \},
\label{vec:kappapH}\\
&&  \vec{\upsilon}^H_i = \{ k_{ii'},k_{ij},k_{ij'},k_{i'j},k_{i'j'},k_{jj'},n_{i'},n_j,n_{j'} \}.
\label{vec:kappaH}
\end{eqnarray}
This means that \textbf{everywhere when we use the operator $\hat{\mathcal{S}}$ which acts on $\vec{\upsilon}'_i$ of the
initial value (as in (\ref{brecurrencesol}),(\ref{def:SoperatorGen})), it acts on the 
corresponding components of $\vec{k}'$, $\vec{n}$ in $\vec{\upsilon}'_i$, even if we write the initial value
in terms of $\vec{k}'$, $\vec{n}$} like in the algorithm below.

Then we can write the equation~(\ref{eq:irrHBicoefs}) in the form of~(\ref{brecurrence}) and find
the solution as a chain of steps:
\begin{enumerate}
\item We start from the initial value and apply the operator~(\ref{def:SoperatorGen}) to obtain
$ \mathbf{h}^{\vec{k}'}_{\vec{n}}$ with $n_{i',j,j'}=0$ and only one nonzero $n_i$:
\begin{eqnarray}
 \mathbf{h}^{\vec{k}'}_{i;\; \vec{0}}  &=& \frac{\delta_{\vec{k}'\vec{k}}}{\mathbf{A}^{\vec{k}}_{i;\; \vec{0}}},\quad
 \mathbf{h}^{\vec{k}'}_{i;\; \vec{n}}\Big|^{n_{i',j,j'}=0}_{n_i\neq 0} =
\hat{\mathcal{S}}^{\;\vec{\zeta}^{\,H},\; \mathcal{M}^H_{\Delta}}_{10,11,12}
\left[ \mathbf{h}^{\vec{k}'}_{i;\; \vec{0}} \right]
,\nonumber\\
 \mathbf{h}^{\vec{k}'}_{\vec{n}}\Big|^{n_{i',j,j'}=0}_{n_i\neq 0} &=&
\mathbf{h}^{\vec{k}'}_{i;\; \vec{n}}\Big|^{n_{i',j,j'}=0}_{n_i\neq 0} 
\mathbf{A}^{\vec{k}'}_{i;\; \vec{n}}\Big|^{n_{i',j,j'}=0}_{n_i\neq 0}
=\nonumber\\
&=& 
\frac{\mathbf{A}^{\vec{k}'}_{i;\; \vec{n}}\Big|^{n_{i',j,j'}=0}_{n_i\neq 0}}{\mathbf{A}^{\vec{k}'}_{i;\; \vec{0}}}
\hat{\mathcal{S}}^{\;\vec{\zeta}^{\,H},\; \mathcal{M}^H_{\Delta}}_{10,11,12}
\left[ \delta_{\vec{k}'\vec{k}} \right]
\label{inih:step1}
\end{eqnarray}
If we fix $i$, it is also possible to obtain
\begin{eqnarray}
&& \mathbf{h}^{\vec{k}'}_{r;\; \vec{n}}\Big|^{n_{i',j,j'}=0}_{n_i\neq 0}=
\frac{
\mathbf{A}^{\vec{k}'}_{i;\; \vec{n}}\Big|^{n_{i',j,j'}=0}_{n_i\neq 0}
}{\mathbf{A}^{\vec{k}'}_{r;\; \vec{n}}\Big|^{n_{i',j,j'}=0}_{n_i\neq 0}\mathbf{A}^{\vec{k}'}_{i;\; \vec{0}}}
 \; \hat{\mathcal{S}}^{\;\vec{\zeta}^{\,H},\; \mathcal{M}^H_{\Delta}}_{10,11,12}
\left[ \delta_{\vec{k}'\vec{k}} \right],\nonumber\\
&& r=i',j,j';\, j\equiv i^{\star},\, j'\equiv{i^{\star}}', \label{eq:hrzeroi}
\end{eqnarray}
without other equations for $i',j,j'$ since they are similar.

\item Now we have to obtain our coefficients with two nonzero indexes, say $r_1$ and $r_2$. For this we take
two equations (like for coefficients of $\mathcal{F}$), and two coefficients from the previous step, which we use as initial values:
\begin{eqnarray}
 \mathbf{h}^{\vec{k}'}_{r_i;\; \vec{n}}\Big|^{n_a=0,\; a\neq r_{1,2}}_{n_{r_1,r_2}\neq 0} &=&
\hat{\mathcal{S}}^{\;\vec{\zeta}^{\,H},\; \mathcal{M}^H_{\Delta}}_{a_1,a_2}
\left[ \mathbf{h}^{\vec{k}'}_{r_i;\; \vec{n}}\Big|^{n_a=0,\; a\neq r_{1,2}}_{n_{r_j}\neq 0, n_{r_i} = 0}
 \right],\quad
\{i,j\} =\{1,2\}, \{2,1\},\nonumber\\
 \{a_1,a_2\}&=&
\begin{cases}
& \{ 11,12\},\; r_j=r_i',\\
& \{ 10,12\},\; r_j=r^*_i,\\
& \{ 10,11\},\; r_j={r^*_i}'.
\end{cases}
\label{inih:step2},\\
 \mathbf{h}^{\vec{k}'}_{\vec{n}}\Big|^{n_a=0, a\neq r_{1,2}}_{n_{r_1,r_2}\neq 0} &=&
 \frac{1}{2}
\sum\limits_{i\in\{1,2\}}
\mathbf{h}^{\vec{k}'}_{r_i;\; \vec{n}}
\Big|^{n_a=0, a\neq r_{1,2}}_{n_{r_1,r_2}\neq 0}
\mathbf{A}^{\vec{k}'}_{r_i;\; \vec{n}}
\Big|^{n_a=0, a\neq r_{1,2}}_{n_{r_1,r_2}\neq 0}
\label{inih:step2a}
\end{eqnarray}

\item The next step is
to obtain coefficients with three nonzero indexes, say $r_{1,2,3}$. For this we take
three equations, and three coefficients from the previous step, which we use as initial values:

\begin{equation}
 \mathbf{h}^{\vec{k}'}_{r_i;\; \vec{n}}\Big|^{n_a=0, a\neq r_{j,k}}_{n_{r_j,r_k}\neq 0}=
\frac{\mathbf{h}^{\vec{k}'}_{\vec{n}}\Big|^{n_a=0, a\neq r_{j,k}}_{n_{r_j,r_k}\neq 0}}{
\mathbf{A}^{\vec{k}'}_{r_i;\; \vec{n}}
\Big|^{n_a=0, a\neq r_{j,k}}_{n_{r_j,r_k}\neq 0}}
,\quad r_i\neq r_{j,k},. \label{eq:hrzerono12}
\end{equation}
where $ \{i,j,k\} = \{1,2,3\}, \{2,1,3\}, \{3,1,2\}$. 

\begin{eqnarray}
 \mathbf{h}^{\vec{k}'}_{r_i;\; \vec{n}}\Big|^{n_a=0,\; a\neq r_{1,2,3}}_{n_{r_1,r_2,r_3}\neq 0} &=&
\hat{\mathcal{S}}^{\;\vec{\zeta}^{\,H},\; \mathcal{M}^H_{\Delta}}_{a_1}
\left[ \mathbf{h}^{\vec{k}'}_{r_i;\; \vec{n}}\Big|^{n_a=0,\; a\neq r_{i,j,k}}_{n_{r_j,r_k}\neq 0, n_{r_i} = 0}
 \right],\nonumber\\
 a_1&=&
\begin{cases}
& 12,\; r_j=r'_i,\; r_k=r^*_i \\
& 11,\; r_j=r'_i,\; r_k={r^*_i}',\\
& 10,\; r_j=r^*_i,\; r_k={r^*_i}'.
\end{cases}
\label{inih:step3},\\
 \mathbf{h}^{\vec{k}'}_{\vec{n}}\Big|^{n_a=0, a\neq r_{1,2,3}}_{n_{r_1,r_2,r_3}\neq 0} 
&=& \frac{1}{3}
\sum\limits_{i\in\{1,2,3\}}
\mathbf{h}^{\vec{k}'}_{r_i;\; \vec{n}}
\Big|^{n_a=0, a\neq r_{1,2,3}}_{n_{r_1,r_2,r_3}\neq 0}
\mathbf{A}^{\vec{k}'}_{r_i;\; \vec{n}}
\Big|^{n_a=0, a\neq r_{1,2,3}}_{n_{r_1,r_2,r_3}\neq 0}
\label{inih:step3a}
\end{eqnarray}

\item  The last step is
to obtain coefficients with all nonzero indexes in $\vec{n}$. For this we take all the
equations, and four coefficients from the previous step, which we use as initial values:

\begin{eqnarray}
&& \mathbf{h}^{\vec{k}'}_{r;\; \vec{n}}\Big|^{n_r=0}_{n_a\neq 0, a\neq r}=
\frac{\mathbf{h}^{\vec{k}'}_{\vec{n}}\Big|^{n_r=0}_{n_a\neq 0, a\neq r}}{
\mathbf{A}^{\vec{k}'}_{r;\; \vec{n}}
\Big|^{n_r=0}_{n_a\neq 0, a\neq r}}
,\; r\in\Omega_H, \label{eq:hrzerono123}\\
&& 
 \mathbf{h}^{\vec{k}'}_{r;\; \vec{n}} =
\hat{\mathcal{S}}^{\;\vec{\zeta}^{\,H},\; \mathcal{M}^H_{\Delta}}
\left[ \mathbf{h}^{\vec{k}'}_{r;\; \vec{n}}\Big|^{n_r=0}_{n_a\neq 0, a\neq r}
 \right],
\label{inih:step4}\\
&& \mathbf{h}^{\vec{k}'}_{\vec{n}} 
= \frac{1}{4}
\sum\limits_{r\in\Omega_H}
\mathbf{h}^{\vec{k}'}_{r;\; \vec{n}}
\mathbf{A}^{\vec{k}'}_{r;\; \vec{n}}
\label{inih:step4a}
\end{eqnarray}

\end{enumerate}

\subsection{\bf Algorithm to solve recurrent equations for $\mathcal{Y}$ in standard basis}
\label{subsection:Ysolve}


In this subsection we solve the recurrence~(\ref{eq:irrYcoefs}). As in the previous subsection, we begin
with changing variables:

\begin{eqnarray}
 \hspace*{-8mm}y^{\vec{k}'}_{\vec{n}} &\equiv&  y^{\vec{k}'\,(\vec{k};\,\vec{J})}_{\vec{n}}
 = \frac{2^{|\vec{k}|-|\vec{k}'|-|\vec{n}|}
 	}{
(c^{\vec{J}})_{\vec{n}}
}
\mathbf{y}^{\vec{k}'}_{\vec{n}}
= 2^{|\vec{k}|-|\vec{k}'|}
\mathcal{\upsilon}^{\vec{J}}_{\vec{n}}\cdot\mathbf{y}^{\vec{k}'}_{\vec{n}}
,\; \label{eq:irrYcoefsChVars}\\
\mathbf{y}^{\vec{k}}_{\vec{n}}&=&1,\quad
\mathbf{y}^{\vec{k}'}_{\vec{0}}=\delta_{\vec{k}'\vec{k}}.
\label{eq:irrYcoefsChVarsInitial}\\
\mathbf{y}^{\vec{k}'}_{\vec{n}} &=& \mathbf{y}^{\vec{k}'}_{i;\; \vec{n}}  \mathbf{A}^{\vec{k}'}_{i;\; \vec{n}},\quad
 \mathbf{A}^{\vec{k}'}_{i;\; \vec{n}} = \!\!\!
\prod_{r\neq i\atop r\in\Omega_Y}\!\!\!
J^*_r! n_r! \left(c^{J_r}\right)_{n_r}\!\!\!\!\cdot
k'_{ss'}!,\quad
i,s,s'\in\Omega_Y,\, i\neq s\neq s'.
\label{eq:irrYcoefsChVars2}
\end{eqnarray}
Coefficients can be found in the text: $J^*_r$~(\ref{def:Jbar}), $\Omega_Y$~(\ref{def:qgpownonconserv}), 
$c^J$~(\ref{eq:irrVcoefs}), $\mathcal{\upsilon}^J_n$~(\ref{eq:irrVcoefsSolution}).

After that we get new equations
\begin{eqnarray}
&& \mathbf{y}^{\vec{k}'}_{i;\; \vec{n}} =  \left(  \hspace*{-0.12cm}
P_i^2 \mathbf{y}^{\vec{k}'}_{i;\; \vec{n}-\vec{e}_i}
 + \!\left[
\mathbf{y}^{\vec{k}'+\vec{e}_{ir}+\vec{e}_{is}-\vec{e}_{rs}}_{i;\; \vec{n}-\vec{e}_i}
  +
\frac{\hat{\lambda}^i_{rs}}{2} \mathbf{y}^{\vec{k}'+\vec{e}_{ir}+\vec{e}_{is}}_{i;\; \vec{n}-\vec{e}_i}
\right] \!+ \!\!\! \phantom{\sum_{j\neq i\atop j\in \Omega_Y} }
\right.\nonumber\\
&& \left. \phantom{\frac{\hat{\lambda}^i_{rs}}{2}}\hspace*{1cm}
+ \sum_{j\neq i\atop j\in \Omega_Y}  
\left[ 
\bar{\chi}_{ij} \mathbf{y}^{\vec{k}'+\vec{e}_{ij}}_{i;\; \vec{n}-\vec{e}_i}
+
\mathbf{y}^{\vec{k}'+2\vec{e}_{ij}}_{i;\; \vec{n}-\vec{e}_i-\vec{e}_j}
\!-\!\frac{\tilde{\lambda}^j_{ij}}{4}
\mathbf{y}^{\vec{k}'+2\vec{e}_{ij}}_{i;\; \vec{n}-\vec{e}_i}
\right]  \hspace*{-0.12cm}
\right)\nonumber\\
&&\hspace*{-0.2cm}  r,s,i\in\Omega_Y;\quad i\neq r\neq s,\quad j\in\{r,s\},
\label{eq:irrYBicoefs}
\end{eqnarray}
and $\gamma=P_i^2$ for the operator $\hat{\mathcal{S}}$~(\ref{def:SoperatorGen}) of the general solution.

Let us introduce some auxiliary operators and vectors to use the algorithm of the previous subsection for the solution of~(\ref{eq:irrYBicoefs})
with $N=5$ and $L=8$.

\begin{equation}
\vec{\zeta}^{\,Y}_i = \{1, \hat{\lambda}^i_{rs}/2, \bar{\chi}_{ir}, -\tilde{\lambda}^j_{ir}/4, \bar{\chi}_{is},
-\tilde{\lambda}^j_{is}/4,1,1\},
\label{def:omegaY}
\end{equation}
\begin{equation}
\mathcal{M}^Y_{\Delta} = 
\begin{pmatrix}
 1 & 1 & -1 & 0 & 0  \\
 1 & 1 & 0 & 0 & 0  \\
 1 & 0 & 0 & 0 & 0  \\
 2 & 0 & 0 & 0 & 0  \\
 0 & 1 & 0 & 0 & 0  \\
 0 & 2 & 0 & 0 & 0  \\
 2 & 0 & 0 & -1 & 0 \\
 0 & 2 & 0 & 0 & -1 \\
\end{pmatrix}
\label{def:DeltaMatrixY}
\end{equation}

We define the transformation rule to obtain $\vec{\upsilon}_i$, $\vec{\upsilon}'_i$ from $\vec{k}$, $\vec{k}'$, $\vec{n}$:
\begin{equation}
 \vec{\upsilon}'^Y_i = \{ k'_{ir},k'_{is},k'_{rs},n_r,n_s \},\,
  \vec{\upsilon}^Y_i = \{ k_{ir},k_{is},k_{rs},n_r,n_s \},\,
i,r,s\in\Omega_Y,\, i\neq r\neq s.  
\label{vec:kappaY}
\end{equation}
This means that \textbf{everywhere when we use the operator $\hat{\mathcal{S}}$ which acts on $\vec{\upsilon}'_i$ of the
initial value (as in (\ref{brecurrencesol}),(\ref{def:SoperatorGen})), it acts on the 
corresponding components of $\vec{k}'$, $\vec{n}$ in $\vec{\upsilon}'_i$, even if we write the initial value
in terms of $\vec{k}'$, $\vec{n}$} like in the algorithm below.

Then we can write the equation~(\ref{eq:irrYBicoefs}) in the form of~(\ref{brecurrence}) and find
the solution as a chain of steps:
\begin{enumerate}
\item We start from the initial value and apply the operator~(\ref{def:SoperatorGen}) to obtain
$ \mathbf{y}^{\vec{k}'}_{\vec{n}}$ with $n_{r,s}=0$ and only one nonzero $n_i$:
\begin{eqnarray}
 \mathbf{y}^{\vec{k}'}_{i;\; \vec{0}}  &=& \frac{\delta_{\vec{k}'\vec{k}}}{\mathbf{A}^{\vec{k}}_{i;\; \vec{0}}},\,
 \mathbf{y}^{\vec{k}'}_{i;\; \vec{n}}\Big|^{n_{r,s}=0}_{n_i\neq 0} =
\hat{\mathcal{S}}^{\;\vec{\zeta}^{\,Y}_i,\; \mathcal{M}^Y_{\Delta}}_{4,5}
\left[ \mathbf{y}^{\vec{k}'}_{i;\; \vec{0}} \right]
,\nonumber\\
 \mathbf{y}^{\vec{k}'}_{\vec{n}}\Big|^{n_{r,s}=0}_{n_i\neq 0} &=&
\mathbf{y}^{\vec{k}'}_{i;\; \vec{n}}\Big|^{n_{r,s}=0}_{n_i\neq 0} 
\mathbf{A}^{\vec{k}'}_{i;\; \vec{n}}\Big|^{n_{r,s}=0}_{n_i\neq 0}
= 
\frac{\mathbf{A}^{\vec{k}'}_{i;\; \vec{n}}\Big|^{n_{r,s}=0}_{n_i\neq 0}}{\mathbf{A}^{\vec{k}'}_{i;\; \vec{0}}}
\hat{\mathcal{S}}^{\;\vec{\zeta}^{\,Y}_i,\; \mathcal{M}^Y_{\Delta}}_{4,5}
\left[ \delta_{\vec{k}'\vec{k}} \right]
\label{inih:step1Y}
\end{eqnarray}
If we fix $i$, it is also possible to obtain
\begin{eqnarray}
&& \mathbf{y}^{\vec{k}'}_{j;\; \vec{n}}\Big|^{n_{r,s}=0}_{n_i\neq 0}=
\frac{
\mathbf{A}^{\vec{k}'}_{i;\; \vec{n}}\Big|^{n_{r,s}=0}_{n_i\neq 0}
}{\mathbf{A}^{\vec{k}'}_{j;\; \vec{n}}\Big|^{n_{r,s}=0}_{n_i\neq 0}\mathbf{A}^{\vec{k}'}_{i;\; \vec{0}}}
 \hat{\mathcal{S}}^{\;\vec{\zeta}^{\,Y}_i,\; \mathcal{M}^Y_{\Delta}}_{4,5}
\left[ \delta_{\vec{k}'\vec{k}} \right],\nonumber\\
&& j=r,s;\, i,r,s\in\Omega_Y,\, i\neq r\neq s,
\quad \{i,r,s\} \in \{1,2,3\}_P \label{eq:hrzeroiY}
\end{eqnarray}
without other equations for $r,s$ since they are similar. 

\item Now we have to obtain our coefficients with two nonzero indexes, say $i$ and $j$. For this we take
two equations and two coefficients from the previous step, which we use as initial values:
\begin{eqnarray}
 \mathbf{y}^{\vec{k}'}_{i;\; \vec{n}}\Big|^{n_a=0,\, a\neq i,j}_{i,j\neq 0} &=&
\hat{\mathcal{S}}^{\;\vec{\zeta}^{\,Y}_i,\; \mathcal{M}^Y_{\Delta}}_{a_1}
\left[ \mathbf{y}^{\vec{k}'}_{i;\; \vec{n}}\Big|^{n_a=0,\, a\neq i,j}_{n_{j}\neq 0, n_i = 0}
 \right],\nonumber\\
 a_1&=&
\begin{cases}
& 4,\; a<j\\ 
& 5,\; a>j\\ 
\end{cases}
\label{inih:step2Y}\\
 \mathbf{y}^{\vec{k}'}_{\vec{n}}\Big|^{n_a=0, a\neq i,j}_{n_{i,j}\neq 0} 
 &=& \frac{1}{2}
\sum\limits_{\bar{j}=i,j}
\mathbf{y}^{\vec{k}'}_{\bar{j};\; \vec{n}}
\Big|^{n_a=0, a\neq i,j}_{n_{i,j}\neq 0}
\mathbf{A}^{\vec{k}'}_{\bar{j};\; \vec{n}}
\Big|^{n_a=0, a\neq i,j}_{n_{i,j}\neq 0}\nonumber\\
\{i,j\} &=&\{1,2\}, \{2,1\}, \{1,3\},\{3,1\}, \{2,3\}, \{3,2\}.
\label{inih:step2aY}
\end{eqnarray}

\item  The last step is
to obtain coefficients with all nonzero indexes in $\vec{n}$. For this we take all the
equations, and four coefficients from the previous step, which we use as initial values:

\begin{eqnarray}
&& \mathbf{y}^{\vec{k}'}_{i;\; \vec{n}}\Big|^{n_i=0}_{n_{r,s}\neq 0}=
\frac{\mathbf{y}^{\vec{k}'}_{\vec{n}}\Big|^{n_i=0}_{n_{r,s}\neq 0}}{
\mathbf{A}^{\vec{k}'}_{i;\; \vec{n}}
\Big|^{n_i=0}_{n_{r,s}\neq 0}}
,\quad
i,r,s\in\Omega_Y,\, i\neq r\neq s, \label{eq:hrzerono123Y}\\
&& 
 \mathbf{y}^{\vec{k}'}_{i;\; \vec{n}} =
\hat{\mathcal{S}}^{\;\vec{\zeta}^{\,Y}_i,\; \mathcal{M}^Y_{\Delta}}
\left[ \mathbf{y}^{\vec{k}'}_{i;\; \vec{n}}\Big|^{n_i=0}_{n_{r,s}\neq 0}
 \right],
\label{inih:step4Y}\\
&& \mathbf{y}^{\vec{k}'}_{\vec{n}} 
= \frac{1}{3}
\sum\limits_{i\in\Omega_Y}
\mathbf{y}^{\vec{k}'}_{i;\; \vec{n}}
\mathbf{A}^{\vec{k}'}_{i;\; \vec{n}}
\label{inih:step4aY}
\end{eqnarray}

\end{enumerate}

\subsection{\bf Other possible methods to find solutions}
\label{subsection:othermethods}

Here we consider alternative ways to obtain irreducible tensors. 
The first way is to apply the method of generating functions directly to
the equations for 
$\mathrm{f}^l_{n_1 n_2}$~(\ref{eq:irrFcoefsChVars}), $\mathrm{h}^{\vec{k}'}_{\vec{n}}$~(\ref{eq:irrHcoefsChVars}),
$\mathrm{y}^{\vec{k}'}_{\vec{n}}$~(\ref{eq:irrYcoefsChVars}) using the tricks
from Appendix~\ref{appendixD}. In this case we obtain a system of complicated hyper\-geo\-met\-ric-like equations, which could be
solved somehow. But changes of variables in 
Appendices~\ref{subsection:FWsolve},\ref{subsection:Hsolve},\ref{subsection:Ysolve} to obtain solutions are more simple.

The second way is to use symmetric structures like 
(\ref{def:SF}),(\ref{def:SW}),(\ref{def:SY}),(\ref{def:SH}) with $\vec{n}= \vec{0}$ contracted with 
$\mathcal{P}^J$~(\ref{eq:propagatorJ}):
\begin{eqnarray}
&&\mathcal{T}^{*\,\vec{J}}_{\vec{k}\; \{(\alpha^{(r)})_{J_r}\}} = 
S^{T\; \vec{J}}_{\vec{k}\,\vec{0}\; \{(\beta^{(r)})_{J_r}\}}\otimes
\prod_{r\in\Omega_T}
 \mathcal{P}^{J_r}_{(\alpha^{(r)})_{J_r},(\beta^{(r)})_{J_r}},\nonumber\\
&& \{(\alpha^{(r)})_{J_r}\} =\{(\mu)_{J_1},...\} ,\quad 
  \{(\beta^{(r)})_{J_r}\} =\{(\beta^{(1)})_{J_1},...\} 
\label{eq:othermethods}
\end{eqnarray}
This automatically gives all the coefficients, but this case 
(as well as the method of helicity amplitudes) appears 
to be as complicated as the method discussed 
in the Appendix~\ref{appendixB}.


\section{Expansion of amplitudes in terms of irreducible tensors}
\label{appendixC}

Here we obtain coefficients of expansions of nonconserved tensors in terms of irreducible tensors. For this task it is
enough to consider coefficients $\tilde{\upsilon}_{u}^{J}$ in the expansion~(\ref{def:TSTnonconserv}). For this task
we use the same "tensor extraction trick" as in~(\ref{eq:Spur1V}),(\ref{eq:Vtensorout}).

\begin{eqnarray}
  Sp_1  \left( \mathcal{V}^{J-r} q_.^{\otimes r-2u} g_{..}^{\otimes u} \right) &=&
  g_{\mu_1\mu_2} 
\left[\phantom{\frac{X^X}{X^X}}\hspace*{-6mm}
q_{\mu_1}q_{\mu_2}   \left( \mathcal{V}^{J-r} q_.^{\otimes r-2u-2} g_{..}^{\otimes u} \right)+
\right. \nonumber\\
&+& \left.
 \sum_i \left( q_{\mu_1}  g_{\mu_2\mu_i} +
q_{\mu_2}  g_{\mu_1\mu_i}
\right)   \left( \mathcal{V}^{J-r} q_.^{\otimes r-2u-1} g_{..}^{\otimes u-1} \right)+
\right. \nonumber\\
&+& \left. 
\sum_{i<j}  g_{\mu_1\mu_i} g_{\mu_2\mu_j}  \left( \mathcal{V}^{J-r} q_.^{\otimes r-2u} g_{..}^{\otimes u-2} \right)  
+
\right.\nonumber\\
&+& \left. g_{\mu_1\mu_2}   
\left( \mathcal{V}^{J-r} q_.^{\otimes r-2u} g_{..}^{\otimes u-1} \right)
+
\right.\nonumber\\
&+& 
\left. 
 \sum_i\left( 
g_{\mu_1\mu_i}   
\left( \mathcal{V}^{J-r}_{...\mu_2...} q_.^{\otimes r-2u} g_{..}^{\otimes u-1} \right)
+( \mu_1 \leftrightarrow \mu_2)
\right)
\right]=\nonumber\\
&=& \left( \mathcal{V}^{J-r} q_.^{\otimes r-2u-2} g_{..}^{\otimes u} \right)\! q^2 + 
\left( \mathcal{V}^{J-r} q_.^{\otimes r-2u} g_{..}^{\otimes u-1} \right) \times\nonumber\\
&\times& \left(\phantom{\frac{X}{X}}\hspace*{-5mm}
 2(r-4u)+2(u-1)+D+2(J-r+2u)
 \right)\!
,\label{eq:Spur1Vq}
\end{eqnarray}
which leads to the equation
\begin{equation}
\tilde{\upsilon}_{u-1}^{J}
- 2
\tilde{\upsilon}_{u}^{J}
(\tilde{c}^J+u-1)=0,
\label{eq:Vtensoroutq}
\end{equation}
where
\begin{equation}
\tilde{c}^J=-(J+(D-4)/2).
\label{def:tildecJ}
\end{equation}

The solution is
\begin{equation}
\label{eq:solVtensoroutq}
\tilde{\upsilon}_{u}^{J}=
\frac{q^{2u}}{2^{u}\left( \tilde{c}^J\right)_{u}}.
\end{equation}

It is easy to show that other coefficients 
factorize:
\begin{equation}
\tilde{\tau}_{\vec{u}}^{\vec{J}} =
\prod_{r\in\Omega_T}\tilde{\upsilon}_{u_r}^{J_r}. 
\label{UtildeAll}
\end{equation}

\section{Useful relations}
\label{appendixD}

\subsection{\bf Coefficients and number of terms in tensor structures}

If one introduce the following tensor structure
\begin{equation}
S_1 =\left( 
\left( a_{\mu_1...\mu_{l_a}}^{k_a} \right) 
 \left( b_{\nu_1...\nu_{l_b}}^{k_b} \right)
...
 \left( c_{\rho_1...\rho_{l_c}}^{k_c} \right)  
\right)
\label{d:S1}
\end{equation}
then the number of terms
\begin{equation}
\mathcal{N}(S_1) = \frac{(\sum\limits_i l_i k_i)!}{\prod\limits_j (l_j!)^{k_j} k_j!}
=
\prod_j \mathcal{N}(S_2^{(j)}) \mathbb{C}_{\sum\limits_i l_i k_i}^{l_j k_j}.
\label{d:NS1}
\end{equation}
Here 
\begin{equation}
S_2= \left( a_{\mu_1...\mu_{l_a}}^{k_a} \right)
\label{d:S2}
\end{equation}
means the  product 
of $k_a$ identical tensors with $k_a l_a$ different Lorentz indexes, which is
symmetrized in all indexes (but not divided by number of terms). 
\begin{equation}
\mathcal{N}(S_2) = \frac{(l_a k_a)!}{(l_a!)^{k_a} k_a!}
\label{d:NS2}
\end{equation}
And for the structure $S_3$ like $S_2$ but where $$\mu_1,...,\mu_{l_a}\in \{\mu_1...\mu_J \},$$ $l_a k_a\le J$ we have
\begin{equation}
\mathcal{N}(S_3) = \frac{J!}{(l_a!)^{k_a} k_a! (J- l_a k_a)!}.
\label{d:NS3}
\end{equation}

\subsection{\bf Transcendent functions}

In this article we use usual notations for transcendent functions:
\begin{itemize}
\item
Pochhammer symbol and its properties
\begin{equation}
(c)_n=
\begin{cases}
& c\cdot ...\cdot (c+n-1),\; n>0,\\
& 1,\; n=0.
\end{cases}
\label{pochhammer}
\end{equation}
\begin{equation}
(-N)_{2s}=2^{2s} (-N/2)_s ((1-N)/2)_s
\label{pochhammer1}
\end{equation}

\item Hypergeometric series
\begin{equation}
{}_pF_q(a_1,...,a_p;\, b_1,...,b_q;\, z)=\sum\limits_{n=0}^{\infty}
\frac{(a_1)_n...(a_p)_n}{(b_1)_n...(b_q)_n} \frac{z^n}{n!}.
\label{hypergeomseries}
\end{equation}

\item
Special case of Srivastava-Daoust function:
\begin{eqnarray}
F_{p}^{q} &&\hspace*{-4mm}
\left(
\begin{array}{c}
(a_1): \{A_1^{(1)}, \dots, A_1^{(N)}\} ,..., (a_q): \{A_q^{(1)},\dots A_q^{(N)}\}; \\
(b_1): \{B_1^{(1)}, \dots, B_1^{(N)}\} ,..., (b_p): \{B_p^{(1)},\dots B_p^{(N)}\};
\end{array}
\Bigg| z_1, \dots, z_N
\right) = \nonumber\\
&=& \sum_{m_1=0}^{\infty} \dots \sum_{m_N=0}^{\infty}
\frac{
	\prod_{j=1}^q (a_j)_{\sum_{k=1}^N A_j^{(k)} m_k}
}{
\prod_{j=1}^p (b_j)_{\sum_{k=1}^N B_j^{(k)} m_k}
} \prod_{k=1}^N
\frac{z_k^{m_k}}{m_k!}
\label{srivastavadaoust}
\end{eqnarray}

\begin{equation}
1+\sum_{j=1}^p B_j^{(k)}
-\sum_{j=1}^q A_j^{(k)} > 0
\end{equation}

\end{itemize}

\subsection{\bf Tricks for generating functions} 

If we have some generating function 
$$
f(x) = \sum_{n=0}^{\infty} a_n x^n
$$
then we can obtain useful relations which someone can apply to different recurrent relations:
\begin{eqnarray}
&& \sum_{n=0}^{\infty} a_{n+k} x^n = \frac{f(x) - \sum\limits_{i=0}^k a_i x^i}{x^k}\equiv f^+_k(x),\quad
 \sum_{n=k}^{\infty} a_{n-k} x^n = x^k f(x) \equiv f^-_k(x), \nonumber\\
&&  \sum_{n=0}^{\infty} n^m a_{n+k} x^n = \left\{
\sum\limits_{i=1}^m  \left\{ m\atop i\right\} x^i \frac{\partial^i}{\partial x^i} 
\right\} f^+_k(x), \nonumber\\
&&  \sum_{n=k}^{\infty} n^m a_{n-k} x^n = \left\{
\sum\limits_{i=1}^m  \left\{ m\atop i\right\} x^i \frac{\partial^i}{\partial x^i} 
\right\} f^-_k(x), \nonumber\\
&&  \left\{ m\atop i\right\} = 
\sum\limits_{j=0}^i \frac{(-1)^{i-j}j^m}{(m-j)!j!},\quad  \left\{ 0\atop 0\right\} =1,\nonumber\\
&& n^m=\sum\limits_{i=0}^m  \left\{ m\atop i\right\} (n)_m,\quad (n)_m =\frac{n!}{(n-m)!}
, \label{d:gentricks}
\end{eqnarray}
where $\left\{ m\atop i\right\}$ are Stirling numbers of the second kind.

\section{Examples for fixed spins}
\label{appendixE}

In the standard basis for $J_{1,2}=2$ we have
\begin{eqnarray}
 &\phantom{=}&\hspace*{-26mm}
 \mathcal{F}^{2,2}=
 \hat{f}^{2,2}_{0} \left( S^{F \{2,2\}}_{0\,\{0,0\}} 
     - \frac{S^{F \{2,2\}}_{0\,\{0,1\}}}{D-1}
     - \frac{S^{F \{2,2\}}_{0\,\{1,0\}}}{D-1}
     + \frac{S^{F \{2,2\}}_{0\,\{1,1\}}}{(D-1)^2} \right) + \nonumber \\
  &\phantom{=}&\hspace*{-18mm} +\hat{f}^{2,2}_{1} \left( 
     S^{F \{2,2\}}_{1\,\{0,0\}}- \frac{4 \chi_{m}}{D-1}\left(
      S^{F \{2,2\}}_{0\,\{0,1\}}+
     S^{F \{2,2\}}_{0\,\{1,0\}}\right) + \frac{4 \chi_{m}S^{F \{2,2\}}_{0\,\{1,1\}}}{(D-1)^2}
      \right) +\nonumber \\
  &\phantom{=}&\hspace*{-18mm} + \hat{f}^{2,2}_{2} \left( 
     S^{F \{2,2\}}_{2\,\{0,0\}}+
     \frac{2 \lambda_{m}}{D-1}\left(
 S^{F \{2,2\}}_{0\,\{0,1\}}+S^{F \{2,2\}}_{0\,\{1,0\}}
\right) 
    - \frac{2 (D-1 + \lambda_{m}) S^{F \{2,2\}}_{0\,\{1,1\}}}{(D-1)^2}
      \right) \hspace*{-2cm} \label{eq:calF22} 
\end{eqnarray}

\begin{eqnarray}
 \mathcal{W}^{2,2}&=&\lim_{\chi_m\to 1} \mathcal{F}^{2,2}\Big|_{F\to W}=\nonumber\\
&=& \hat{w}^{2,2}_{0} \left( S^{W \{2,2\}}_{0\,\{0,0\}} 
     - \frac{S^{W \{2,2\}}_{0\,\{0,1\}}}{D-1}
     - \frac{S^{W \{2,2\}}_{0\,\{1,0\}}}{D-1}
     + \frac{S^{W \{2,2\}}_{0\,\{1,1\}}}{(D-1)^2} \right) + \nonumber \\
  &+& \hat{w}^{2,2}_{1} \left( 
     S^{W \{2,2\}}_{1\,\{0,0\}}- \frac{4}{D-1}\left(
      S^{W \{2,2\}}_{0\,\{0,1\}}+
     S^{W \{2,2\}}_{0\,\{1,0\}}\right) + \frac{4 S^{W \{2,2\}}_{0\,\{1,1\}}}{(D-1)^2}
      \right)+ \nonumber \\
  &+& \hat{w}^{2,2}_{2} \left( 
     S^{W \{2,2\}}_{2\,\{0,0\}}
    - \frac{2 S^{W \{2,2\}}_{0\,\{1,1\}}}{D-1}
      \right) \label{eq:calW22} 
\end{eqnarray}

Here we omit arguments of $\mathcal{F}$, $\mathcal{W}$, $\hat{f}$, $\hat{w}$ for brevity.

In the harmonical basis we have the same expression for $\mathcal{F}$ and $\mathcal{W}$ 
\begin{eqnarray}
\mathcal{F}^{2,2}&=&
\hat{f}^{2,2}_{0} \left( S^{F \{2,2\}}_{0\,\{0,0\}} 
- \frac{S^{F \{2,2\}}_{0\,\{0,1\}}}{D-2}
- \frac{S^{F \{2,2\}}_{0\,\{1,0\}}}{D-2}
+ \frac{S^{F \{2,2\}}_{0\,\{1,1\}}}{(D-2)^2} \right) + 
\hat{f}^{2,2}_{1}  
S^{F \{2,2\}}_{1\,\{0,0\}} +\nonumber \\
&+& \hat{f}^{2,2}_{2} \left( 
S^{F \{2,2\}}_{2\,\{0,0\}}+
\frac{2}{(D-1)^2}\times\right.\nonumber\\
&&\phantom{+ \hat{f}^{2,2}_{2}
	()}
\times\left.\left[
(D-2) S^{F \{2,2\}}_{0\,\{0,0\}}-
S^{F \{2,2\}}_{0\,\{0,1\}}-S^{F \{2,2\}}_{0\,\{1,0\}}
-  D \,S^{F \{2,2\}}_{0\,\{1,1\}}
\right]
\phantom{\frac{2}{(D-1)^2}}\hspace*{-1.5cm}
\right)
\label{eq:calF22HARM} 
\end{eqnarray}
Let us remind you that structures $S^{T \vec{J}}_{k\,\vec{n}}$ are different in
different basises~(\ref{def:SW}).



\section*{Aknowledgements}

Author thanks to V.~A.~Petrov for useful discussions.


%
%

\end{document}